\documentclass[aps,prb,twocolumn,superscriptaddress,showpacs]{revtex4-1}
\usepackage{graphicx}
\usepackage{dcolumn}
\usepackage{bm}
\usepackage{amsmath}
\usepackage{amsfonts}
\usepackage{amssymb}
\usepackage{subfigure}
\usepackage{leftidx}

\usepackage{color}

\begin{document}

\title{Orbital-resolved vortex core states in FeSe Superconductors: calculation based on a three-orbital model}

\author{Q. E. Wang}
\affiliation{Department of Physics and Center of Theoretical and Computational Physics, The University of Hong Kong, Hong Kong, China}

\author{F. C. Zhang}
\email{fuchun@hku.hk}
\affiliation{Department of Physics and Center of Theoretical and Computational Physics, The University of Hong Kong, Hong Kong, China}
\affiliation{Department of Physics, Zhejiang University, Hangzhou, China}
\affiliation{Collaborative Innovation Center of Advanced Microstructures, Nanjing 210093, China}

\date{\today}

\begin{abstract}

We study electronic structure of vortex core states of FeSe superconductors based on a t$_{2g}$ three-orbital model by solving the Bogoliubov-de Gennes(BdG) equation self-consistently. The orbital-resolved vortex core states of different pairing symmetries manifest themselves as distinguishable structures due to different quasi-particle wavefunctions. The obtained vortices are classified in terms of the invariant subgroups of the symmetry group of the mean-field Hamiltonian in the presence of magnetic field. Isotropic $s$ and anisotropic $s$ wave vortices have $G_5$ symmetry for each orbital, whereas $d_{x^2-y^2}$ wave vortices show $G^{*}_{6}$ symmetry for $d_{xz/yz}$ orbitals and $G^{*}_{5}$ symmetry for $d_{xy}$ orbital. In the case of $d_{x^2-y^2}$ wave vortices, hybridized-pairing between $d_{xz}$ and $d_{yz}$ orbitals gives rise to a relative phase difference in terms of gauge transformed pairing order parameters between $d_{xz/yz}$ and $d_{xy}$ orbitals, which is essentially caused by a transformation of co-representation of $G^{*}_{5}$ and $G^{*}_{6}$ subgroup. The calculated local density of states(LDOS) of $d_{x^2-y^2}$ wave vortices show qualitatively similar pattern with experiment results. The phase difference of $\frac{\pi}{4}$ between $d_{xz/yz}$ and $d_{xy}$ orbital-resolved $d_{x^2-y^2}$ wave vortices can be verified by further experiment observation.

\end{abstract}


\maketitle

\section{Introduction}

Quantized vortices, as stable topological defects, observed in a variety of quantum systems such as superconductor and superfluid, are characterized by their nature of soliton solutions of dynamical systems\cite{N.D.Mermin}. Electronic structures of vortices in cuprate superconductors exhibit charging effects\cite{D.I.Khomskii, T.Nagaoka, Y.Chen} and are anisotropic due to $d_{x^2-y^2}$ wave pairing symmetry\cite{G. Blatter}. Earlier theoretical works have investigated the vortex line states based on microscopic models\cite{F. Gygi,Y.D.Zhu,Y.Wang,M.Takigawa}. In iron-based superconductors, band structure and multi-orbital pairings play an important role and the vortex structures may be richer due to multi-orbital dependency. Vortex core states of two-fold rotational symmetry, which is proposed to be attributed to the orbital-dependent reconstruction in FeSe superconductors\cite{F.C.Hsu}, have been reported by C. L. Song \emph{et al.} from scanning tunneling microscopy(STM) experiment \cite{C. L. Song}.

In most of iron-based superconductors, the Fermi surfaces consist of both electron pockets around M point at the corners and hole pockets around $\Gamma$ point of the folded Brilliouin zone (BZ).  An $s_{\pm}$ wave superconducting (SC) pairing symmetry has been proposed, where the pairing order parameters at the electron and hole pockets have opposite signs\cite{K.Kuroki,I.I.Mazin,P.J.Hirschfeld,A.Chubukov,Fengjie Ma,Qimiao Si,Wei-Qiang Chen,Kangjun Seo,Fa Wang,Xianhui Chen}. FeSe superconductor is interesting for its unique electronic structure in which only electron pockets are found and the hole pockets are well below the Fermi level, as angle-resolved photoemission spectroscopy(ARPES) shows\cite{D.F.Liu}. When the hole pocket at $\Gamma$ vanishes, we don't have $s_{\pm}$ wave pairing state anymore and the s and $d_{x^2-y^2}$ wave pairing states should be considered\cite{R.Yu,Y.Zhou}. The iron-based superconductivity without hole pocket is a great challenge to the weak-coupling theory where the superconductivity is proposed to be driven by the nesting of the electron and hole Fermi surfaces. The absence of the hole pocket in FeSe makes the argument difficult.

Vortex structures in iron-based superconductors have been studied by a number of authors\cite{X.Hu,T.Zhou,H.H.Huang,D.Wang,Araujo}. These studies are mainly based on band structures having hole pocket at $\Gamma$. We expect the vortex structure be affected by the Fermi surface topology. In this work we use a three-orbital model to study vortex structure of FeSe SC state. The three-orbital microscopic model reproduces qualitatively the correct Fermi surface with only electron pockets.  We solve the BdG equation self-consistently to study orbital-resolved vortex core states for various SC pairing symmetries: isotropic s(on-site pairing), anisotropic s(next nearest neighbor site pairing), and $d_{x^2-y^2}$(nearest neighbor site pairing) waves. We compare results of calculations with that observed from the recent STM experiment on FeSe vortex and suggest that the pairing symmetry to be $d_{x^2-y^2}$ wave. We predict that there is a relative phase difference about $\frac{\pi}{4}$ between pairing order parameters defined on $d_{xz/yz}$ and $d_{xy}$ orbitals in the case of $d_{x^2-y^2}$ orbital-resolved vortices, while such a phase difference is trivial in the case of isotropic $s$ and anisotropic $s$ wave vortices. The paper is organized as follows. An introduction of magnetic translation group and classification of vortex solutions are given in Section \ref{sec2}. In section \ref{sec3} we present the three-orbital model and the self-consistent BdG approach. In Section \ref{sec4}, we discuss properties of the vortex core states for different pairing symmetries and compare our results with experimental observations. Finally, a summary is given in Section \ref{sec5}.

\section{Magnetic translational symmetry and winding structures of single vortex}\label{sec2}

From a theoretical point of view, vortex lattice in mixed states of type II superconductors is ground state of fermionic system which is characterized by interaction between a homogenous magnetic field with $C_{\infty}$ symmetry and Cooper pairs with a definite SC pairing symmetry\cite{Abrikosov}. In iron-based superconductors, situation becomes complicated because of orbital degrees of freedom. Consequently, the crystal symmetry, band structure, and SC pairing symmetry, determine the electronic structure of vortices. Among these constraints of symmetry, the vortex structures are mainly dominated by magnetic translation invariance, whose generator are crystal momentum and vector potential of magnetic field\cite{E.Brown}. However, such conventional magnetic translation group defines a magnetic unit cell containing two vortices. It is not the symmetry group of Abrikosov lattice in which only single vortex is stabilized within one magnetic unit cell. Breakthrough of this difficulty was presented by M. Ozaki \emph{et al.}\cite{M.Ozaki1}. In their work the magnetic translation group describing single vortex was discovered to be a subgroup of direct product of conventional magnetic translation group and gauge transformation group U(1). Therefore, stable vortex structure can be solved numerically in one magnetic unit cell taking advantages of nontrivial winding boundary conditions derived from properties of magnetic translation group\cite{M.Ozaki2}.

Instead of doing calculations of two vortices in one magnetic unit cell, we follow the method given by M. Ozaki \emph{et al.}\cite{M.Ozaki1,M.Ozaki2}, in which only single vortex structures are calculated in one magnetic unit cell, so that the calculated results can be classified by irreducible representations of magnetic translation group. The numerical calculations in previous works, as mentioned above\cite{X.Hu,T.Zhou,H.H.Huang}, are mostly carried out for two vortices in one magnetic unit cell. These vortex states, however, can not be identified by invariant subgroups of magnetic translation group because they belong to the irreducible representations of conventional magnetic translation group. Furthermore, two vortices in one magnetic unit cell are not independent because the induction of interaction between them. It is well-known that the topological defects in unconventional superconductors and superfluids with certain symmetry breaking behave distinguishably from the conventional singular(hard core) vortices\cite{M.M.Salomaa}. For instance, a vortex in ${^3}$He has a finite amplitude of order parameters in the soft core region whose size is larger than the coherent length, whereas the winding structure is non-trivial. Therefore in our numerical calculation, we concentrate on winding structures of vortices for each orbitals, although the vortices in iron-based superconductors are mostly of hard core feature, and classify the vortex structures of isotropic $s$, anisotropic $s$, and $d_{x^2-y^2}$ wave pairing symmetries in terms of invariant subgroups of magnetic translation group. Special attention will be paid to the difference of vortex states defined between $A_{1g}$(isotropic $s$ and anisotropic $s$ wave) and $B_{1g}$($d_{x^2-y^2}$ wave) irreducible unitary representations of $D_{4}$ group.

The Hamiltonian of the SC system in the presence of a homogeneous magnetic field along $\hat{z}$ direction is obtained from its zero-field form by modifying the hopping and pairing terms with Peierls phase\cite{Peierls}, respectively, which is of the following form
\begin{flalign}\label{eq1}
\begin{split}
H&=H_{0}+H_{pair}\\
H_{0}&=\sum_{i,j,\alpha,\beta,\sigma}[\tilde{t}_{\sigma\sigma}(i\alpha,j\beta)-\mu\delta_{ij}\delta_{\alpha\beta}]a_{i\alpha\sigma}^{\dag}a_{j\beta\sigma}\\
H_{pair}&=\sum_{i,j,\alpha,\beta}[\tilde{\Delta}_{\uparrow\downarrow}(i\alpha,j\beta)a_{i\alpha\uparrow}^{\dag}a_{j\beta\downarrow}^{\dag}+h.c.]
\end{split}
\end{flalign}
in which
\begin{flalign}\label{eq2}
\begin{split}
\tilde{t}_{\sigma\sigma}(i\alpha,j\beta)&=t_{\sigma\sigma}(i\alpha,j\beta)\exp[\frac{i e}{\hbar c}\int^{i}_{j}\vec{A}(\vec{r})\cdot
d\vec{r}]\\
\tilde{\Delta}_{\uparrow\downarrow}(i\alpha,j\beta)&=\Delta_{\uparrow\downarrow}(i\alpha,j\beta)\exp[i\phi(i,j)]
\end{split}
\end{flalign}
where $a_{i\alpha\sigma}^{\dag}$($a_{i\alpha\sigma}$) denotes the creation(annihilation) operator of electrons with spin $\sigma =\uparrow,\downarrow$ and orbital $\alpha$ at site $i$. $t_{\sigma\sigma}(i\alpha,j\beta)$ are hopping integrals and $\mu$ is the chemical potential. We assume that the screening magnetic field inside the superconductor can be neglected except for that the magnetic field is close to the upper critical field. The SC pairing mechanism has been proposed to be of magnetic origin. In this paper, however, we shall focus on the vortex core state and start from an extended attractive Hubbard model for simplicity. The SC order parameter stemming from the mean-field decoupling of the paired scattering term is expressed as $\Delta_{\uparrow\downarrow}(i\alpha,j\beta)=V_{\uparrow\downarrow}(i\alpha,j\beta)\langle a^{\ }_{j\beta\downarrow}a^{\ }_{i\alpha\uparrow}\rangle$ for singlet pairing channel. The Peierls phase\cite{Peierls} in hopping terms comes from the fact that the Lagrangian of electron in a magnetic field contains a dynamical term $\frac{e}{c}\vec{v}\cdot\vec{A}$, which gives rise to the phase accumulation in the propagator of electron describing the hopping process between two lattice sites. The modification of pairing order parameters accounts for eliminating the mixing of different pairing states under the action of magnetic translation group. The mathematical interpretation of doing this is essentially searching for gauge transformed order parameters, which span a representation of magnetic translation group\cite{M.Ozaki1, M.Ozaki2}. The gauge transformation, carried out by phase $\phi(i,j)$, has different definition with respect to anisotropic $s$ and $d_{x^2-y^2}$ wave pairing states, whereas in the case of isotropic $s$ wave pairing it is trivial. The gauge transformed order parameter for $d_{x^2-y^2}$ wave pairing has been derived by means of group theoretical analysis\cite{M.Ozaki1,M.Ozaki2}. The magnetic translation operator takes following form in symmetric gauge $\vec{A}=-\frac{1}{2}\vec{r}\times\vec{B}$, when it acts on creation operators\cite{E.Brown, M.Ozaki1}
\begin{flalign}\label{eq3}
\begin{split}
L(\vec{R}_{\lambda})a_{i\alpha\sigma}^{\dag}&=e^{i\frac{\pi}{2}(\mathcal{N}_{v}\lambda_{x}\lambda_{y})}T(\vec{R}_{\lambda})a_{i\alpha\sigma}^{\dag} \\
&=e^{i\frac{\pi}{2}\mathcal{N}_{v}[\lambda_{x}\lambda_{y}+\frac{1}{N}
(\lambda_{x}i_{y}-\lambda_{y}i_{x})]}a_{i+\lambda,\alpha\sigma}^{\dag}
\end{split}
\end{flalign}
and the resultant transformation of order parameter is
\begin{flalign}\label{eq4}
&\langle a^{\ }_{j+\lambda,\beta\downarrow} a^{\
}_{i+\lambda,\alpha\uparrow}\rangle \\ \nonumber
&=e^{i\pi\mathcal{N}_{v}[\lambda_{x}\lambda_{y}+\frac{1}{2N}
\lambda_{x}(i_{y}+j_{y})-\frac{1}{2N}\lambda_{y}(i_{x}+j_{x})]}\langle
a^{\ }_{j\beta\downarrow}a^{\ }_{i\alpha\uparrow}\rangle
\end{flalign}
where $\vec{R}_{\lambda}=\lambda_{x}N\hat{x}+\lambda_{x}N\hat{y}$ is the basis vector of magnetic unit cell containing N lattice sites and $\mathcal{N}_{v}$ is the number of vortices within one magnetic unit cell. We have restricted ourselves to the cases of square vortex lattice with lattice constant set to unity. Eq. \eqref{eq3} defines actions of magnetic translation group $\{L(\vec{R}_{\lambda})\}$ on field operators, and all of the operations form a group in representation space spanned by gauge transformed order parameters, provided that certain group condition is satisfied. Note that the gauge transformation, as an internal symmetry transformation, takes its complex conjugate form when acts on annihilation operators. Different from the situation for conventional magnetic translation group\cite{E.Brown} $\{T(\vec{R}_{\lambda})\}$: $\mathcal{N}_{v}=2$, the group condition of magnetic translation group, which is the symmetry group of Abrikosov lattice, is that only single magnetic flux $\varphi_{0}=\frac{\emph{h}c}{2e}$ is contained in one magnetic unit cell\cite{M.Ozaki1,M.Ozaki2}, i.e., $\mathcal{N}_{v}=1$. It has been pointed out that $d_{x^2-y^2}$ $\sim\cos(k_x)-\cos(k_y)$ wave order parameters will mix with extended $s^{*}$ $\sim\cos(k_x)+\cos(k_y) $, $p_x$ $\sim i\sin(k_x)$, and $p_y$ $\sim i\sin(k_y)$ wave order parameters under operation of magnetic translation group\cite{M.Ozaki1,M.Ozaki2}. Such a mixing originates from the fact the symmetry group of normal state Hamiltonian contains a local gauge transformation generated by the vector potential of a magnetic field. The re-defined SC gauge transformed order parameters transforming according to invariant subgroups of $D_{4}$ group without any gauge component, as order parameters do in the absence of magnetic field, are obtained by generating all of them with the action of a conjugate rotation subgroup $\{C_{4z}^{k}(i_{x},j_{y}), k= 1,2,3,4\}$ on one of the pairing bonds of every local order parameters accompanied by a Peierls phase factor\cite{Peierls}. The generator of conjugate rotation subgroup is defined as
\begin{flalign}\label{eq5}
C_{4z}(i_{x},j_{y})=T(i_{x},j_{y})C_{4z}T^{-1}(i_{x},j_{y})
\end{flalign}
where $C_{4z}$ is 4-fold rotation around the origin of the coordinate system. Therefore the mixing of order parameters under magnetic translation is eliminated by re-defining rotations of all local order parameters at different sites back to origin. The $d_{x^2-y^2}$ wave gauge transformed order parameter is consequently re-defined as
\begin{flalign}\label{eq6}
&\tilde{\Delta}^{d_{x^2-y^2}}_{\uparrow\downarrow}(i\alpha,j\beta)\\
\nonumber &=\frac{V_{\uparrow\downarrow}(j\beta,i\alpha)}{2}\langle
a_{i\alpha\downarrow}a_{j\beta\uparrow}\rangle(e^{\pm
iKi_{y}}\delta_{i\pm\hat{x},j}-e^{\mp
iKi_{x}}\delta_{i\pm\hat{y},j})
\end{flalign}
where $K=\frac{\pi\mathcal{N}_{v}}{2N^{2}}$ and $\hat{x}$($\hat{y}$) denote the unit vectors of two-dimensional lattice. Note that for
singlet pairing the order parameters are symmetric under exchange of site-orbital quantum number.

Here we follow method given by M. Ozaki \emph{et al.}\cite{M.Ozaki1,M.Ozaki2} to derive the gauge transformed order parameters for anisotropic $s \sim \cos(k_x)\cdot\cos(k_y)$ wave pairing symmetry. The results of action of conjugate rotation subgroup on pairing bond along $\hat{x}+\hat{y}$ direction are
\begin{flalign}\label{eq7}
\begin{split}
&C_{4z}(i_x,i_y)\langle
a_{i\alpha\downarrow}a_{i+\hat{x}+\hat{y},\beta\uparrow} \rangle \\
&=e^{-2iKi_{y}}\langle
a_{i\alpha\downarrow}a_{i-\hat{x}+\hat{y},\beta\uparrow} \rangle \\
&C_{2z}(i_x,i_y)\langle
a_{i\alpha\downarrow}a_{i+\hat{x}+\hat{y},\beta\uparrow} \rangle \\
&=e^{2iK(i_{x}-i_{y})}\langle
a_{i\alpha\downarrow}a_{i-\hat{x}-\hat{y},\beta\uparrow} \rangle \\
&C^{3}_{4z}(i_x,i_y)\langle
a_{i\alpha\downarrow}a_{i+\hat{x}+\hat{y},\beta\uparrow} \rangle \\
&=e^{2iKi_{x}}\langle
a_{i\alpha\downarrow}a_{i+\hat{x}-\hat{y},\beta\uparrow} \rangle
\end{split}
\end{flalign}
then a symmetric phase rearrangement can be made by multiplying a Peierls phase $e^{iK(i_y-i_x)}$ to regain the magnetic translational symmetry as following

\begin{flalign}\label{eq8}
\begin{split}
&\tilde{\Delta}^{anis. \ s}_{\uparrow\downarrow}(i\alpha,j\beta)\\
&=\frac{V_{\uparrow\downarrow}(j\beta,i\alpha)}{4}\langle
a_{i\alpha\downarrow}a_{j\beta\uparrow}\rangle[e^{iK(i_{y}-i_{x})}\delta_{i+\hat{x}+\hat{y},j}\\
&+e^{-iK(i_{x}+i_{y})}\delta_{i-\hat{x}+\hat{y},j}\\
&+e^{-iK(i_{y}-i_{x})}\delta_{i-\hat{x}-\hat{y},j}+e^{iK(i_{x}+i_{y})}\delta_{i+\hat{x}-\hat{y},j}]
\end{split}
\end{flalign}
The magnetic translation property of gauge transformed order parameters for anisotropic $s$ wave pairing state, which is consistent with $d_{x^2-y^2}$ wave, is
\begin{flalign}\label{eq9}
\begin{split}
&\tilde{\Delta}^{anis. \ s}_{\uparrow\downarrow}(i+\lambda,\alpha,j+\lambda,\beta) \\
&=e^{i\pi\mathcal{N}_{v}[\lambda_{x}\lambda_{y}+\frac{1}{N}
(\lambda_{x}i_{y}-\lambda_{y}i_{x})]}\tilde{\Delta}^{anis. \ s}_{\uparrow\downarrow}(i\alpha,j\beta)
\end{split}
\end{flalign}
where $j$ is always related to $i$ as next nearest neighbor site pairing. Compare this expression with Eq. \eqref{eq4}, it is obvious that the gauge transformed order parameters(referring to order parameters thereafter) now form a basis of representation of magnetic translation group and the mixing between anisotropic $s$ and $d_{xy}$ wave pairing states under action of magnetic translation group has been eliminated.

\begin{figure}[!ht]
\begin{center}
\subfigure[]{\includegraphics[width=100pt]{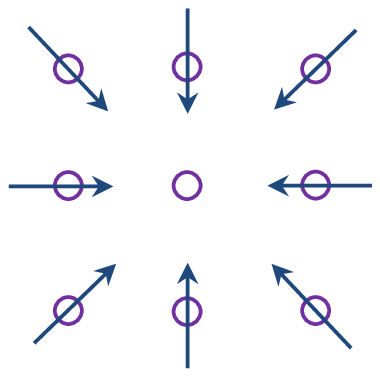}}
\subfigure[]{\includegraphics[width=100pt]{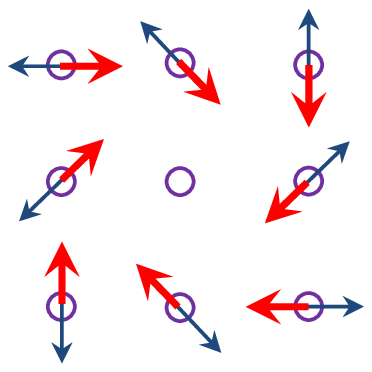}}
\caption{(color online) Schematic pictures showing the phase difference between $G^{*}_{5}$ (a) and $G^{*}_{6}$ (b) winding structures in the vicinity of the vortex core center\cite{M.Ozaki1}. The purple circles depict lattice sites on which the SC order parameters are defined and the arrows show the phase distribution of $\Delta_{\uparrow\downarrow}(i\alpha,j\beta)$. The blue arrows in (b) denote phase difference of $-\frac{3\pi}{4}$ and red arrows $\frac{\pi}{4}$ between $G^{*}_{6}$ and $G^{*}_{5}$ winding structures.} \label{figs}
\end{center}
\end{figure}

The SC ground states, in the absence of magnetic field, can be classified by finding all the invariant subgroups of the symmetry group $D_{4}\otimes U(1)$, which have a one-to-one correspondence to the irreducible unitary representations of the symmetry group of normal state Hamiltonian\cite{M.Sigrist, G.E.Volovik}. In the case of $D_{4}$ point group symmetry, such a classification is obtained by the fact that $D_{4}$ has three invariant subgroups of index 2, and the two dimensional cyclic group, as a subgroup of $U(1)$, compensate the phase change of order parameters by $e^{i\pi}$ when the elements of coset representative acts on them. In the same manner, the ground state of a vortex structure can also be classified by finding all the invariant subgroups of symmetry group of the Hamiltonian in a magnetic field\cite{M.Ozaki1}, and consequently the winding structure of the vortex core states have symmetry constraints of different classes. The topological characteristics of vortex states are location of pinning center, phase distribution of order parameters, and winding number. It turns out that the winding number of vortices of different symmetry properties, having a structural vanishing region, can be calculated from the symmetry constraints of corresponding maximal little groups. In work of M. Ozaki \emph{et al.}\cite{M.Ozaki1,M.Ozaki2}, winding numbers $\mathcal {W}$ of $s^{*}$ and $d_{x^2-y^2}$ wave vortices have been calculated. Here we calculate $\mathcal {W}$ for anisotropic $s$ and $d_{xy}$ wave states and list all the results in Table \ref{table:wd}, in which
\begin{flalign}\label{eq10}
\begin{split}
&G^{l}=(e+tC_{2x})\tilde{C}^{l}\wedge L\\
&\tilde{C}^{l}=\{e^{-\frac{\pi}{2}l k}C^{k}_{4z}, k=1,2,3,4\}
\end{split}
\end{flalign}
Note that $\tilde{C}^{l}$ always acts on paired field operators rather than single particle operator. The derivation is based on the fact that the generator of $\tilde{C}^{l}$, as a symmetry transformation of order parameters, leaves them invariant\cite{M.Ozaki2}. The winding structures of $G^{*}_{5}$ and $G^{*}_{6}$ vortices, which have been obtained from our numerical calculations, are shown in Fig. \ref{figs}, respectively, where they differ by a co-representation transformation as
\begin{flalign}\label{eq10-1}
\begin{split}
&G^{*}_{6}= \left(\widehat{\frac{3\pi}{4}}\right)^{-1}G^{*}_{5}\widehat{\frac{3\pi}{4}} \\
&G^{*}_{6}= \widehat{\frac{\pi}{4}}G^{*}_{5}\left(\widehat{\frac{\pi}{4}}\right)^{-1}
\end{split}
\end{flalign}
The gauge transformation of field operator is defined as $\widehat{\phi}\cdot a_{i\alpha\sigma}=e^{-i\frac{\phi}{2}}a_{i\alpha\sigma}$\cite{M.Ozaki1}. Note that the global gauge transformation of $-\frac{3\pi}{4}$ or $\frac{\pi}{4}$ are both allowed by group theory. But it turns out from our numerical calculation that the phase difference of $\frac{\pi}{4}$ is more energetically favorable.

\begin{table}[!t]
\caption{Winding number of order parameters for different pairing states. $G_{i},i=1,2,3,4,5,6$ are six maximal little groups. $G^{*}_{5,6}$ differs from $G_{5,6}$ by taking the complex conjugate of gauge transformation. The order parameters transform according to basis functions of $D_{4}$ group as s wave: $\sim const.$, anisotropic s wave: $\sim\cos(k_x)\cos(k_y)$, extended $s^{*}$ wave: $\sim\cos(k_x)+\cos(k_y)$, $d_{x^2-y^2}$ wave: $\sim\cos(k_x)-\cos(k_y)$, and $d_{xy}$ wave: $\sim\sin(k_x)\sin(k_y)$, respectively. The index $l$ is defined in Eq. \eqref{eq10}.}\label{table:wd}
\begin{center}
\begin{tabular}{c|ccccccc}
 \hline\hline
&$l$  ~~&$\mathcal{W}$($s$, anis. $s$, $s^{*}$)  ~~& $\mathcal {W}$($d_{x^2-y^2}$, $d_{xy}$)  \\
 \hline
$G_{1}\sim G_{2}$         ~~&$0$  ~~&$0$       ~~&$2$ or $-2$   \\
$G_{5}\sim G_{6}$         ~~&$1$  ~~&$1$       ~~&$3$ or $-1$   \\
$G^{*}_{5}\sim G^{*}_{6}$ ~~&$-1$ ~~&$-1$      ~~&$1$ or $-3$   \\
$G_{3}\sim G_{4}$         ~~&$2$  ~~&$2$       ~~&$4$ or $0$    \\
\hline\hline
\end{tabular}
\end{center}
\end{table}

\section{Methodology and Band Model}\label{sec3}

\begin{figure}[t]
\begin{center}
\subfigure[]{\includegraphics[width=220pt]{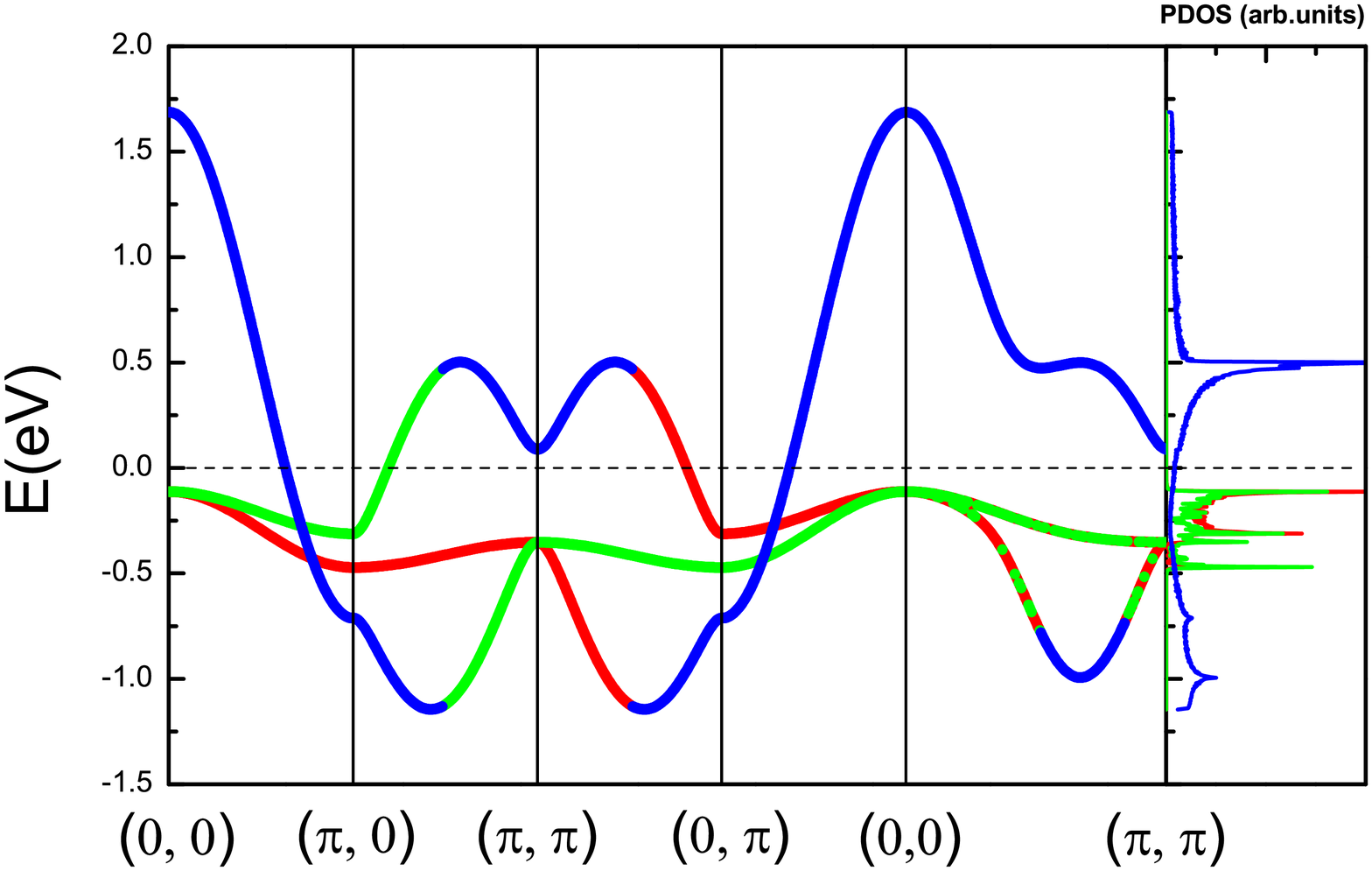}}
\subfigure[]{\includegraphics[width=160pt]{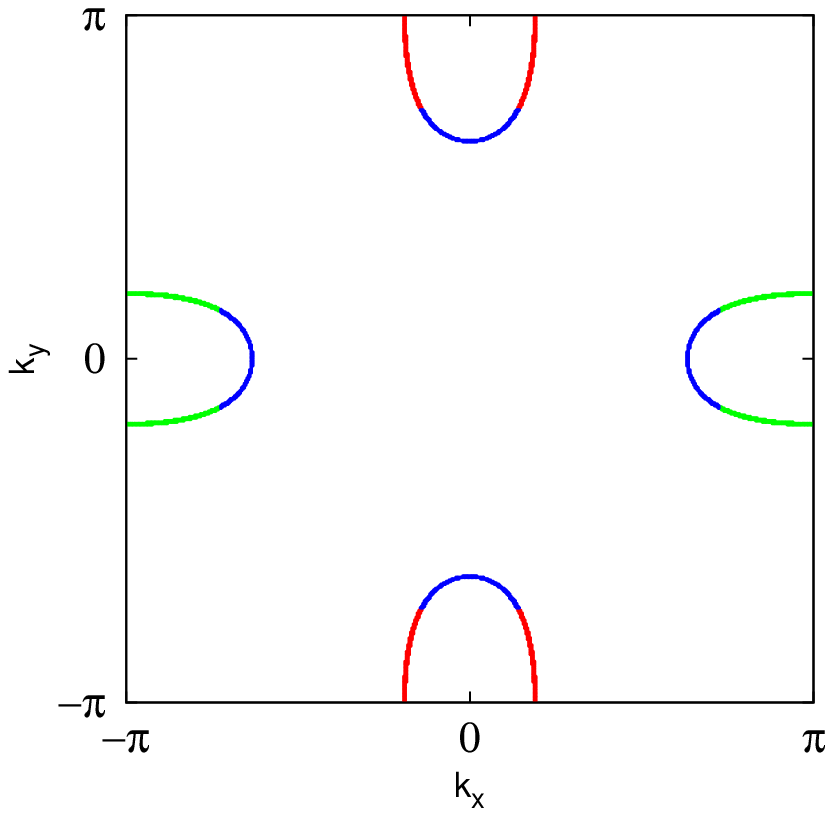}}
\caption{(color online) Orbital-resolved band structure, PDOS (a)
and Fermi Surface (b). The red $(d_{xz})$, green $(d_{yz})$, and
blue $(d_{xy})$ curves represent wight-dominating orbitals. The
Fermi level has been set to zero.} \label{fig1}
\end{center}
\end{figure}

It has been reported that the electronic structure of iron-based superconductors in the vicinity of the Fermi level is dominated by $d_{xz}$, $d_{yz}$, and $d_{xy}$ orbitals from first-principle calculation\cite{I.R.Shein}, therefore it is feasible to calculate the vortex core states based on an effective three-orbital model\cite{C.Fang}. Taking advantage of the 4-fold rotational symmetry, the Bl\"{o}ch Hamiltonian can be written as following
\begin{flalign}\label{eq11}
\begin{split}
H_{0}=&\sum_{k}\psi^{\dag}(k)M(k)\psi(k)\\
M(k)=&K_0+K_{1}e^{ik_x}+C_{4z}K_{1}C_{4z}^{3}e^{ik_y}\\
&+C_{2z}K_{1}C_{2z}e^{-ik_x}+C_{4z}^{3}K_{1}C_{4z}e^{-ik_y}\\
&+K_{2}e^{i(k_x+k_y)}+C_{4z}K_{2}C_{4z}^{3}e^{i(-k_x+k_y)}\\
&+C_{2z}K_{2}C_{2z}e^{i(-k_x-k_y)}+C_{4z}^{3}K_{2}C_{4z}e^{i(k_x-k_y)}
\end{split}
\end{flalign}
where
$\psi^{\dag}(k)=[a_{xz}^{\dag}(k),a_{yz}^{\dag}(k),a_{xy}^{\dag}(k)]$
and the 4-fold rotation is carried out by one of the generators of
$D_{4}$ group
\begin{flalign}\label{eq12}
C_{4z}=\left(
  \begin{array}{ccc}
    0 &-1 & 0 \\
    1 & 0 & 0 \\
    0 & 0 & 1 \\
  \end{array}
\right)
\end{flalign}
The irreducible hopping subsets\cite{F.Wang}(in unit: eV) corresponding to on-site atomic energies, hopping along $\hat{x}$, and $\hat{x}+\hat{y}$ directions are
\begin{flalign}\label{eq13}
\begin{split}
K_{0}&=\textrm{diag}(-\mu,-\mu,0.4-\mu)\\
K_{1}&=\left(
        \begin{array}{ccc}
        0.05& 0.00& -0.20 \\
        0.00& 0.01&  0.00 \\
        0.20& 0.00&  0.20
         \\
        \end{array}
        \right)\\
K_{2}&=\left(
        \begin{array}{ccc}
        0.02& 0.01& 0.10 \\
        0.01& 0.02& 0.10 \\
       -0.10&-0.10& 0.20 \\
        \end{array}
        \right)
\end{split}
\end{flalign}
For simplicity, the spin indices have been dropped. Instead of going along the boundary of the irreducible BZ, an alternative path has been used to show the band structure with dominating orbital weights in Fig. \ref{fig1} (a). The projected density of states(PDOS) reveals strongly-hybridized bands which are composed of $d_{xz}$ and $d_{yz}$ orbitals along the off-diagonal line of the extend BZ below the Fermi level. The Fermi surface (Fig. \ref{fig1} (b)), obtained with a chemical potential $\mu$=0.312 eV corresponding to a filling factor $n$=4.23, has four electron pockets which do not have any SC gap node in cases of anisotropic $s$ and $d_{x^2-y^2}$ wave pairing sates. The absence of electron or hole pockets at $\Gamma$ point is consistent with experimental observation\cite{D.F.Liu}.

The Hamiltonian in Eq. \eqref{eq1} can be diagonalized by conducting the Bogoliubov-Valatin transformation\cite{N.N.Bogoliubov,J.G.Valatin} containing $t_{2g}$ orbital degrees of freedom as
\begin{flalign}\label{eq14}
a_{i\alpha\sigma}=\sum_{\epsilon_{n\uparrow}>0}u^{n}_{i\alpha\sigma\sigma}\gamma_{n\sigma}+\bar{\sigma}v^{n*}_{i\alpha\sigma\bar{\sigma}}\gamma^{\dag}_{n\bar{\sigma}}
\end{flalign}
where the quasiparticle creation operator $\gamma_{n\sigma}^{\dag}$ is the ladder operator of the eigen-spectrum of the Hamiltonian which satisfies $[H,\gamma_{n\sigma}^{\dag}]_{-}=\epsilon_{n\sigma}\gamma_{n\sigma}^{\dag}$. The diagonal condition of the Hamiltonian is the BdG equation
\begin{flalign}\label{eq15}
\sum_{j,\beta}\left[
               \begin{array}{cc}
                  \tilde{h}_{\uparrow\uparrow}(i\alpha,j\beta) & \tilde\Delta_{\uparrow\downarrow}(i\alpha,j\beta) \\
                 \tilde\Delta^{*}_{\uparrow\downarrow}(i\alpha,j\beta) & -\tilde{h}^{*}_{\downarrow\downarrow}(i\alpha,j\beta) \\
               \end{array}
             \right]\left[
                      \begin{array}{c}
                        u_{j\beta\uparrow\uparrow}^n \\
                        v_{j\beta\downarrow\uparrow}^n \\
                      \end{array}
                    \right] =
                    \epsilon_{n\uparrow}\left[
                                          \begin{array}{c}
                                            u_{i\alpha\uparrow\uparrow}^n \\
                                            v_{i\alpha\downarrow\uparrow}^n \\
                                          \end{array}
                                        \right]
\end{flalign}
where $\tilde{h}_{\sigma\sigma}(i\alpha,j\beta)=\tilde{t}_{\sigma\sigma}(i\alpha,j\beta)-\mu\delta_{ij}\delta_{\alpha\beta}$ and the order parameters defined on different orbitals are
\begin{flalign}\label{eq16}
\tilde\Delta_{\uparrow\downarrow}(i\alpha,j\beta)=-\frac{V_{\uparrow\downarrow}(i\alpha,j\beta)}{2}\sum_{\epsilon_{n\uparrow}>0,<0}u_{i\alpha\uparrow\uparrow}^{n}v_{j\beta\downarrow\uparrow}^{n*}\tanh(\frac{\epsilon_{n\uparrow}}{2k_{B}T})
\end{flalign}
Eq. \eqref{eq3} and \eqref{eq14} give a nontrivial winding boundary condition to quasi-particle amplitudes as
\begin{flalign}\label{eq17}
\left[
\begin{array}{c}
u_{i+\lambda,\alpha\uparrow\uparrow}^n \\
v_{i+\lambda,\alpha\downarrow\uparrow}^n \\
\end{array}
\right]=
\left[
\begin{array}{c}
e^{i\frac{\pi}{2}\mathcal{N}_{v}[\lambda_{x}\lambda_{y}+\frac{1}{N}(\lambda_{x}i_{y}-\lambda_{y}i_{x})]}u_{i\alpha\uparrow\uparrow}^n \\
e^{-i\frac{\pi}{2}\mathcal{N}_{v}[\lambda_{x}\lambda_{y}+\frac{1}{N}(\lambda_{x}i_{y}-\lambda_{y}i_{x})]}v_{i\alpha\downarrow\uparrow}^n \\
\end{array}
\right]
\end{flalign}

The order parameters are calculated by BdG equation self-consistently with the above boundary condition, which is assigned to the matrix element $\tilde{h}_{\sigma\sigma}(i\alpha,j\beta)$ and $\tilde\Delta_{\uparrow\downarrow}(i\alpha,j\beta)$ for $\mathcal{N}_{v}=1$. The self-consistent calculation starts with arbitrarily distributed order parameters and the iteration is performed with a
convergence criterion that the order parameters have relative difference less that $10^{-3}$ between two consecutive steps. The particle density are calculated via quasi-particle wavefunctions as

\begin{flalign}\label{eq18}
\begin{split}
\langle n_{i\alpha\uparrow} \rangle=& \frac{1}{2}
\sum_{\epsilon_{n\uparrow}>,<0}|u_{i\alpha\uparrow\uparrow}^{n}|^2[1-\tanh(\frac{\epsilon_{n\uparrow}}{2k_{B}T})]
\\
\langle n_{i\alpha\downarrow} \rangle=& \frac{1}{2}
\sum_{\epsilon_{n\uparrow}>,<0}|v_{i\alpha\downarrow\uparrow}^{n}|^2[1+\tanh(\frac{\epsilon_{n\uparrow}}{2k_{B}T})]
\end{split}
\end{flalign}
The energy spectrum of the quasi-particle, i.e., the LDOS at site $i$ for orbital $\alpha$ is calculated via
\begin{flalign}\label{eq19}
\rho_{i\alpha}(\epsilon)=& \frac{1}{M_{x}M_{y}}\sum_{\vec{k}\in
FBZ}\sum_{\epsilon_{n\uparrow}>,<0}|u_{i\alpha\uparrow\uparrow}^{n}|^2\delta[\epsilon-\epsilon_{n\uparrow}(\vec{k})]
\\ \nonumber
&+|v_{i\alpha\downarrow\uparrow}^{n}|^2\delta[\epsilon+\epsilon_{n\uparrow}(\vec{k})]
\end{flalign}
where the supercell method has been used\cite{J.X.Zhu} for $M_{x}=M_{y}=10$. The Lorentzian smearing method is used to visualize the LDOS with a broadening width $\sigma=0.001$. All the self-consistent calculations are performed on a 28$\times$28 lattice at temperature T$=0.1$K.

\begin{figure}[!t]
\begin{center}
\includegraphics[width=180pt]{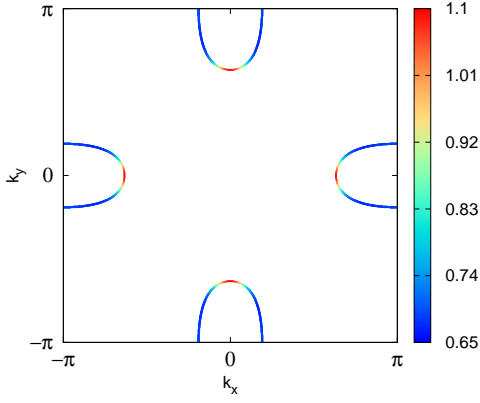}
\caption{(color online) Color mapping of Fermi velocity $\hbar
v_{F}$(in unit eV$\cdot$m).} \label{fig2}
\end{center}
\end{figure}
\begin{figure}[t]
\begin{center}
\subfigure[]{\includegraphics[width=120pt]{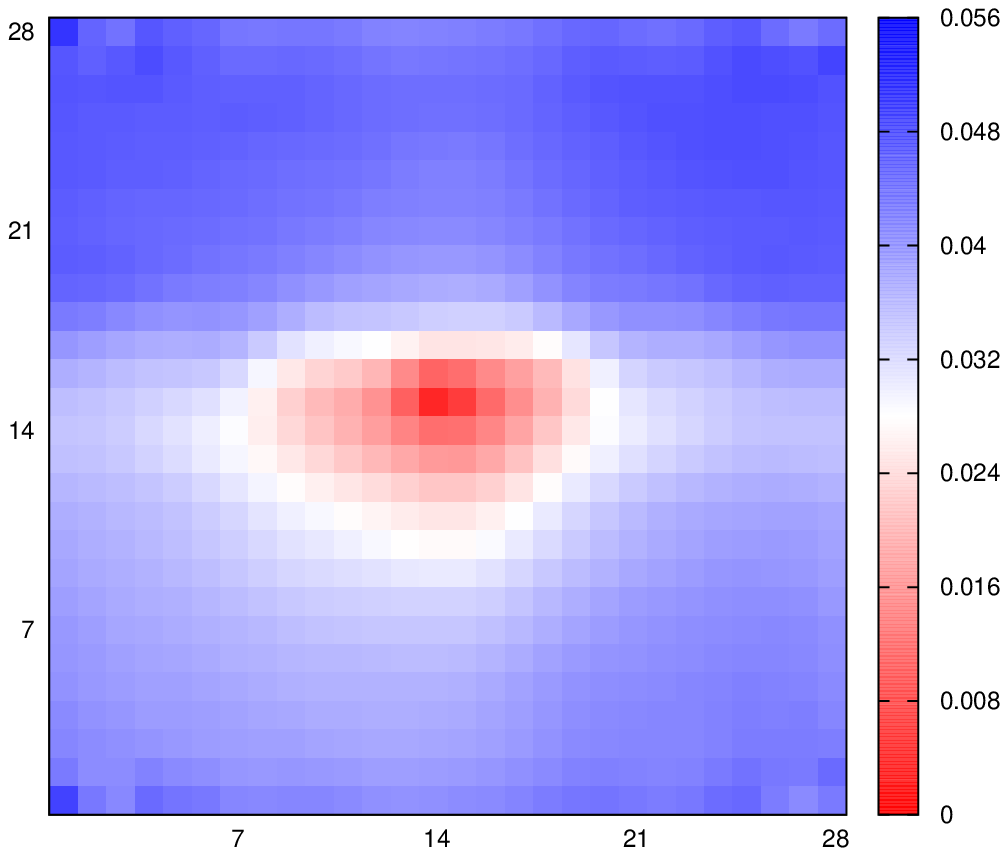}}
\subfigure[]{\includegraphics[width=120pt]{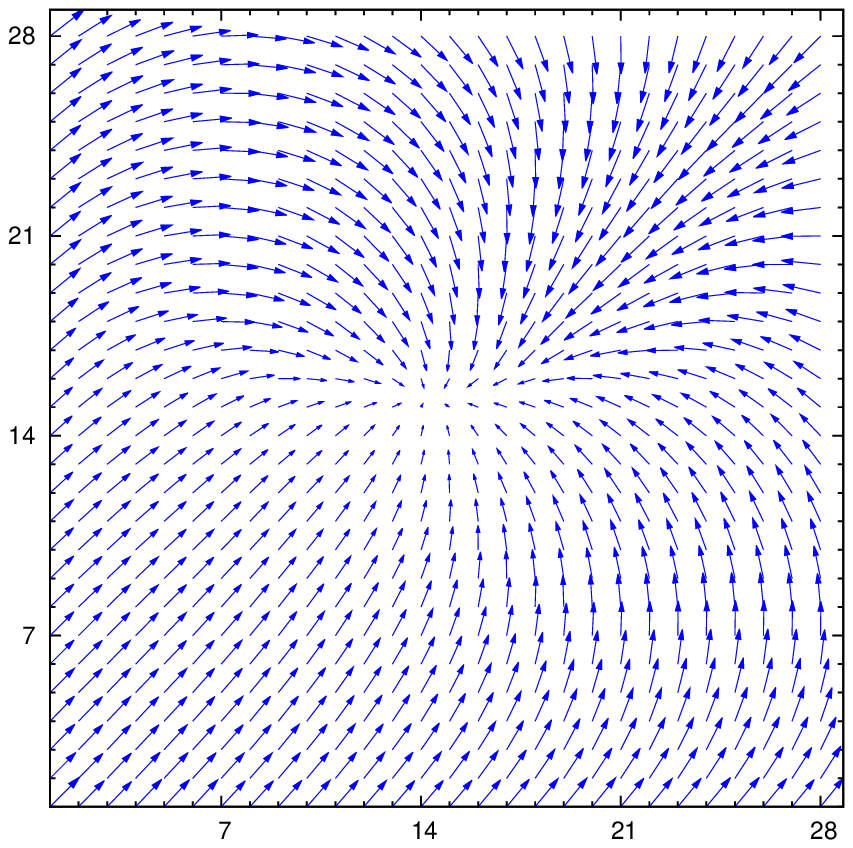}}
\subfigure[]{\includegraphics[width=120pt]{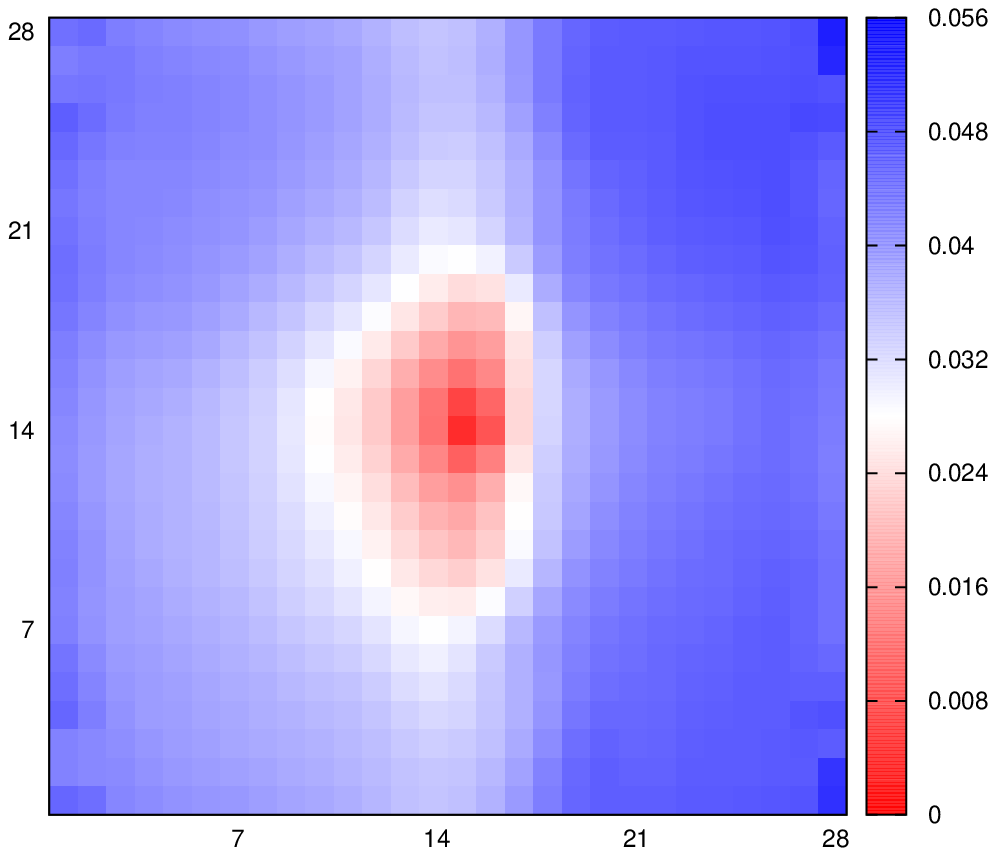}}
\subfigure[]{\includegraphics[width=120pt]{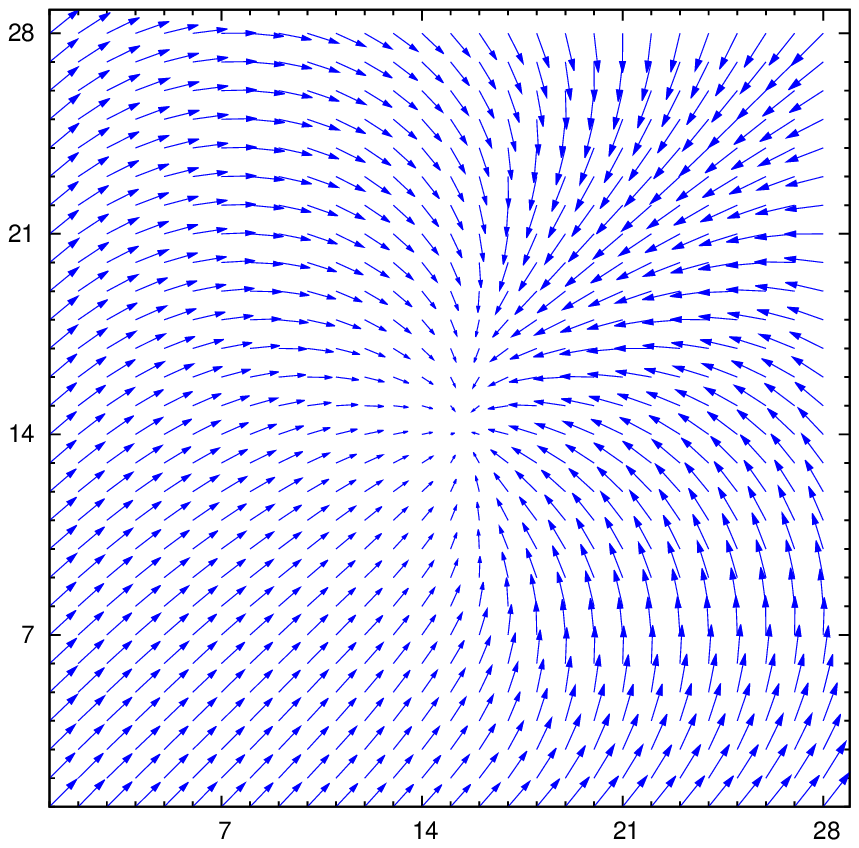}}
\subfigure[]{\includegraphics[width=120pt]{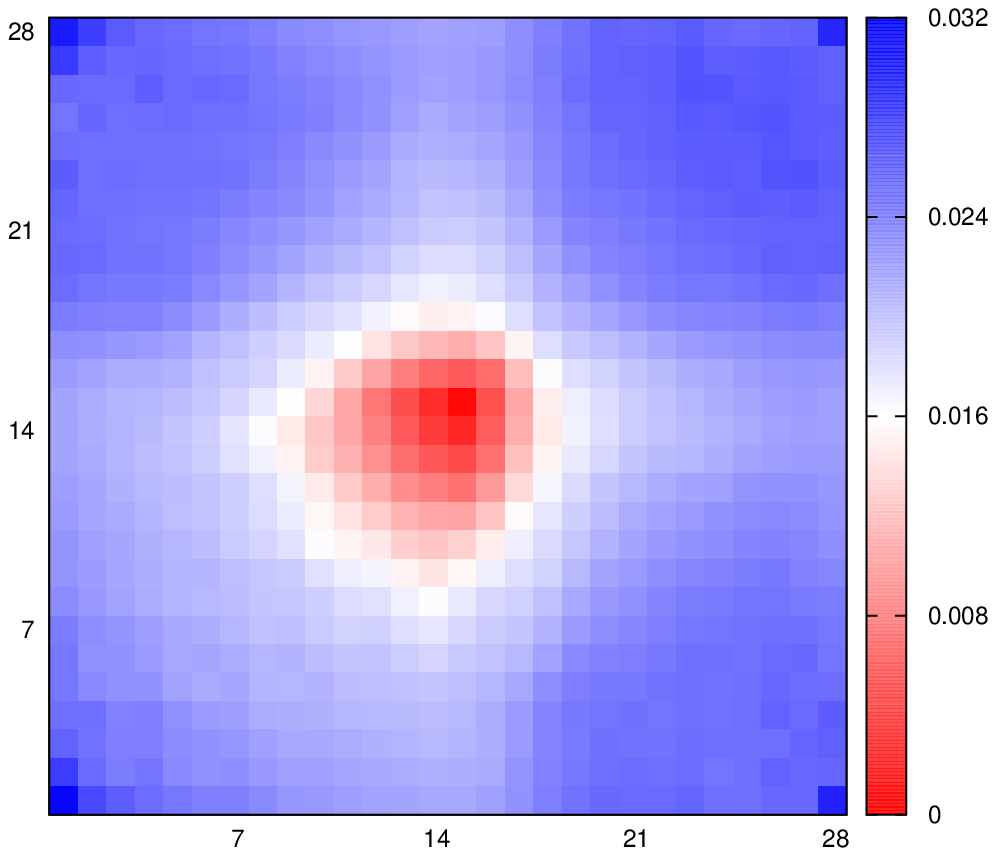}}
\subfigure[]{\includegraphics[width=120pt]{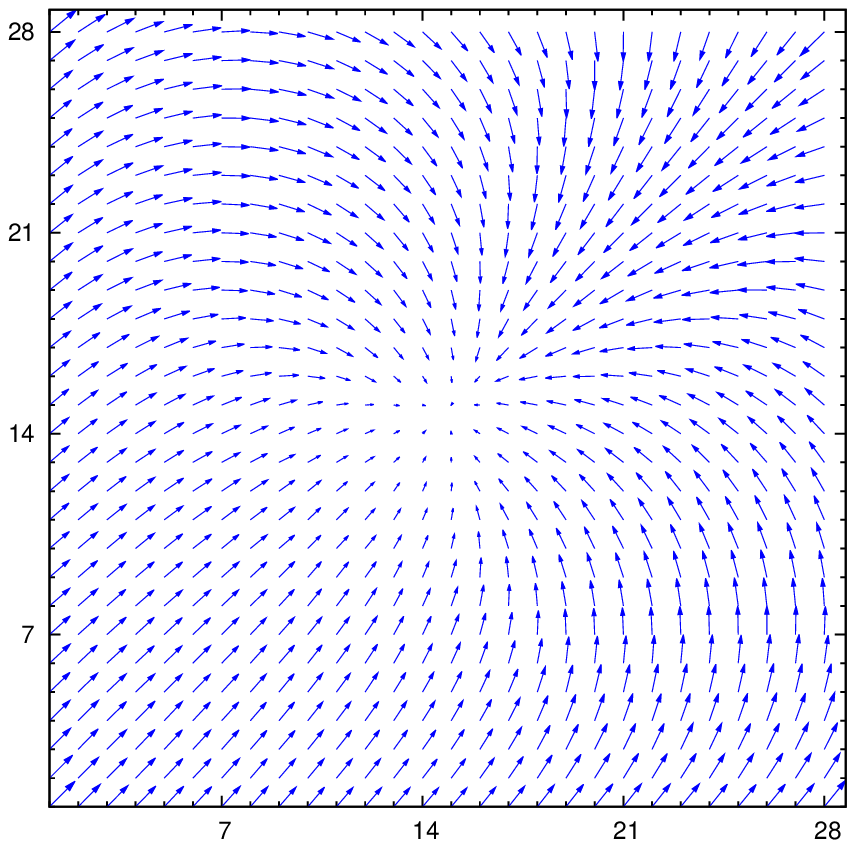}}
\caption{(color online) Amplitudes(color mapping) and phase distribution of order parameters for isotropic $s$ wave pairing state for $d_{xz}$ orbital (a) and (b), $d_{yz}$ orbital (c) and (d), and $d_{xy}$ orbital (e) and (f), respectively. The phase distribution of order parameters have been mapped to a vector field. Length of arrows represent the amplitude of order parameters.}
\label{fig3}
\end{center}
\end{figure}
\begin{figure}[t]
\begin{center}
\includegraphics[width=240pt]{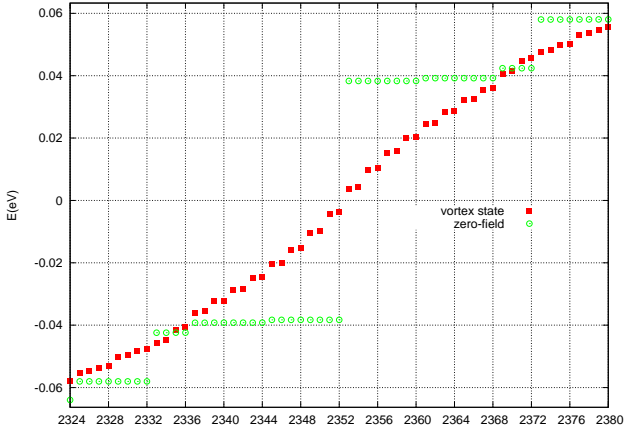}
\caption{(color online) Eigenvalues of BdG equation at around Fermi level in the cases of isotropic $s$ wave pairing state for zero-field states, shown in green circles, and vortex states, shown in red squares, respectively. The eigenvalues are plotted in an ascending sequence in horizontal axis.} \label{fig4}
\end{center}
\end{figure}
\begin{figure}[h]
\begin{center}
\subfigure[]{\includegraphics[width=120pt]{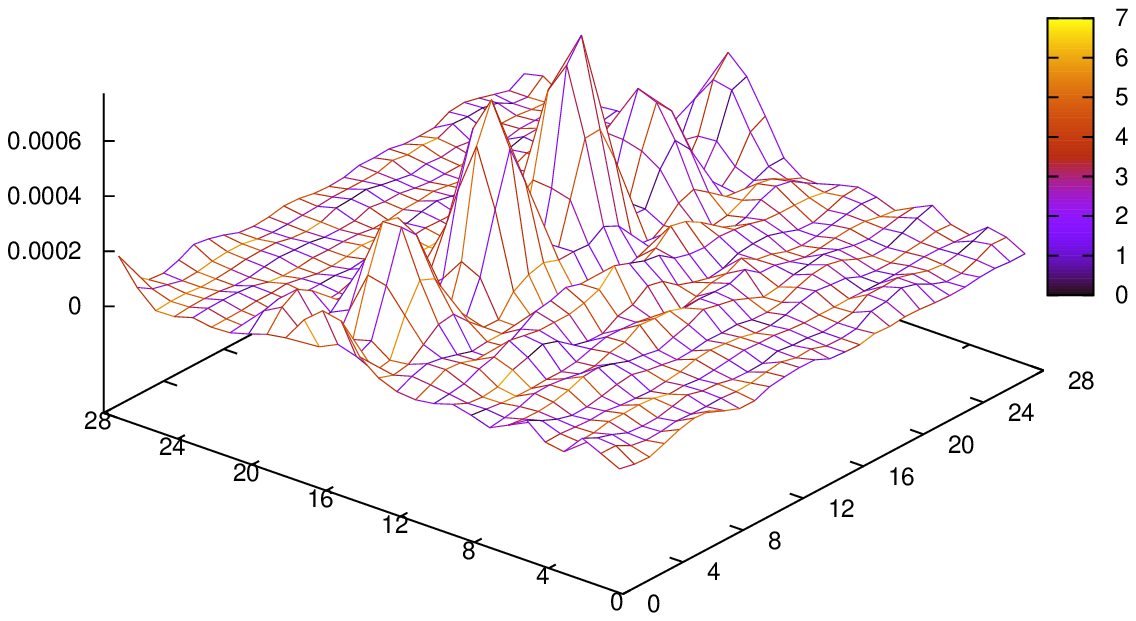}}
\subfigure[]{\includegraphics[width=120pt]{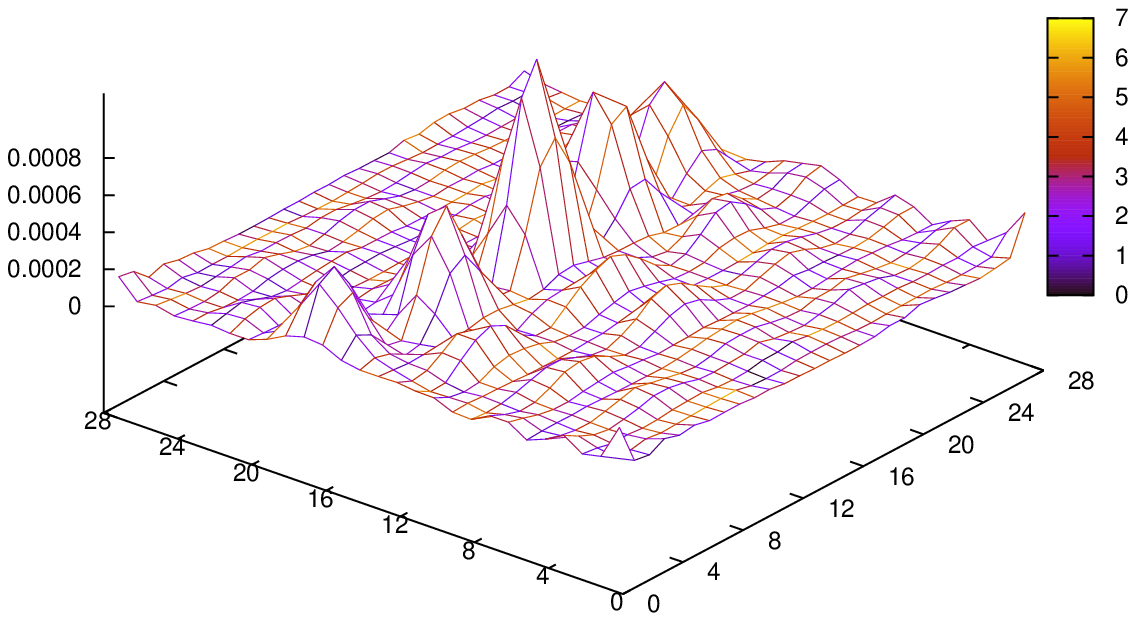}}
\subfigure[]{\includegraphics[width=120pt]{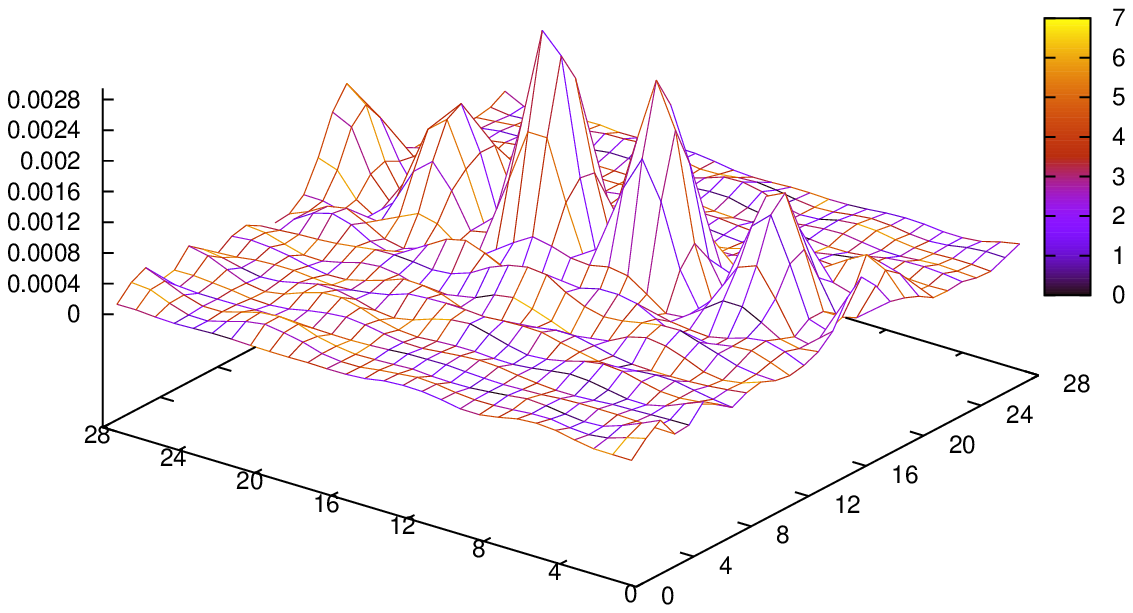}}
\subfigure[]{\includegraphics[width=120pt]{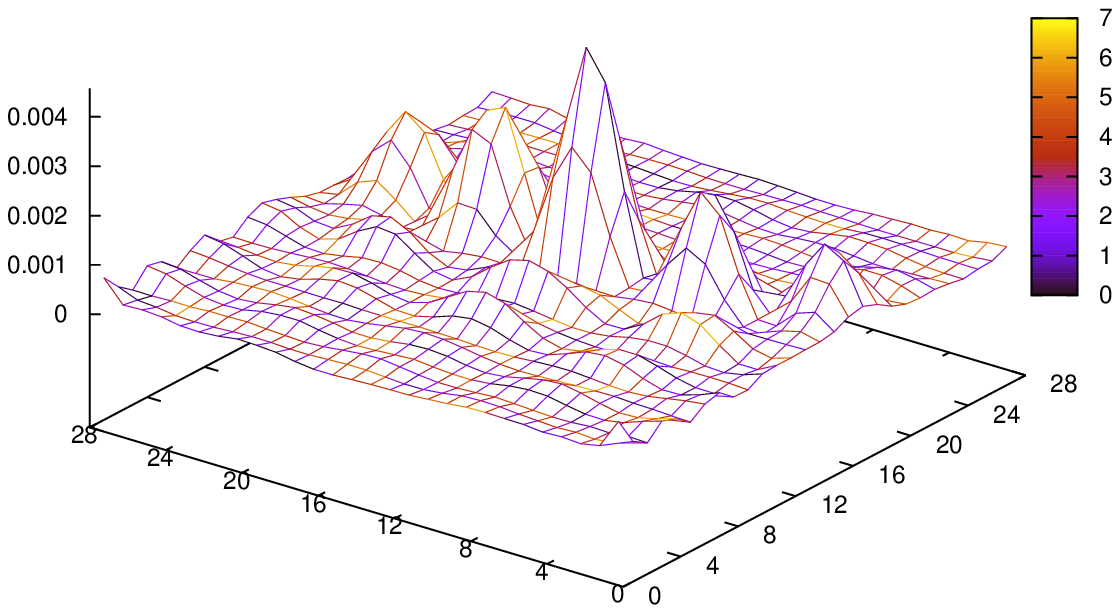}}
\subfigure[]{\includegraphics[width=120pt]{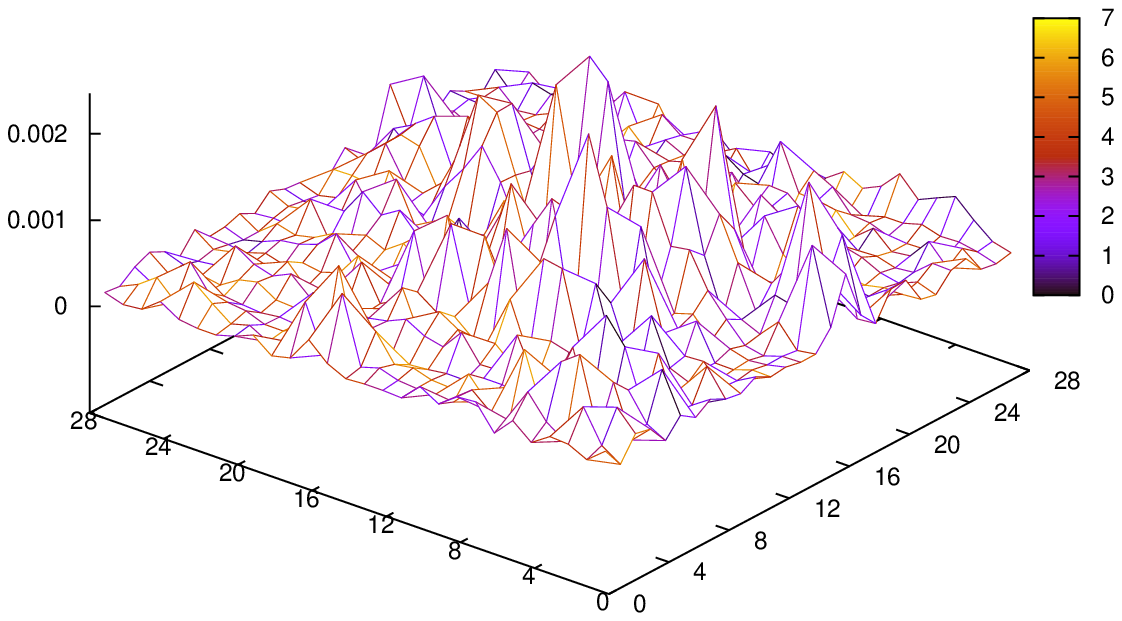}}
\subfigure[]{\includegraphics[width=120pt]{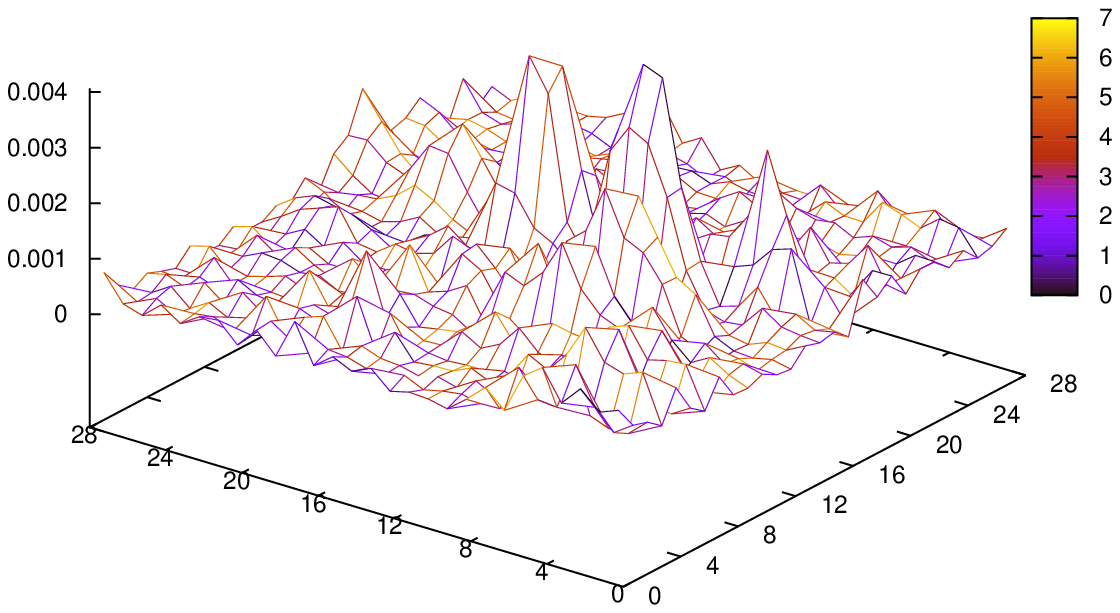}}
\caption{(color online) Amplitudes and phases(color mapping) of
quasi-particle wavefunctions $u^{n}_{i\alpha\uparrow\uparrow}$ and
$v^{n}_{i\alpha\downarrow\uparrow}$ for isotropic $s$ wave pairing
symmetry of index n=2353 for $d_{xz}$ orbital (a) and (b), $d_{yz}$
orbital (c) and (d), and $d_{xy}$ orbital (e) and (f),
respectively.} \label{fig5}
\end{center}
\end{figure}

Calculation of magnetic exchange couplings shows that the leading pairing instability comes from the intra-orbital pairing contribution, whereas the inter-orbital components are found to be significantly small\cite{C.Fang}. Consequently, only intra-orbital pairing potential is considered in our numerical calculation. The SC gap function for a multi-orbital superconductor is generally defined in momentum space as
\begin{flalign}\label{eq20}
\Delta^{i}_{\alpha\beta}(\vec{k})=g^{i}(\vec{k})\Gamma_{\alpha\beta}(i\sigma_2)
\end{flalign}
where $g^{i}(\vec{k})$ is basis of the irreducible unitary representations of $D_{4}$ point group, $i\sigma_{2}$ defines a tensor state for singlet pairing, and $\Gamma_{\alpha\beta}$ is the orbital basis for $D_{4}$ transformation. The transformation properties of band structure determine all the symmetry transformation of SC order parameters \cite{M.Daghofer, C.Fang}. Another reason that the inter-orbital pairing has been omitted in our calculation is that only if $\Gamma_{\alpha\beta}$ transform according to $A_{1g}$ representation, then symmetry of pairing state can be exclusively determined by its spatial component $g^{i}(\vec{k})$, such that the calculated vortex sate has a classification of Table \ref{table:wd}. For isotropic $s$ wave pairing,
\begin{flalign}\label{eq21}
V_{\uparrow\downarrow}(i\alpha,j\alpha)=-g_{0}\delta_{ij}
\end{flalign}
for anisotropic $s$ wave pairing,
\begin{flalign}\label{eq22}
V_{\uparrow\downarrow}(i\alpha,j\alpha)&=-\frac{g_{1}}{4}(\delta_{i+\hat{x}+\hat{y},j}+\delta_{i-\hat{x}+\hat{y},j} \\
&+\delta_{i-\hat{x}-\hat{y},j}+\delta_{i+\hat{x}-\hat{y},j})
\end{flalign}
and for $d_{x^2-y^2}$ wave pairing,
\begin{flalign}\label{eq23}
V_{\uparrow\downarrow}(i\alpha,j\alpha)=-\frac{g_2}{2}(\delta_{i+\hat{x},j}+\delta_{i+\hat{y},j}+\delta_{i-\hat{x},j}+\delta_{i-\hat{y},j})
\end{flalign}
where $g_{0,1,2}$ are pairing amplitudes for each pairing symmetry. Fig. \ref{fig2} shows the Fermi velovity $\hbar \vec{v}_{n}(\vec{k})= \nabla_{\vec{k}}\epsilon_{n}(\vec{k})$ which is used to determine the pairing potential. In order to mimic the intermediate coupling cases for FeSe \cite{F.C.Hsu} and A$_{y}$Fe$_{2-x}$Se$_{2}$ (A=K, Rb, or Cs) \cite{W.Li} superconductors whose coherent length $\xi=\frac{\hbar v_{F}}{\pi\Delta(0)}$ ranges from $4a$ to $12a$, where $a$ is lattice constant, the maximum pairing amplitudes are taken to be $g_0=0.62$, $g_1=2.60$, and $g_2=1.28$, respectively, which result in two SC order parameters(eV) due to orbital anisotropy in zero-field case as for isotropic $s$ wave
\begin{flalign}\label{eq24}
|\Delta^{s}_{xz,yz}(0)| = 0.047; ~|\Delta^{s}_{xy}(0)| = 0.026
\end{flalign}
for anisotropic $s$ wave
\begin{flalign}\label{eq25}
|\Delta^{anis. \ s}_{xz,yz}(0)| = 0.048; ~|\Delta^{anis. \ s}_{xy}(0)| = 0.023
\end{flalign}
and for $d_{x^2-y^2}$ wave
\begin{flalign}\label{eq26}
|\Delta^{d_{x^2-y^2}}_{xz,yz}(0)| = 0.048; ~|\Delta^{d_{x^2-y^2}}_{xy}(0)| = 0.025
\end{flalign}

\section{Results and Discussion}\label{sec4}

\begin{figure}[t]
\begin{center}
\subfigure[]{\includegraphics[width=120pt]{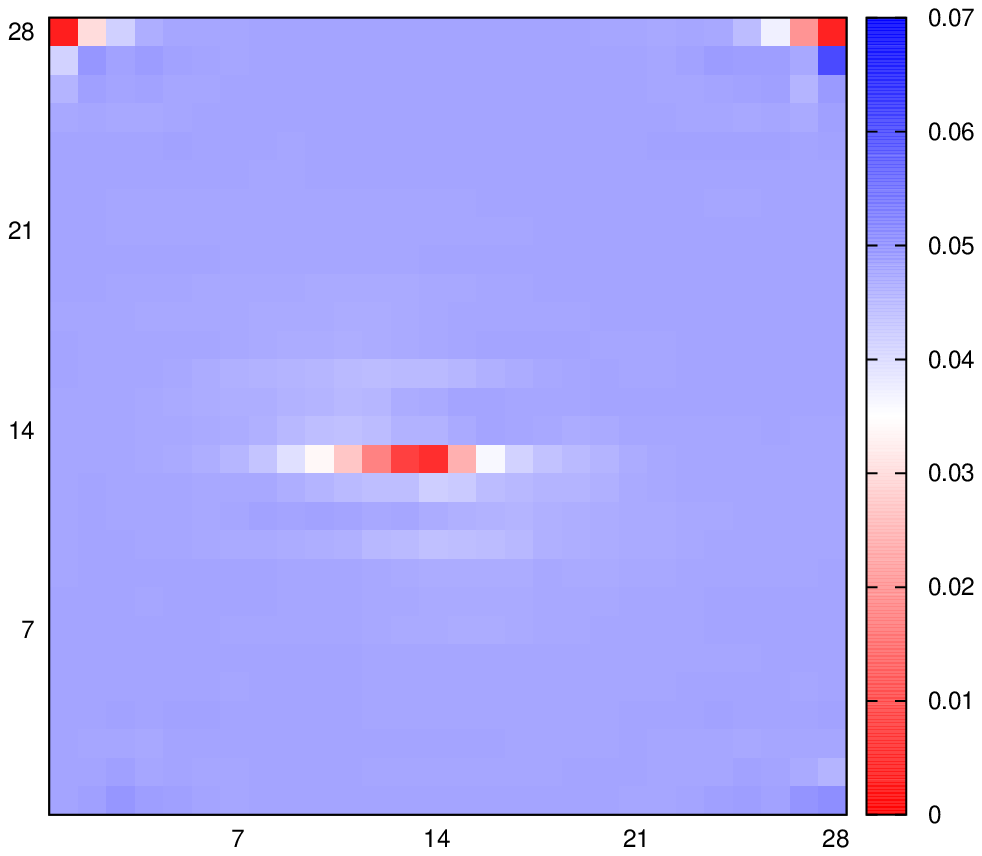}}
\subfigure[]{\includegraphics[width=120pt]{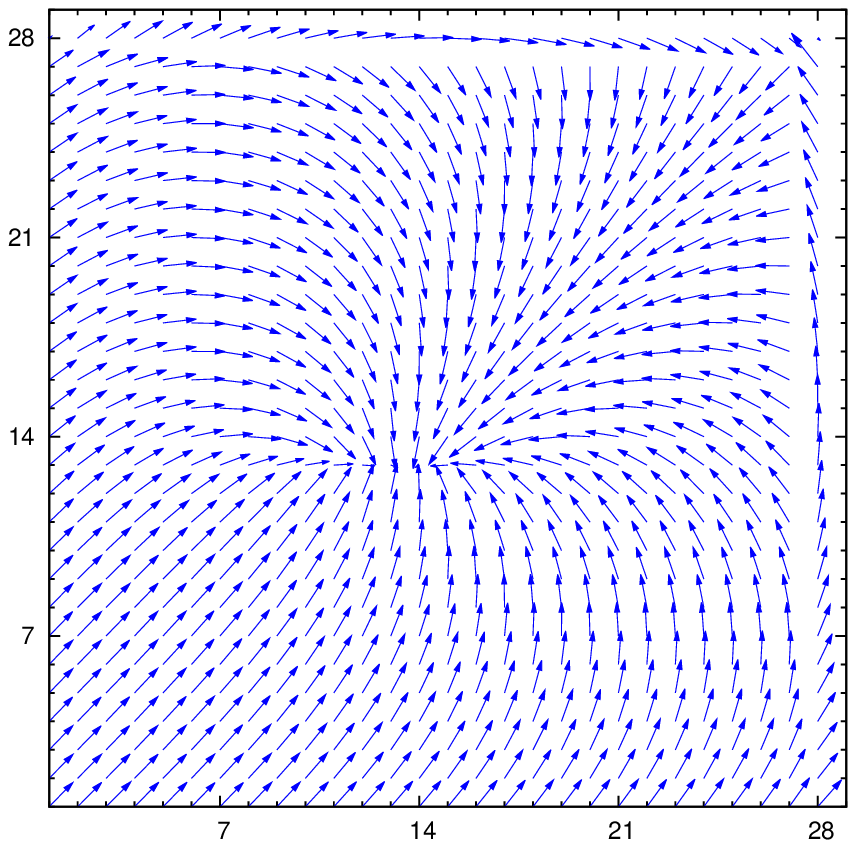}}
\subfigure[]{\includegraphics[width=120pt]{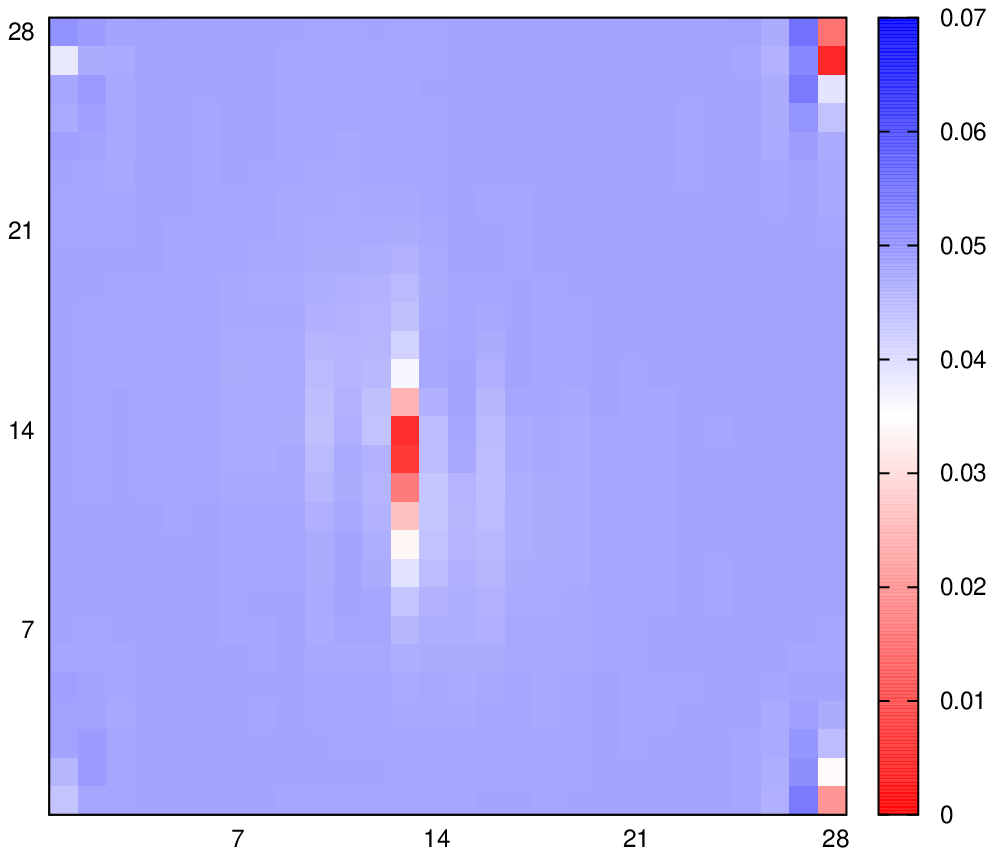}}
\subfigure[]{\includegraphics[width=120pt]{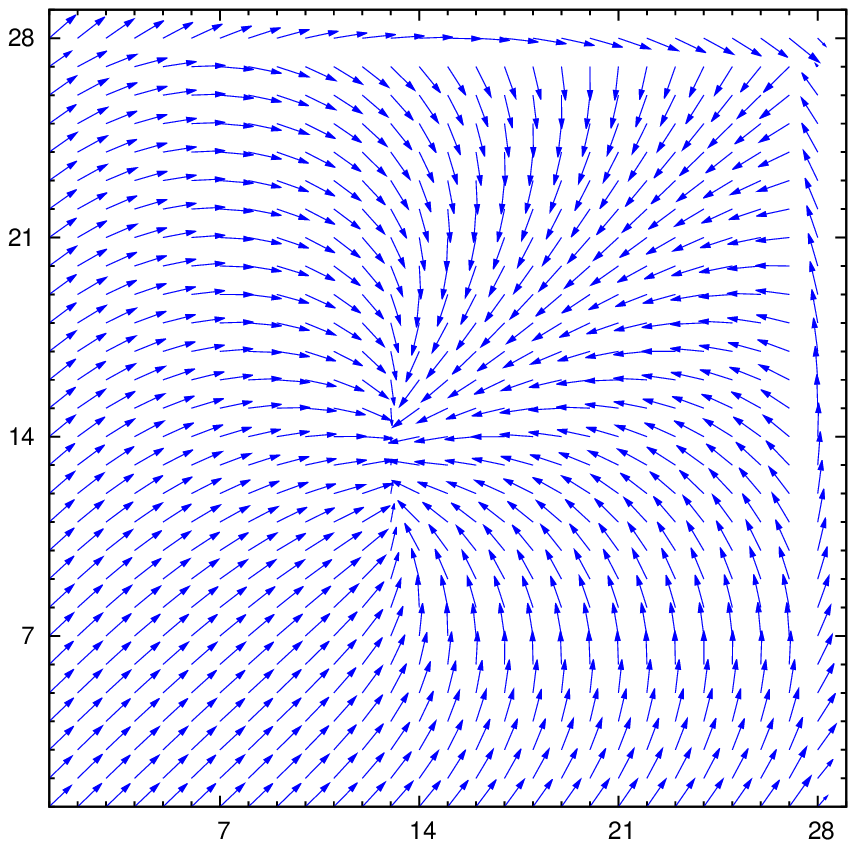}}
\subfigure[]{\includegraphics[width=120pt]{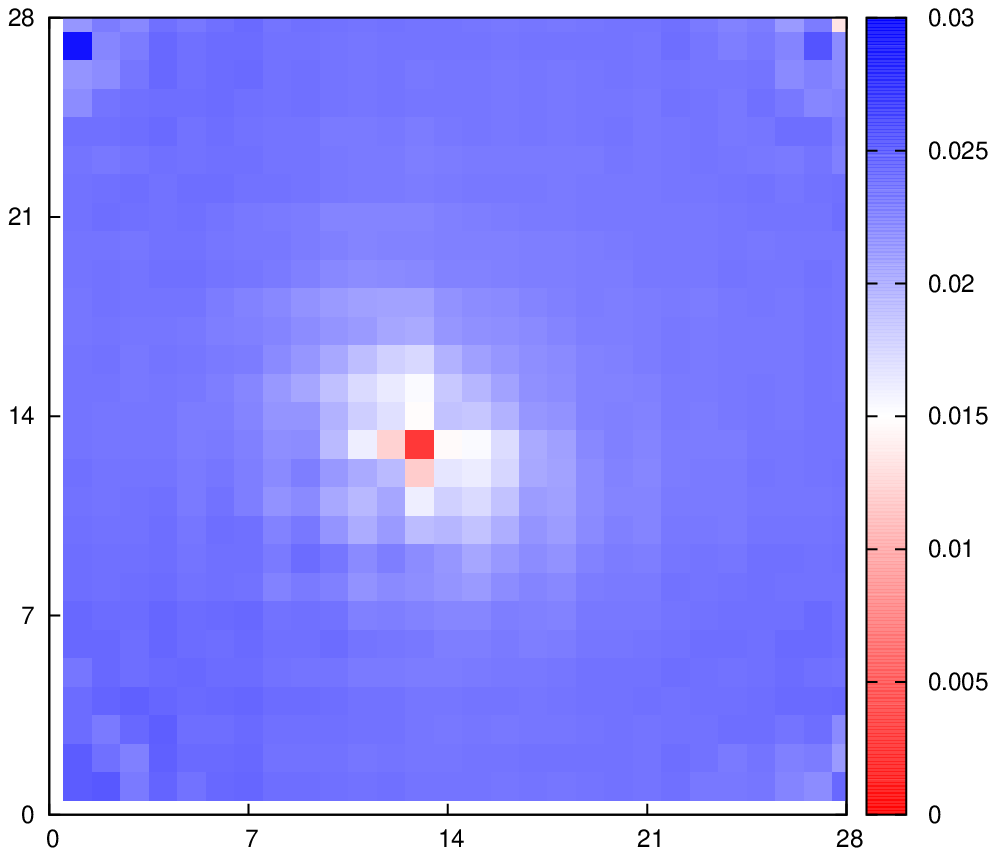}}
\subfigure[]{\includegraphics[width=120pt]{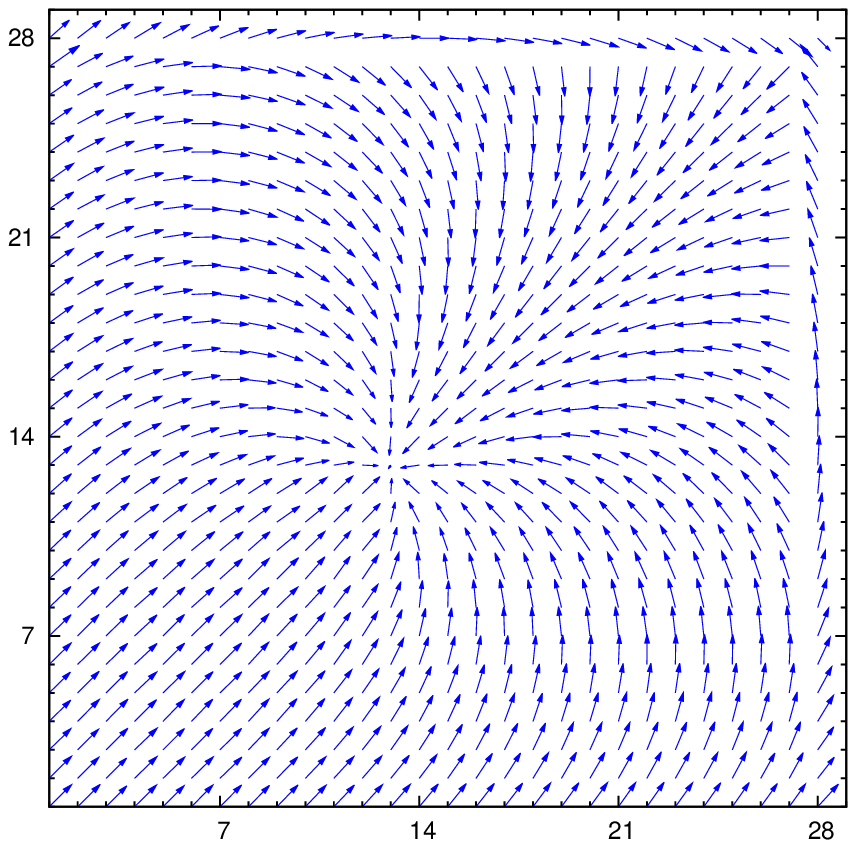}}
\caption{(color online) Amplitudes(color mapping) and phase
distribution of anisotropic $s$ wave pairing bonds along $\hat{x}+\hat{y}$
direction for $d_{xz}$ orbital (a) and (b), $d_{yz}$ orbital (c) and
(d), and $d_{xy}$ orbital (e) and (f), respectively. Results of
pairing bonds along $-\hat{x}+\hat{y}$, $-\hat{x}-\hat{y}$, and $\hat{x}-\hat{y}$ directions are the same with these results.} \label{fig6}
\end{center}
\end{figure}

\begin{figure}[t]
\begin{center}
\subfigure[]{\includegraphics[width=120pt]{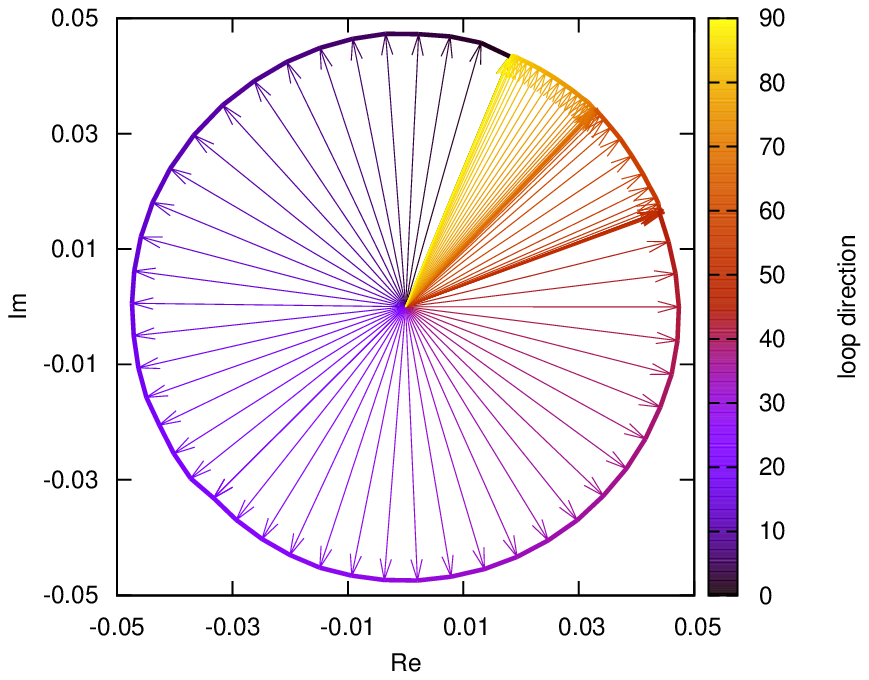}}
\subfigure[]{\includegraphics[width=120pt]{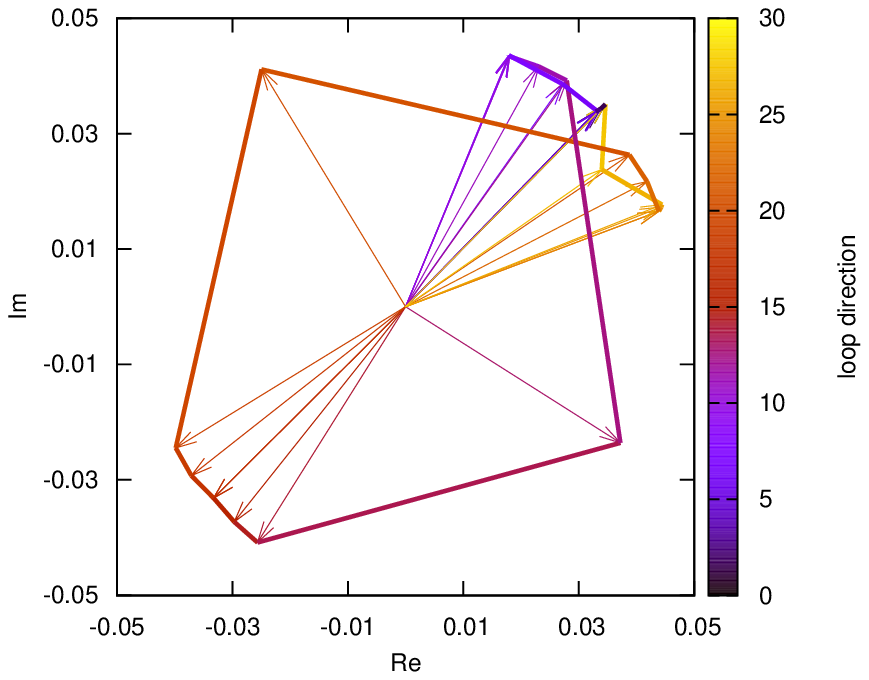}}
\caption{(color online) Phase mapping onto complex plane of
anisotropic $s$ wave pairing bond along $\hat{x}+\hat{y}$ direction. Loop
around center of magnetic unit cell (a) is
$(25,3)\rightarrow(25,25)\rightarrow(3,25)\rightarrow(3,3)\rightarrow(25,3)$ and around corner of magnetic unit cell (b) is
$(3,1)\rightarrow(3,3)\rightarrow(1,3)\rightarrow(28,3)\rightarrow(25,3)
\rightarrow(25,1)\rightarrow(25,28)\rightarrow(25,25)\rightarrow(28,25)
\rightarrow(1,25)\rightarrow(3,25)\rightarrow(3,28)\rightarrow(3,1)$.
The loop direction has been shown by color mapping of each steps.}
\label{fig7}
\end{center}
\end{figure}

\begin{figure}[h]
\begin{center}
\includegraphics[width=240pt]{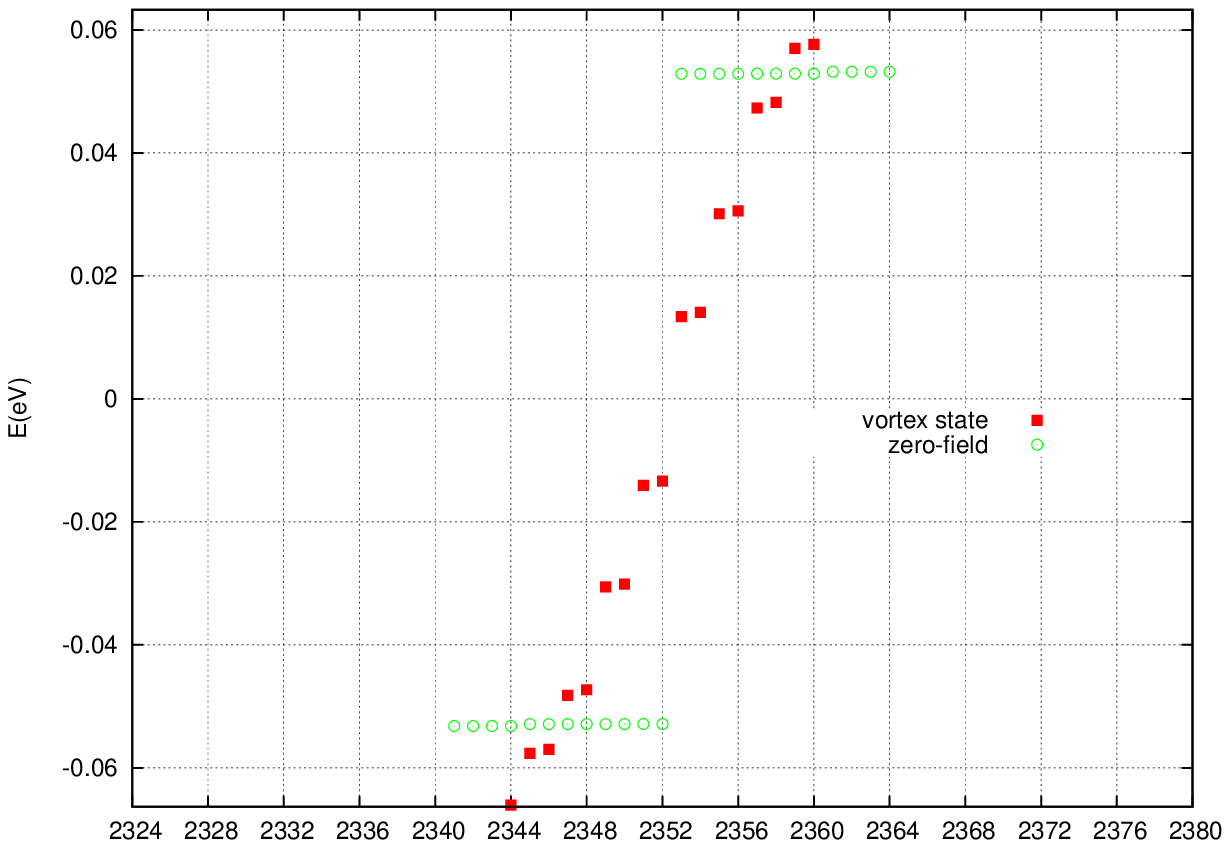}
\caption{(color online) Eigenvalues of BdG equation at around Fermi level in the cases of anisotropic $s$ wave pairing state for zero-field states, shown in green circles, and vortex states, shown in red squares, respectively.} \label{fig8}
\end{center}
\end{figure}

\begin{figure}[t]
\begin{center}
\subfigure[]{\includegraphics[width=120pt]{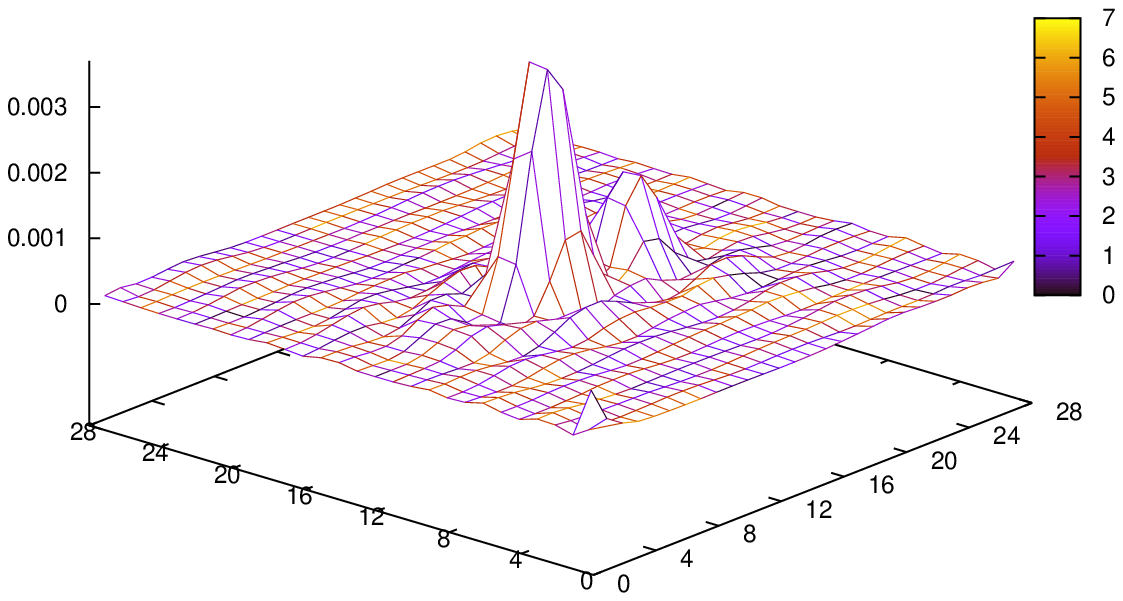}}
\subfigure[]{\includegraphics[width=120pt]{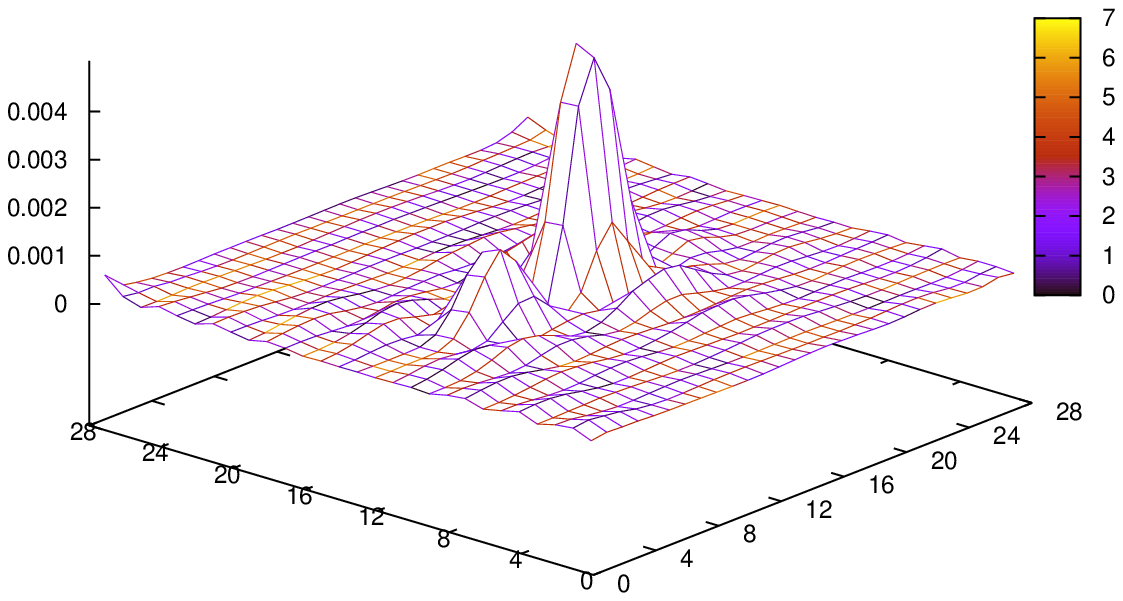}}
\subfigure[]{\includegraphics[width=120pt]{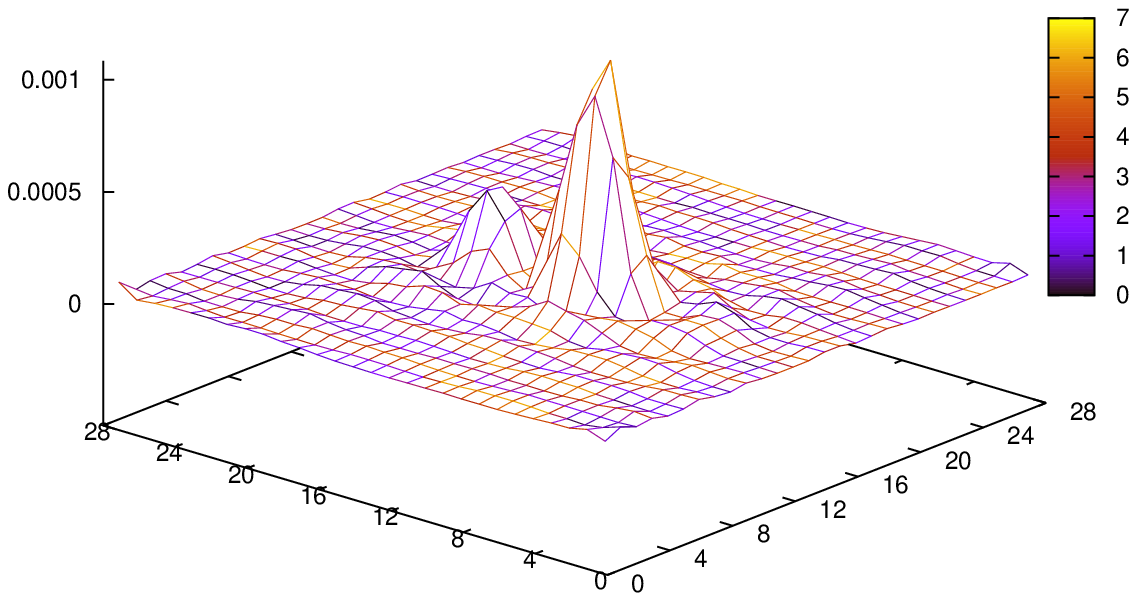}}
\subfigure[]{\includegraphics[width=120pt]{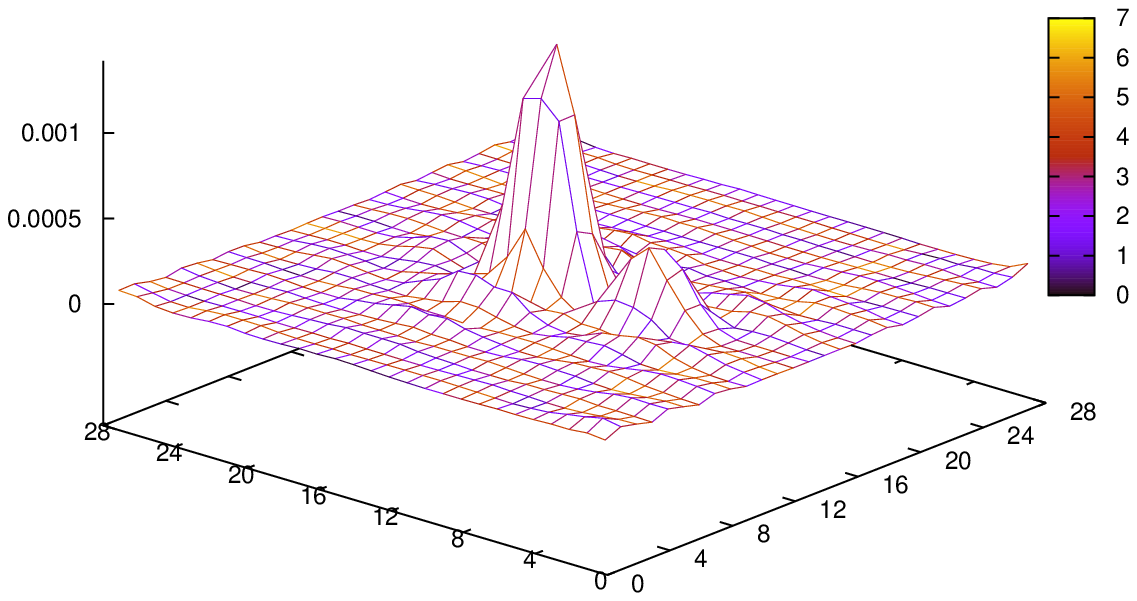}}
\subfigure[]{\includegraphics[width=120pt]{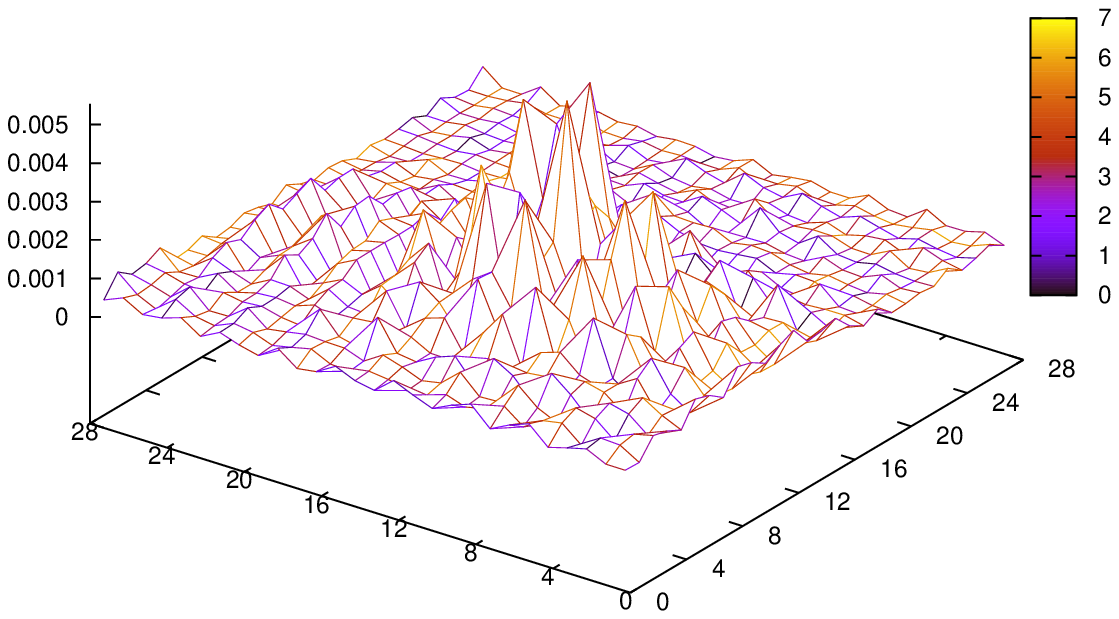}}
\subfigure[]{\includegraphics[width=120pt]{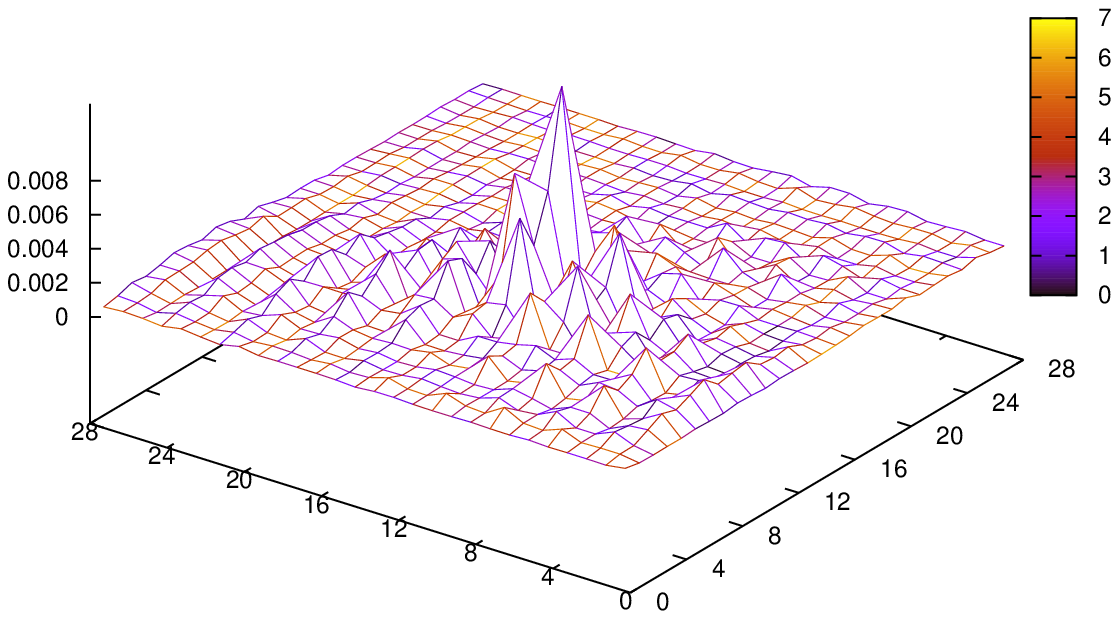}}
\caption{(color online) Amplitudes and phases(color mapping) of
quasi-particle wavefunctions $u^{n}_{i\alpha\uparrow\uparrow}$ and
$v^{n}_{i\alpha\downarrow\uparrow}$ for anisotropic $s$ wave pairing state
of index n=2353 for $d_{xz}$ orbital (a) and (b), $d_{yz}$ orbital
(c) and (d), and $d_{xy}$ orbital (e) and (f), respectively.}
\label{fig9}
\end{center}
\end{figure}

The vortex structures for isotropic $s$ wave pairing state are shown in Fig. \ref{fig3} for different orbitals, respectively. The vortex states exhibit orbital anisotropy. For $d_{xz}$ and $d_{yz}$ orbitals the amplitudes have two plateaus with a difference about 0.005eV along $\hat{y}$ and $\hat{x}$ directions on both sides of the core region and the pinning center deviates slightly from the center of magnetic unit cell. The phase distribution shows a winding number $\mathcal {W}=1$, such that the symmetry subgroup of the vortex structure is $G_5$\cite{M.Ozaki1}. The winding structure of the $s$ wave vortex, as mapped to a vector field, has a sink-type core center. Fig. \ref{fig4} shows the eigenvalues obtained from vortex and zero-field states, where it has been found there are 16 in-gap eigenstates for both positive and negative eigenvalues. We examine the behavior of the quasi-particle wavefunction $u^{n}_{i\alpha\uparrow\uparrow}$ and $v^{n}_{i\alpha\downarrow\uparrow}$ and it turns out that all the 32 in-gap states are extended to the entire magnetic unit cell(Fig. \ref{fig5}, eigenstate $|\epsilon_{2353\uparrow}\rangle$). The orbital anisotropy again appears as for $d_{xz}$ and $d_{yz}$ orbitals, the wavefunction extends to $\hat{x}$ and $\hat{y}$ direction because the spatial orientation of d-orbital harmonics, whereas for $d_{xy}$ orbital, the spreading of wavefunction is symmetric in $\hat{x}$ and $\hat{y}$ directions. These extended wavefunctions amount to large scale variation of order parameters within the entire magnetic unit cell and consequently a relatively large vortex core region.

\begin{figure}[!t]
\begin{center}
\subfigure[]{\includegraphics[width=120pt]{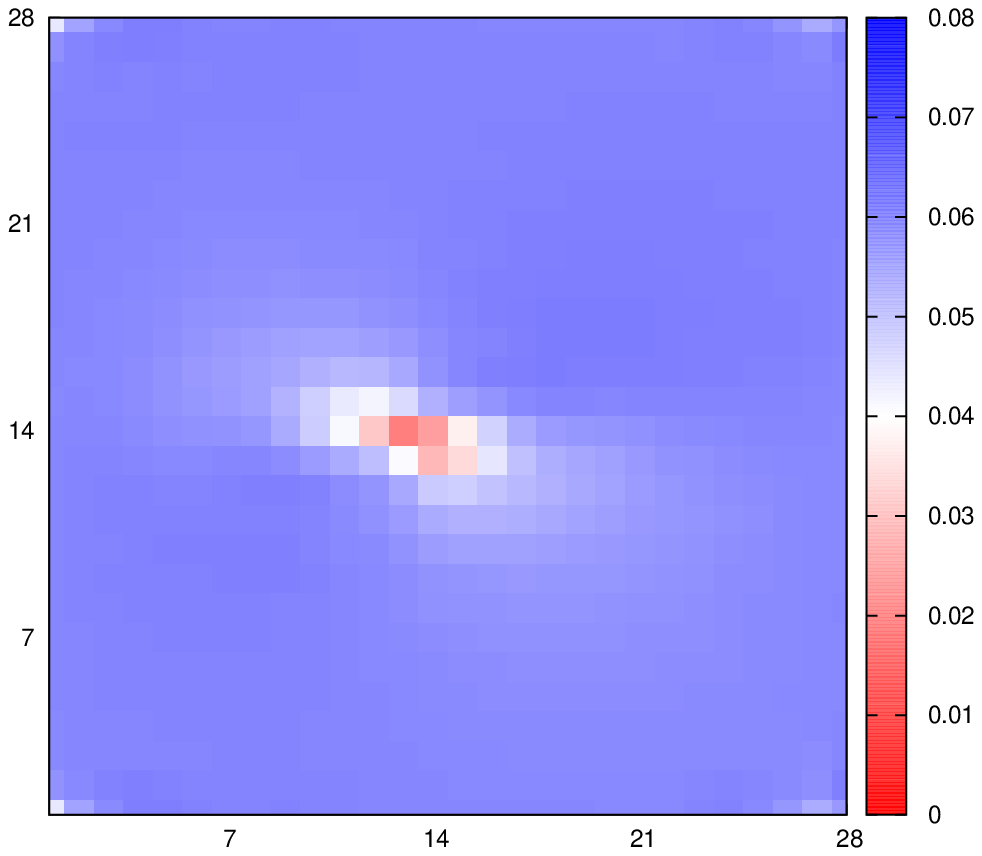}}
\subfigure[]{\includegraphics[width=120pt]{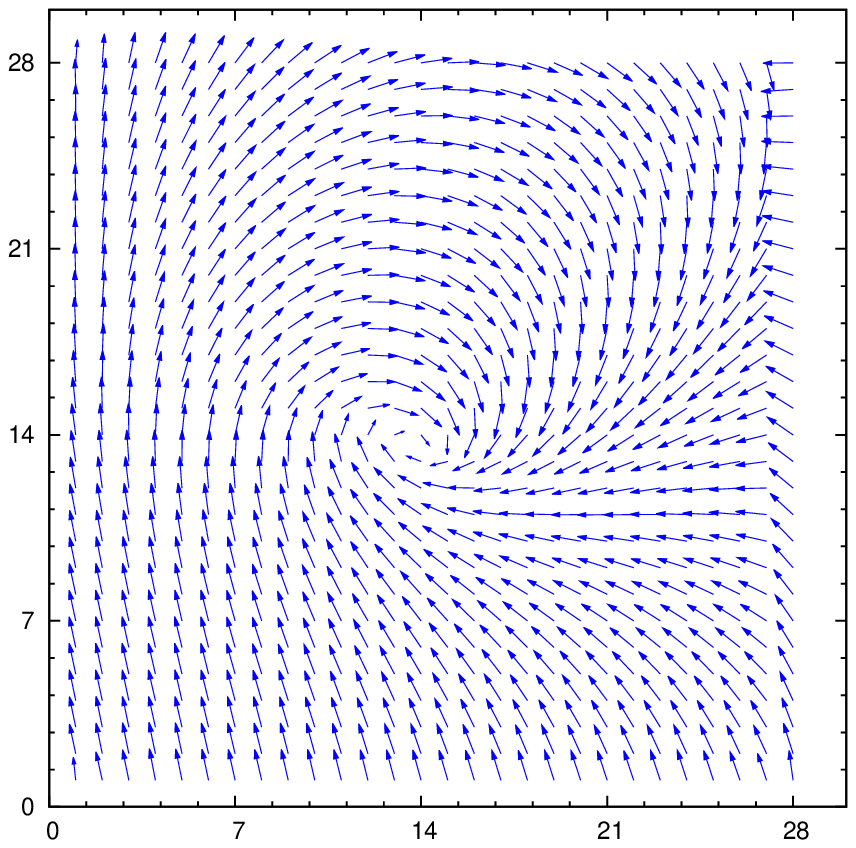}}
\subfigure[]{\includegraphics[width=120pt]{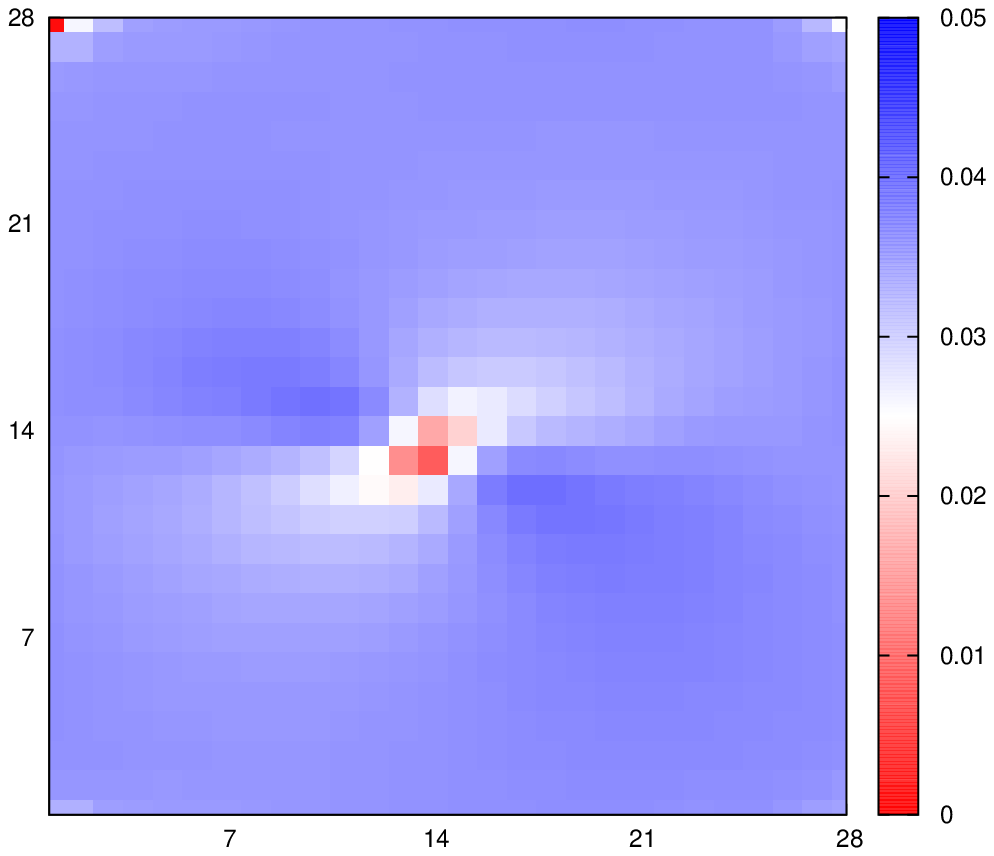}}
\subfigure[]{\includegraphics[width=120pt]{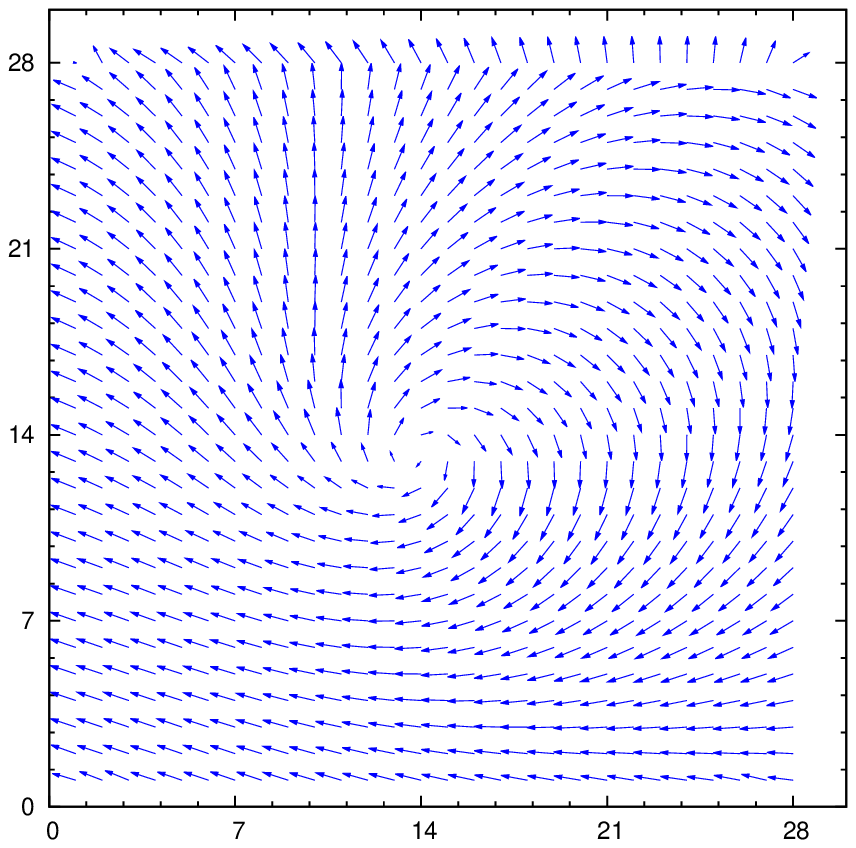}}
\subfigure[]{\includegraphics[width=120pt]{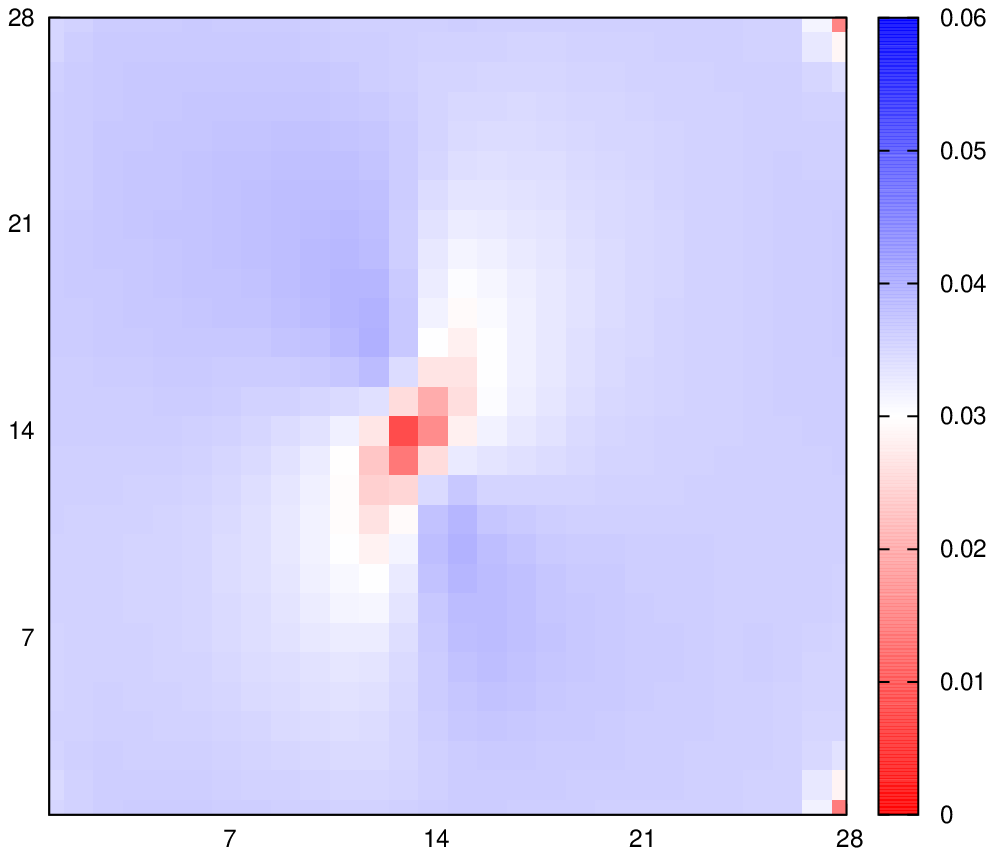}}
\subfigure[]{\includegraphics[width=120pt]{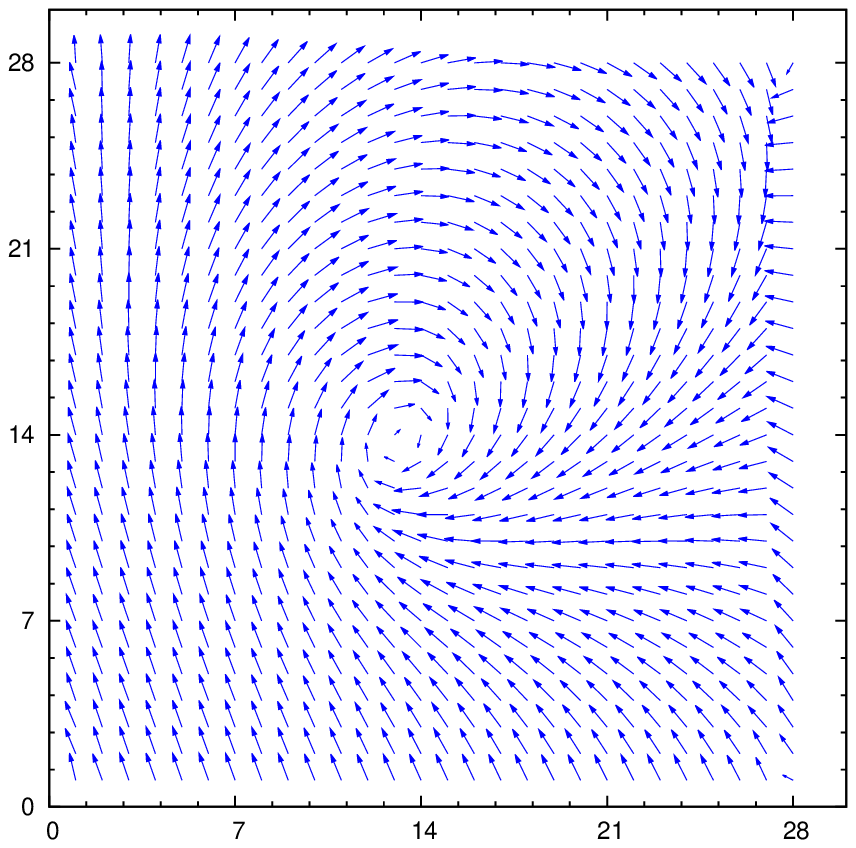}}
\subfigure[]{\includegraphics[width=120pt]{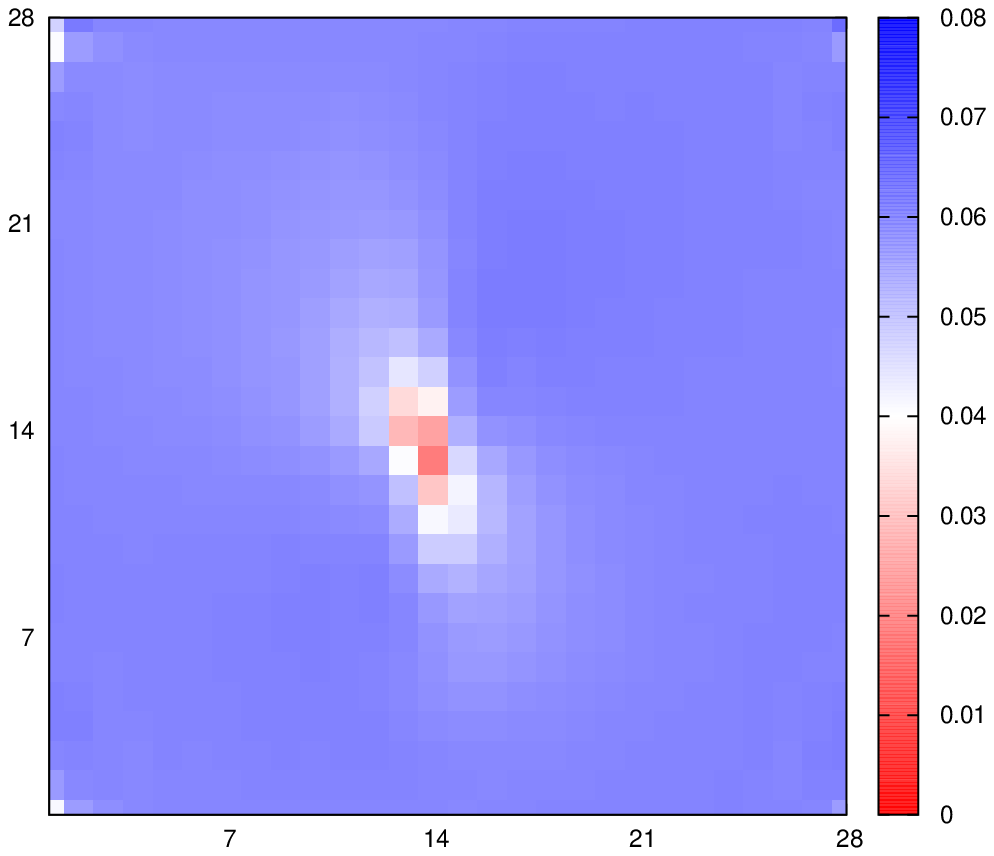}}
\subfigure[]{\includegraphics[width=120pt]{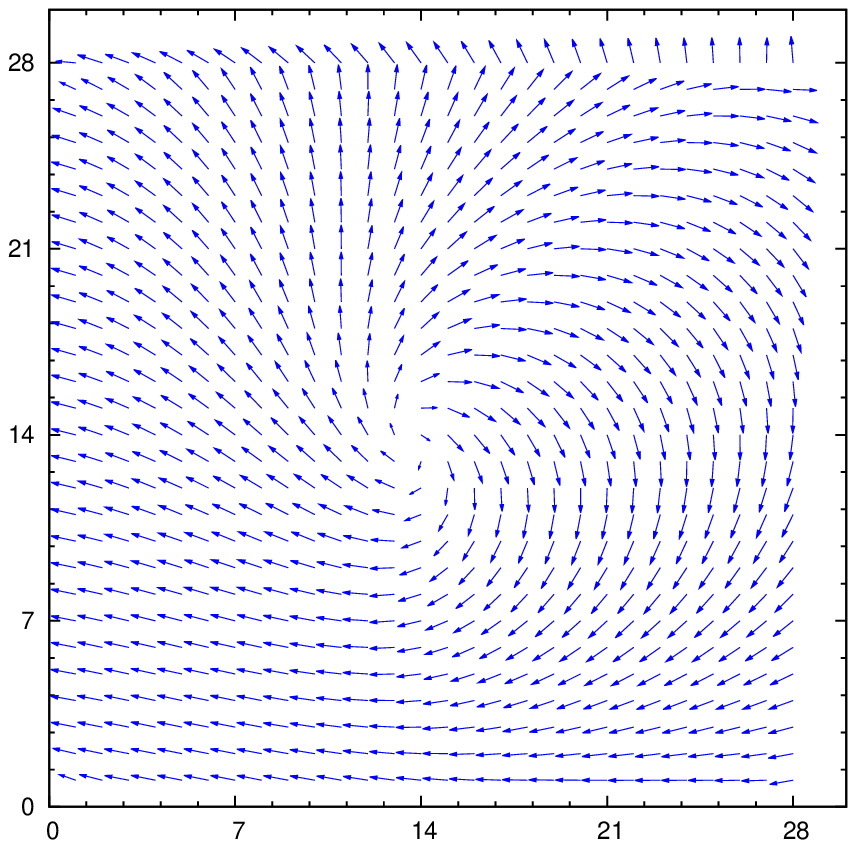}}
\caption{(color online) Amplitudes(color mapping) and phase
distribution of $d_{x^2-y^2}$ wave pairing bonds for $d_{xz}$
orbital along $\hat{x}$ direction (a) and (b), $\hat{y}$ direction
(c) and (d), and for $d_{yz}$ orbital along $\hat{x}$ direction (e)
and (f), $\hat{y}$ direction (g) and (h), respectively. Results of
pairing bonds along the other two directions of next nearest site pairing are same
with these results.} \label{fig10}
\end{center}
\end{figure}

\begin{figure}[]
\begin{center}
\subfigure[]{\includegraphics[width=120pt]{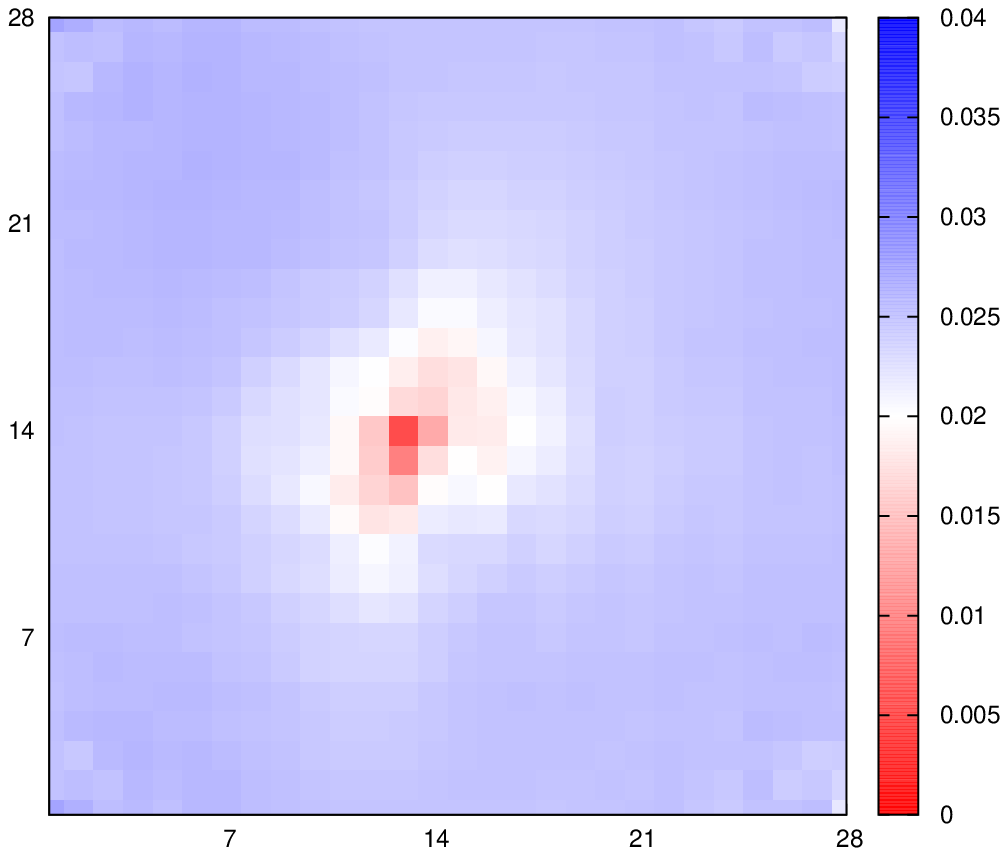}}
\subfigure[]{\includegraphics[width=120pt]{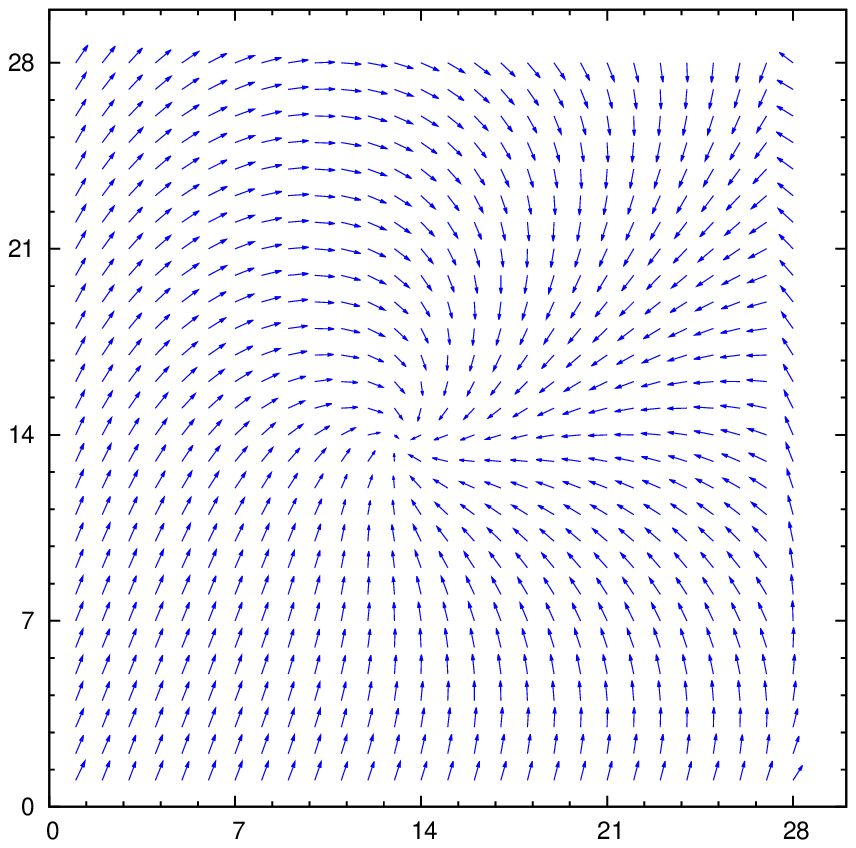}}
\subfigure[]{\includegraphics[width=120pt]{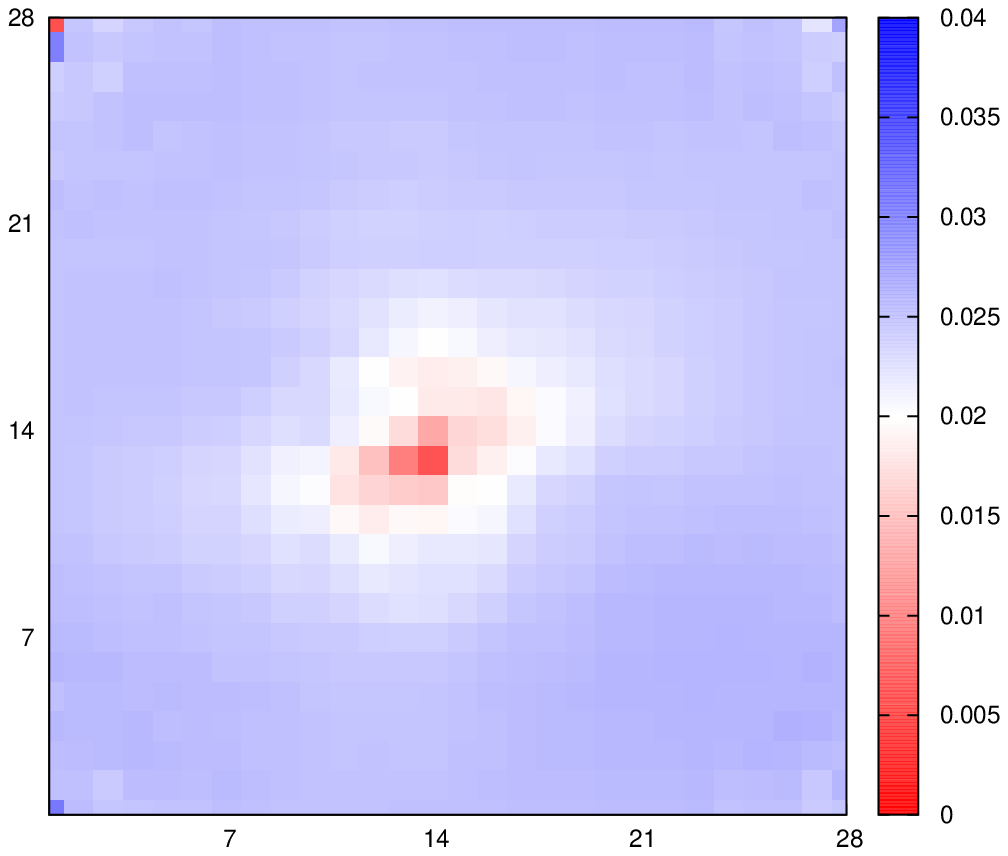}}
\subfigure[]{\includegraphics[width=120pt]{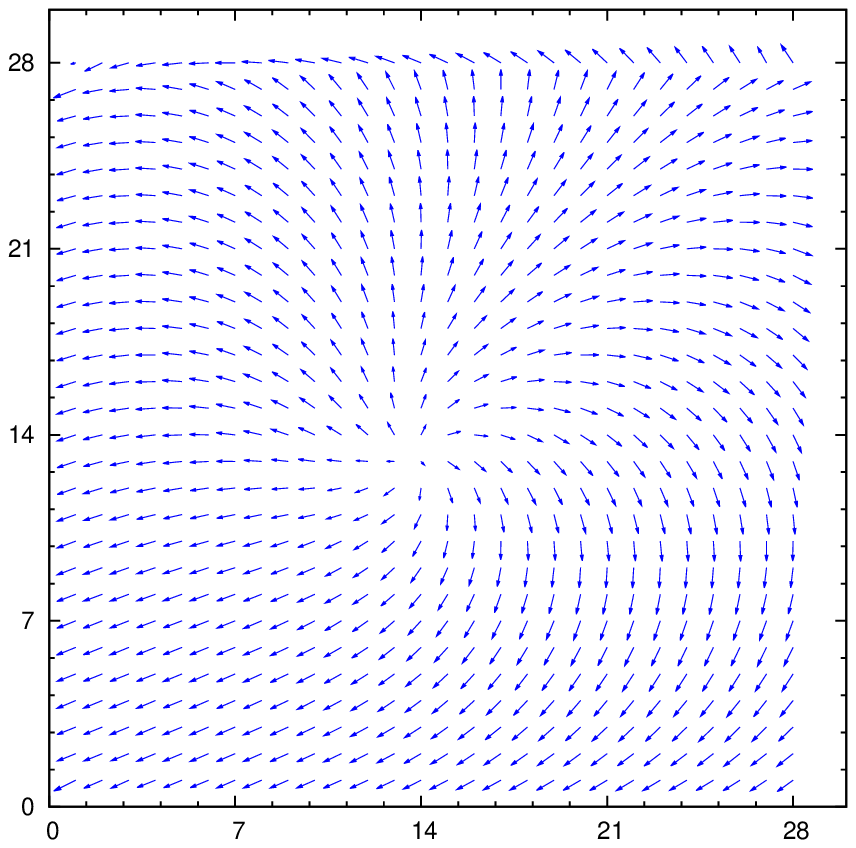}}
\caption{(color online) Amplitudes(color mapping) and phase
distribution of $d_{x^2-y^2}$ wave pairing bond for $d_{xy}$ orbital
along $\hat{x}$ direction (a) and (b), and $\hat{y}$ direction (c)
and (d), respectively. Results of pairing bonds along the other two
directions of next nearest site pairing are same with these results.}
\label{fig11}
\end{center}
\end{figure}

\begin{figure}[t]
\begin{center}
\includegraphics[width=180pt]{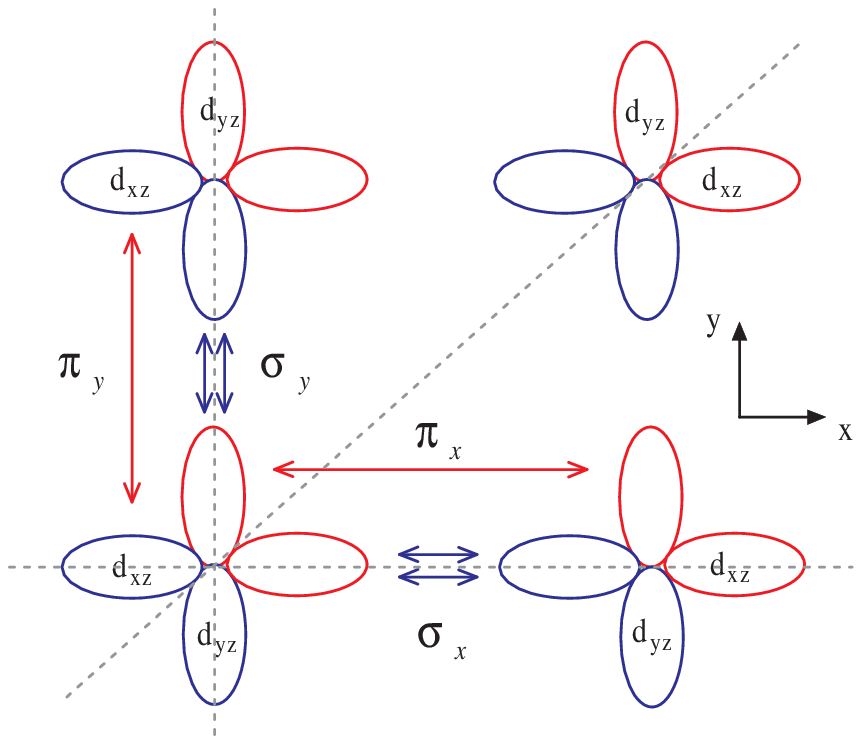} \caption{(color online)
A schematic picture illustrates that $d_{x^2-y^2}$ wave pairing
state is re-defined between $d_{xz}$ and $d_{yz}$ orbitals due to
4-fold rotational symmetry. The red and blue color indicate positive
and negative signs of orbital wavefunctions. The long and short
double-headed arrows corresponding $\pi$ and $\sigma$ pairing bonds
along $\hat{x}$ and $\hat{y}$ directions show the exchange of
orbital states under $C_{4z}$ rotation.} \label{fig12}
\end{center}
\end{figure}

The structures of anisotropic $s$ wave vortices are shown in Fig. \ref{fig6}. The core regions of $d_{xz/yz}$ orbital vortices are not a geometric point any more. Instead, they have been stretched along $\hat{x}$ and $\hat{y}$ directions due to the fact that although the pairing bonds are defined on next nearest neighbor sites, the electrons forming Cooper pairs come from distinguishable oriented orbitals. The symmetry subgroup of anisotropic $s$ wave vortices is still $G_{5}$, but orbital asymmetry results in a line-type topological defect for $d_{xz/yz}$ orbital vortices, whereas $d_{xy}$ orbital vortex is still of sink-type. One special fact worth noting is that there is a suppression of order parameters at corners of magnetic unit cell, which also exists for pairing bond along $-\hat{x}\pm\hat{y}$ and $\hat{x}-\hat{y}$ directions. In order to understand the physical origin of this phenomena, we examine the phase variation along two loops around the center and corner of magnetic unit cell, respectively. The loop around the corner is well-defined in order parameter space because the nontrivial winding periodic boundary condition Eq. \eqref{eq17} has been applied. Since the homotopy group of order parameter space of a vortex state is $\pi_{1}[U(1)]=\mathbb{Z}$ and that the winding number $\mathcal {N}_{v}=1$ has been fixed when the self-consistent calculation is carried out, we expect that the variation along the loop around the corner is definitely not homotopic equivalent to that around vortex at center. Fig. \ref{fig7} (a) shows the phase variation around the vortex core, where the phases change slowly on a number of lattice sites at the very beginning of the loop as shown in Fig. \ref{fig6} (a) in the vicinity of site (25,3). We have deliberately chosen a loop far away from the core region, since a stable topological defect always leaves its signature anywhere arbitrarily away from it \cite{N.D.Mermin}. However, the phase variation of order parameters around the corner of magnetic unit cell exhibits some turning-back points, from which the clockwise increments contribute negative phase winding. Therefore the total winding around the corner is zero, which proves that the suppression of order parameters at corners of magnetic unit cell is not a vortex. Detailed analysis about the phase difference on each lattice sites shows that such singularities at corners is actually caused by the discontinuity of boundary condition of wavefunction of each orbitals when the calculation is carried out on a $N_{x} \times N_{y}$ lattice. From Eq. \eqref{eq17}, we know that the variation of boundary condition along $\hat{x}$ direction for adjacent $(\lambda_{x}=1,\lambda_{y}=0)$ magnetic unit cell is $e^{iKN_{x}i_{y}}$, and it will come back to $e^{i(KN_{x}+2\pi)}$ when the condition $i_{y}=4N_{y}+1$ is satisfied. It is obviously that such a condition cannot be realized in numerical calculation for any given $N_{y}$, therefore the discontinuity, which can be regarded as an impurity induced by winding boundary condition, cannot be avoided. The impurity nature of these singularities can also be recognized as the suppression of order parameters occurs on single site at corners, which is different from a genuine vortex having an effective core region. We also noted that such a singularity does not exist for $\mathcal {N}_{v}=4$, but in this case the vortex states cannot be classified by invariant subgroups of magnetic translation group, which is originally aimed at describing the Abrikosov lattice for $\mathcal {N}_{v}=1$. There are 12 in-gap eigenstates, as shown in Fig. \ref{fig8}, which locate symmetrically on both sides of the Fermi level. The wavefunctions of these states are typically localized for $d_{xz/yz}$ orbitals and extended for $d_{xy}$ orbital, as shown in Fig. \ref{fig9} for eigenstate $|\epsilon_{2353}\rangle$. It has been observed that the wavefunctions for each orbitals show particle-hole asymmetry. Although the difference of vortices between isotropic $s$ and anisotropic $s$ wave pairing states has been observed from the hitherto results, such a difference may rely on the limitation of our model calculation in that since the Hamiltonian is defined on site-orbital representation, there is no well-defined k-space energy cut-off in the vicinity of the Fermi level for the attractive pairing potential. Therefore, pairing electrons may come from the region which is far away from the four electron pockets. Consequently, the absence of pocket at $\Gamma$ point may induce ambiguity for anisotropic $s$ wave pairing state in a framework of BCS-type pairing scheme.
\begin{table}[!b]
\caption{Values(in unit: $10^{-1} eV$) of orbital-resolved
$d_{x^2-y^2}$ wave pairing order parameters(pairing bonds) $\pi_{x, y}$ and $\sigma_{x, y}$ as
defined in Fig. \ref{fig12} for site (3,3) for zero-field SC and
vortex states. The spin and site indices have been omitted.}\label{table:op}
\begin{center}
\begin{tabular}{lllllllllll}
 \hline\hline
&     ~~& Zero-field SC state      ~~& Vortex state  \\
 \hline
&$\Delta_{xz}(\sigma_{x})$         ~~&(0.43,  0.43)      ~~&(-0.12, 0.58)  \\
&$\Delta_{xz}(\pi_{y})$            ~~&(0.032,  0.032)    ~~&(-0.34, 0.11)  \\
&$\Delta_{yz}(\pi_{x})$            ~~&(-0.032, -0.032)  ~~&(-0.12, 0.34)   \\
&$\Delta_{yz}(\sigma_{y})$         ~~&(-0.43,  -0.43)      ~~&(-0.58, 0.11)  \\
&$\Delta_{xy}(\hat{x})$            ~~&(0.17,  0.17)      ~~&(0.091, 0.24)  \\
&$\Delta_{xy}(\hat{y})$            ~~&(-0.17, -0.17)    ~~&(-0.24, 0.096)  \\
\hline\hline
\end{tabular}
\end{center}
\end{table}
\begin{figure}[h]
\begin{center}
\includegraphics[width=240pt]{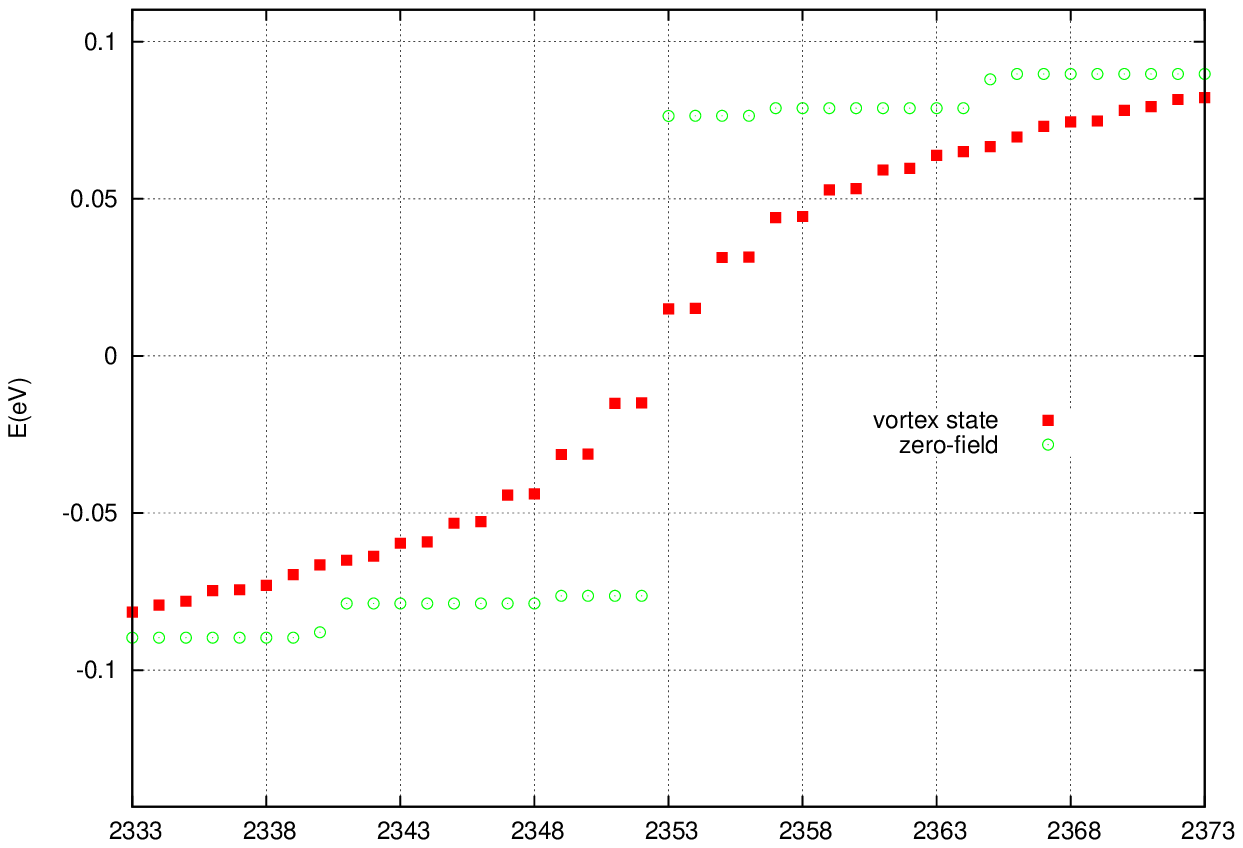}
\caption{(color online) Eigenvalues of BdG equation at around Fermi level in the cases of $d_{x^2-y^2}$ wave pairing state for zero-field states, shown in green circles, and vortex states, shown in red squares, respectively.} \label{fig13}
\end{center}
\end{figure}

\begin{figure}[t]
\begin{center}
\subfigure[]{\includegraphics[width=120pt]{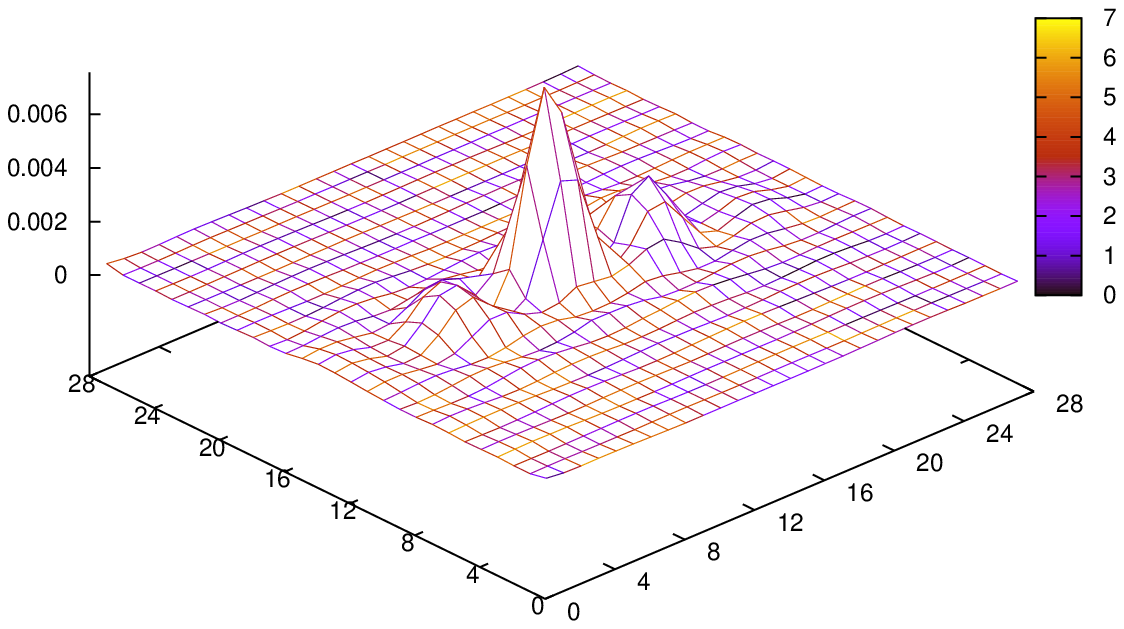}}
\subfigure[]{\includegraphics[width=120pt]{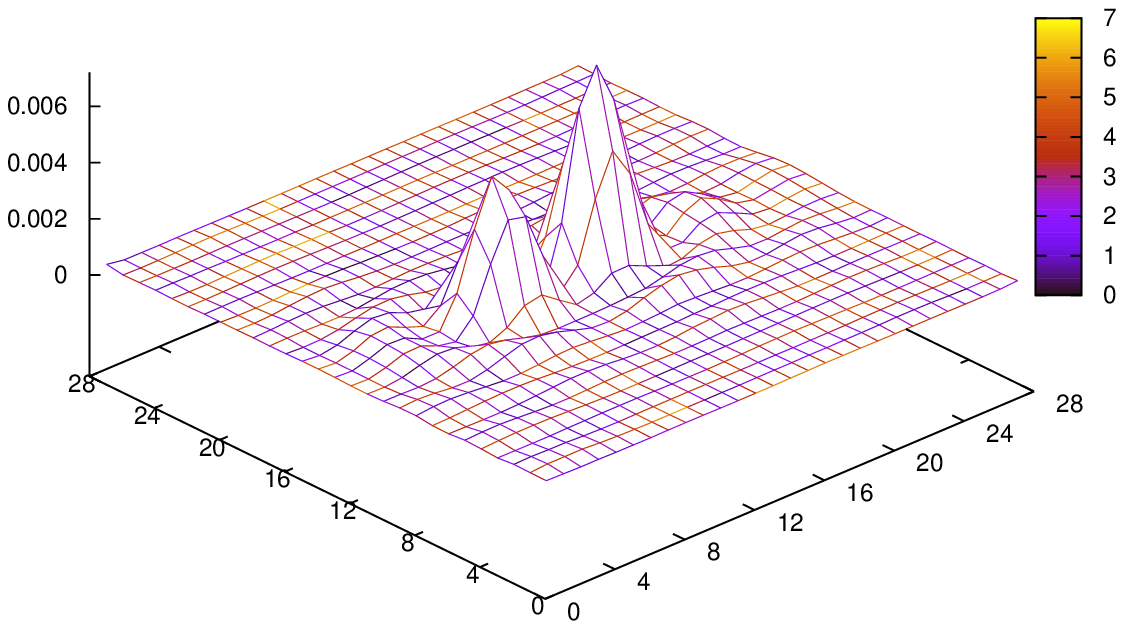}}
\subfigure[]{\includegraphics[width=120pt]{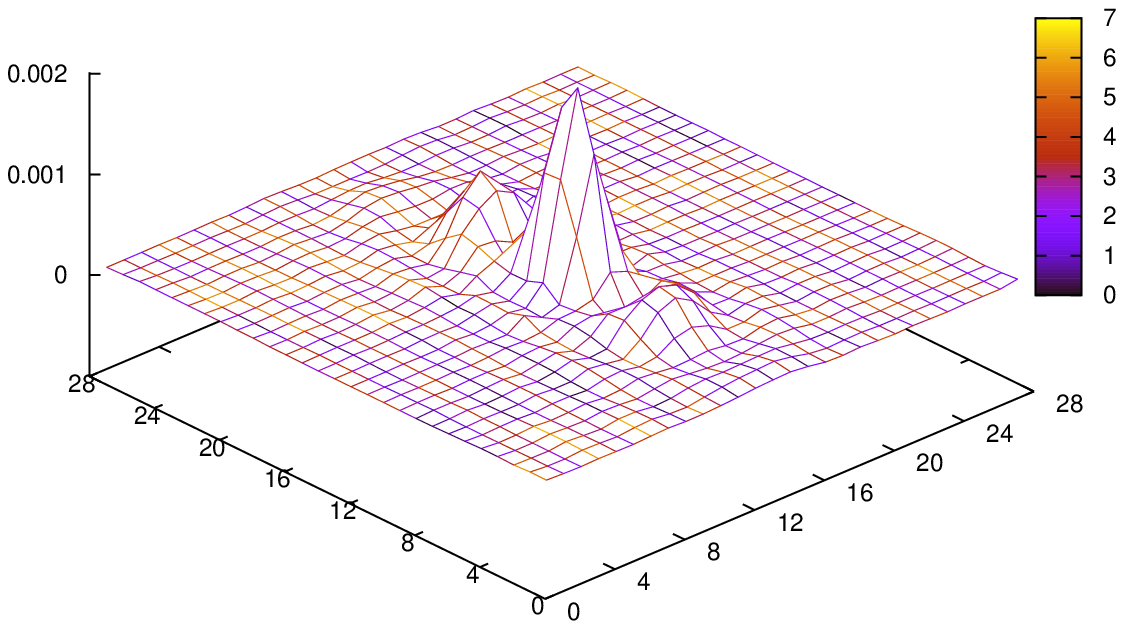}}
\subfigure[]{\includegraphics[width=120pt]{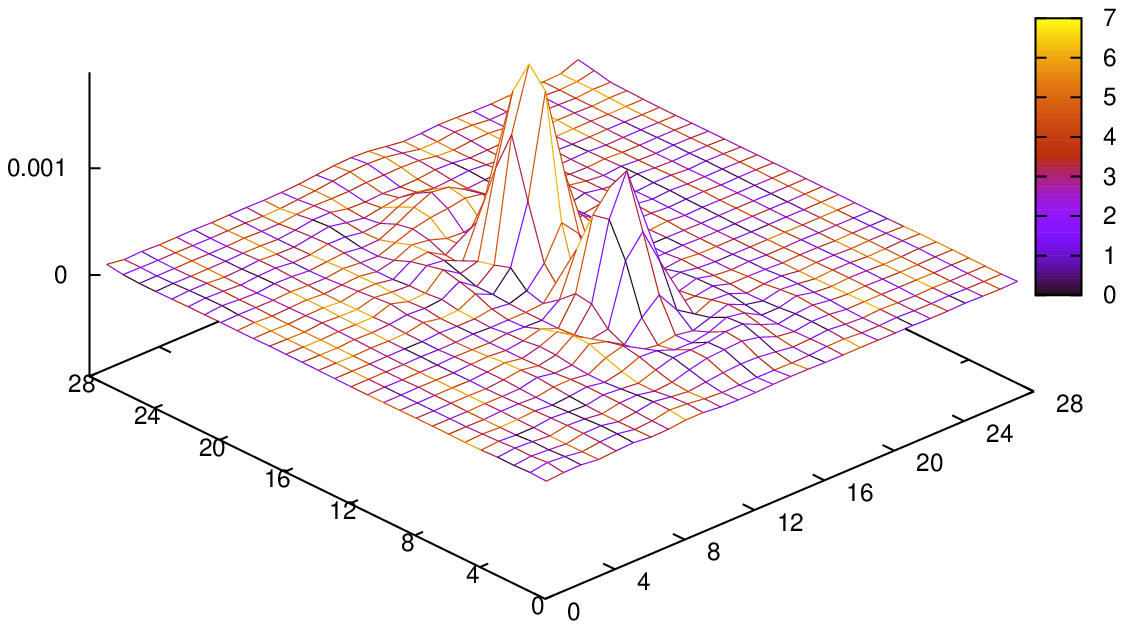}}
\subfigure[]{\includegraphics[width=120pt]{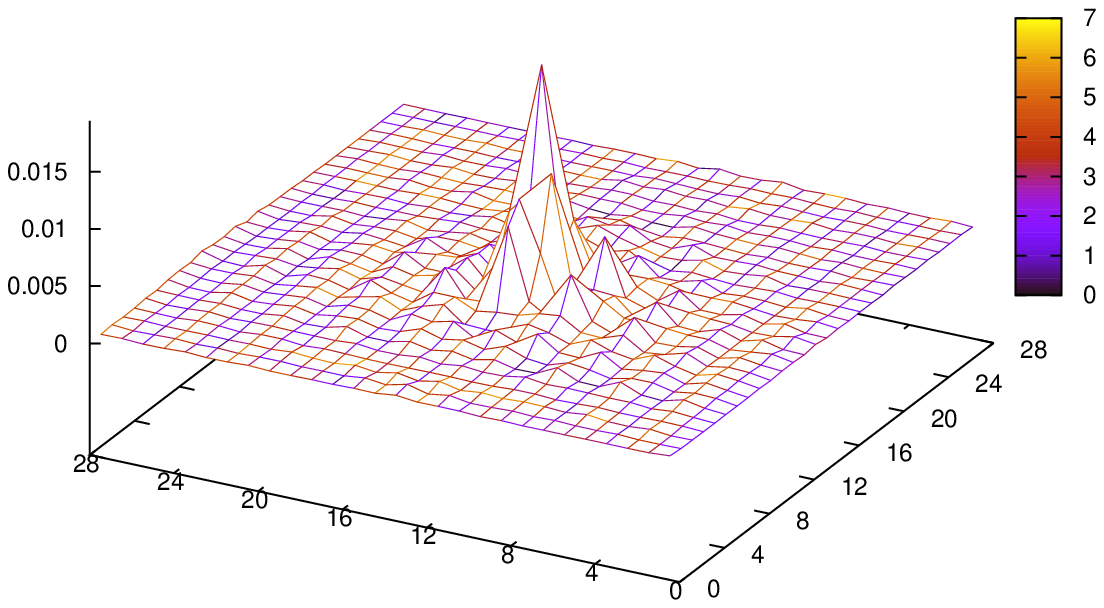}}
\subfigure[]{\includegraphics[width=120pt]{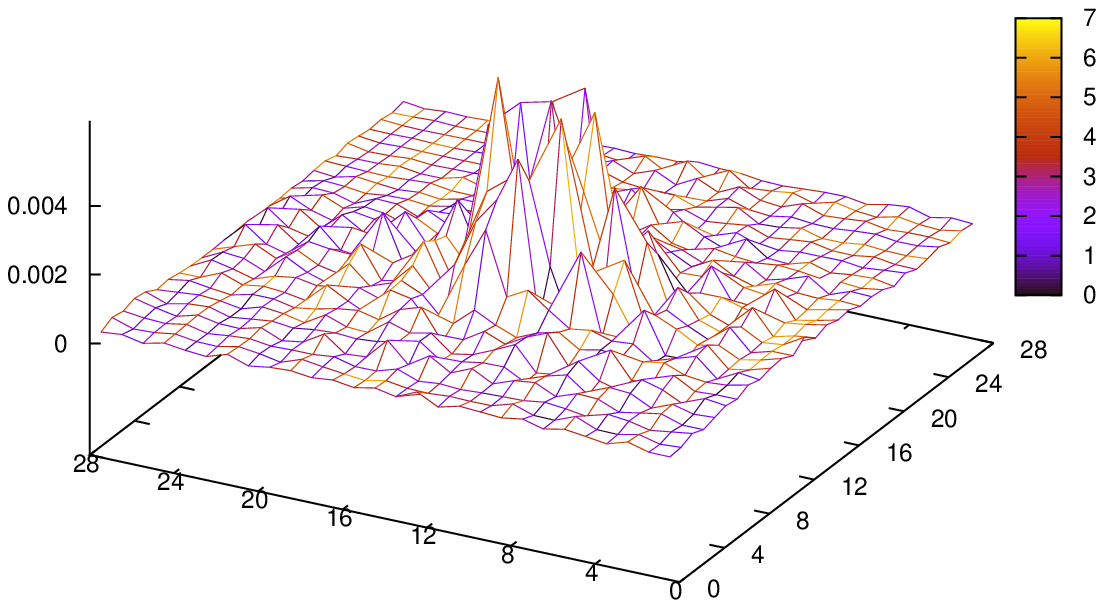}}
\caption{(color online) Amplitudes and phases(color mapping) of
quasi-particle wavefunctions $u^{n}_{i\alpha\uparrow\uparrow}$ and
$v^{n}_{i\alpha\downarrow\uparrow}$ for $d_{x^2-y^2}$ wave pairing
state of index n=2353 for $d_{xz}$ orbital (a) and (b), $d_{yz}$
orbital (c) and (d), and index n=2354 for $d_{xy}$ orbital (e) and
(f), respectively.} \label{fig14}
\end{center}
\end{figure}

\begin{figure*}[]
\begin{center}
\subfigure[]{\includegraphics[width=240pt]{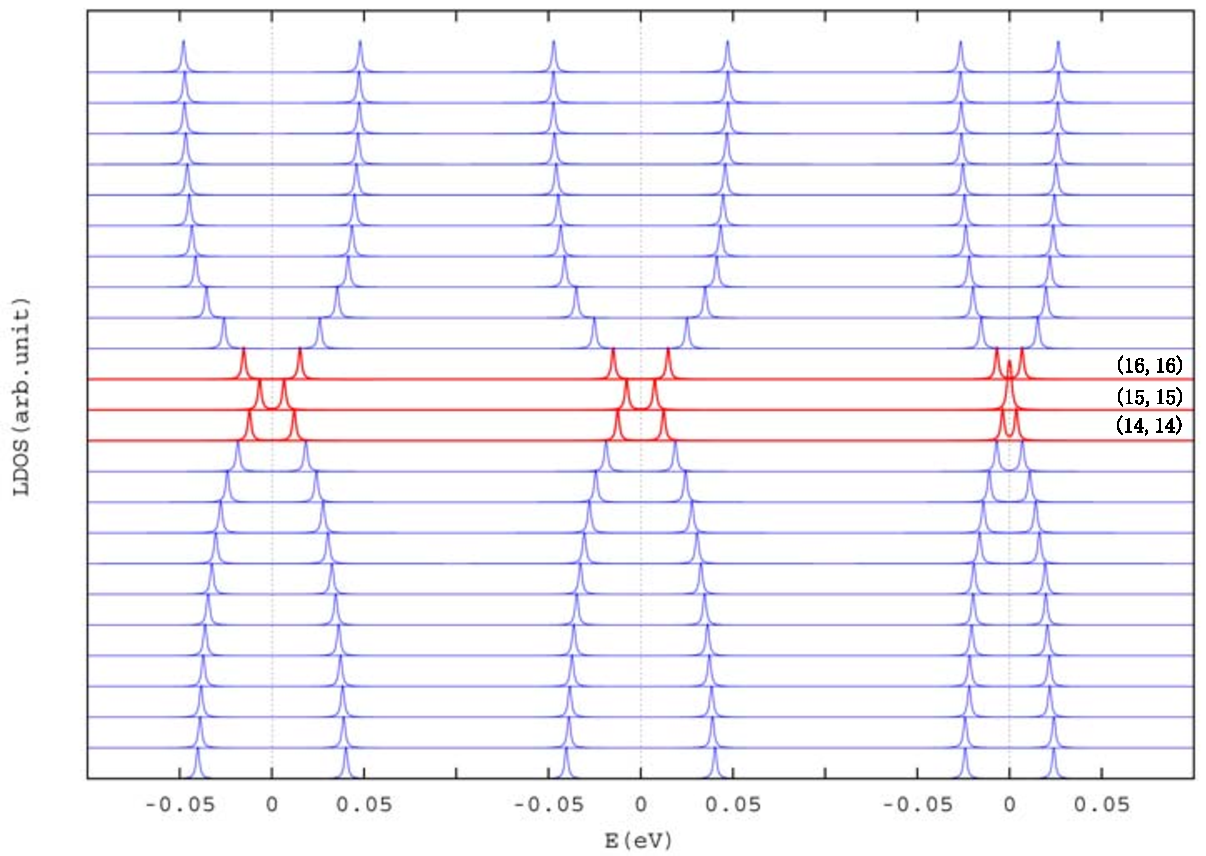}}
\subfigure[]{\includegraphics[width=240pt]{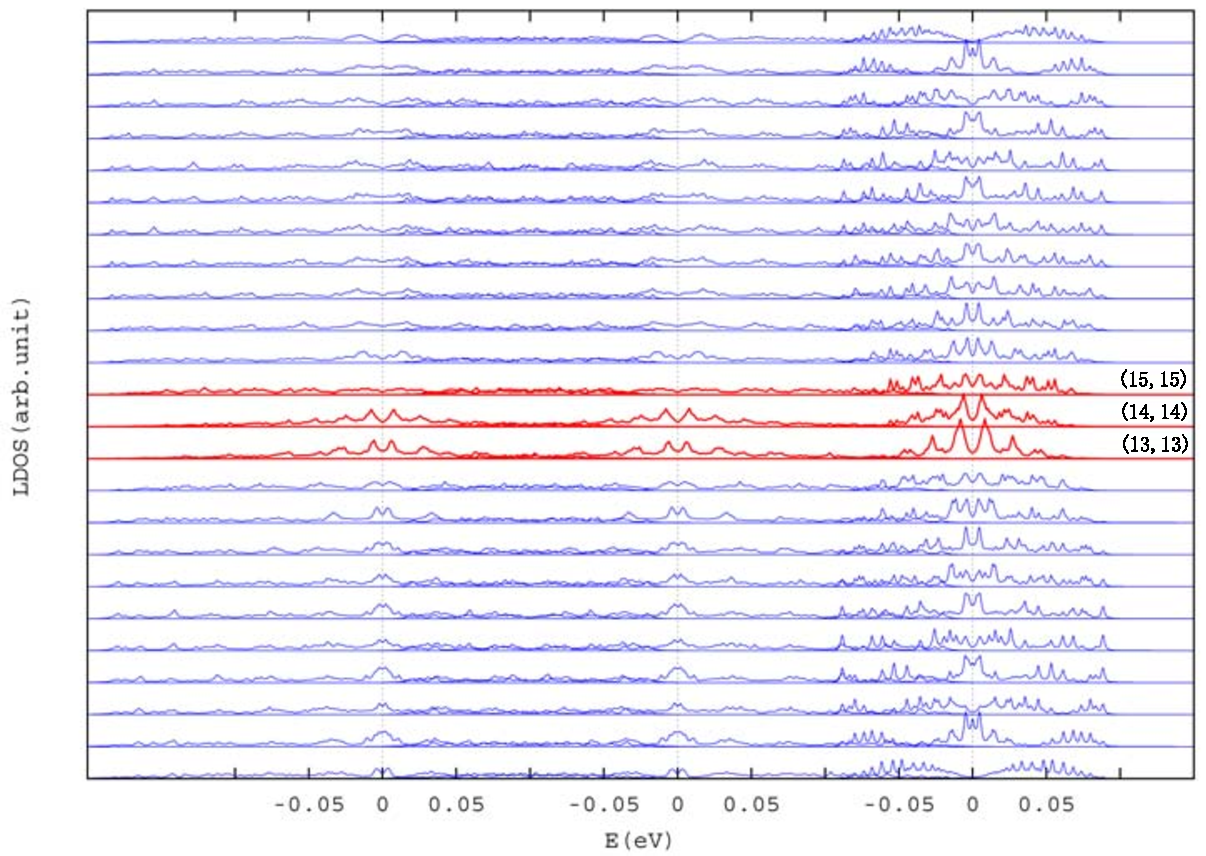}}
\subfigure[]{\includegraphics[width=240pt]{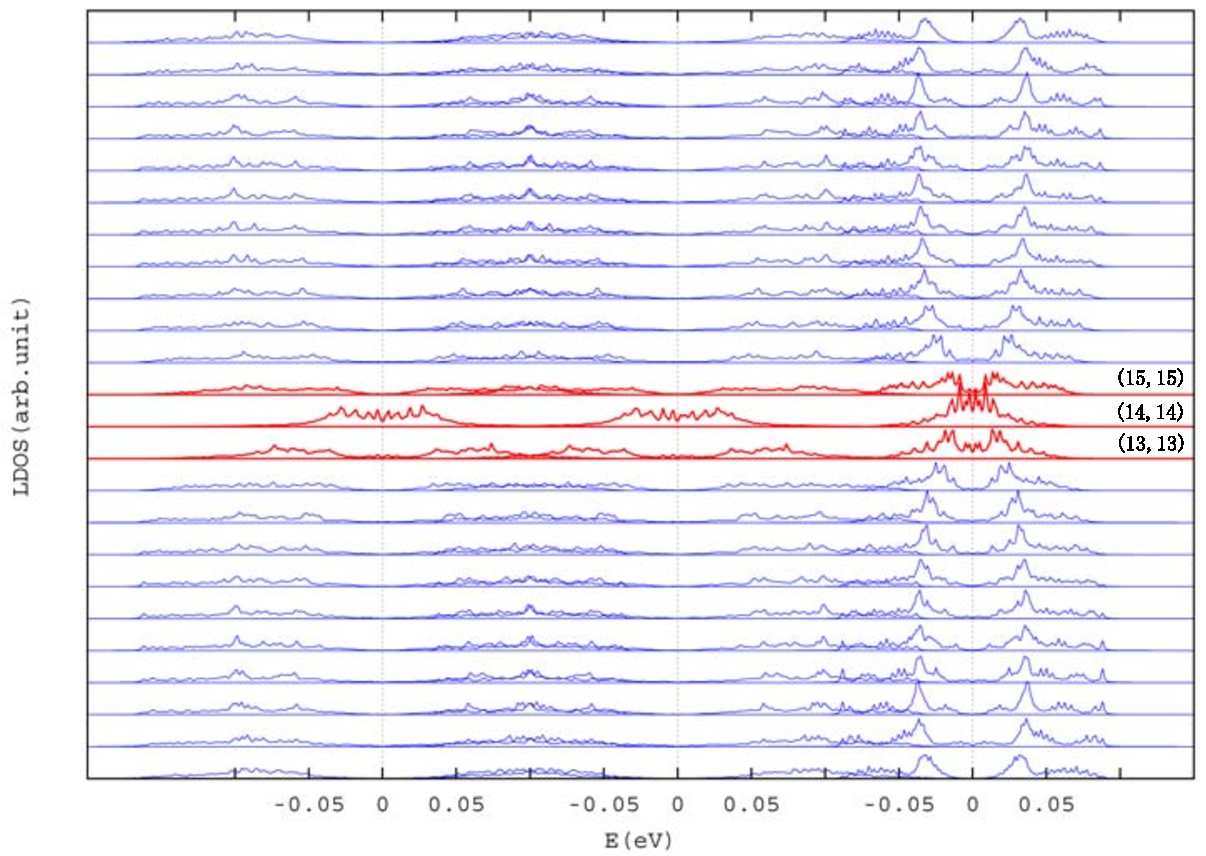}}
\subfigure[]{\includegraphics[width=240pt]{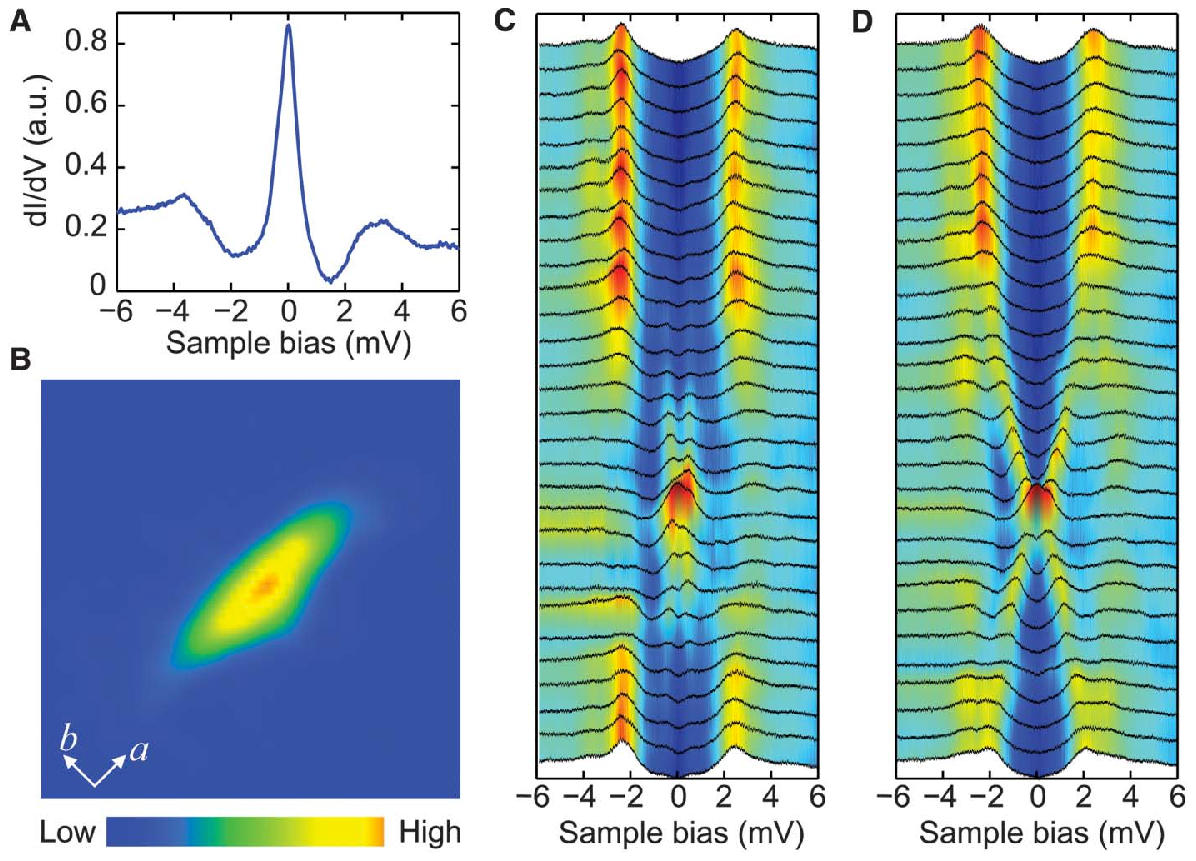}}
\caption{(color online) Orbital-resolved LDOS along off-diagonal
line from site (3,3) $\rightarrow$ (26,26). Each subfigure from left
to right is LDOS for $d_{xz}$, $d_{yz}$, and $d_{xy}$ orbitals in
the cases of isotropic $s$ wave (a), anisotropic $s$ wave (b), and
$d_{x^2-y^2}$ wave (c) pairing states, respectively. The Fermi level
has been set to zero and sites in vortex region have been
highlighted in red. The vortex core states from Scanning tunneling spectroscopy(STS) (d). STS on the center of a vortex core \textbf{A}. Zero-bias conductance map for a single vortex at 0.4 K and 1 T magnetic field \textbf{B}. Tunneling conductance curves measured at equally spaced (2 nm) distances along $\hat{a}$ axis \textbf{C} and $\hat{b}$ axis \textbf{D}. Reprinted figure with permission from C. L. Song \emph{et al.}, Science \textbf{332}, 1410 (2011)\cite{C. L. Song}. Copyright 2011 by American Association for the Advancement of Science (AAAS).} \label{fig15}
\end{center}
\end{figure*}

\begin{figure}[!t]
\begin{center}
\subfigure[]{\includegraphics[width=120pt]{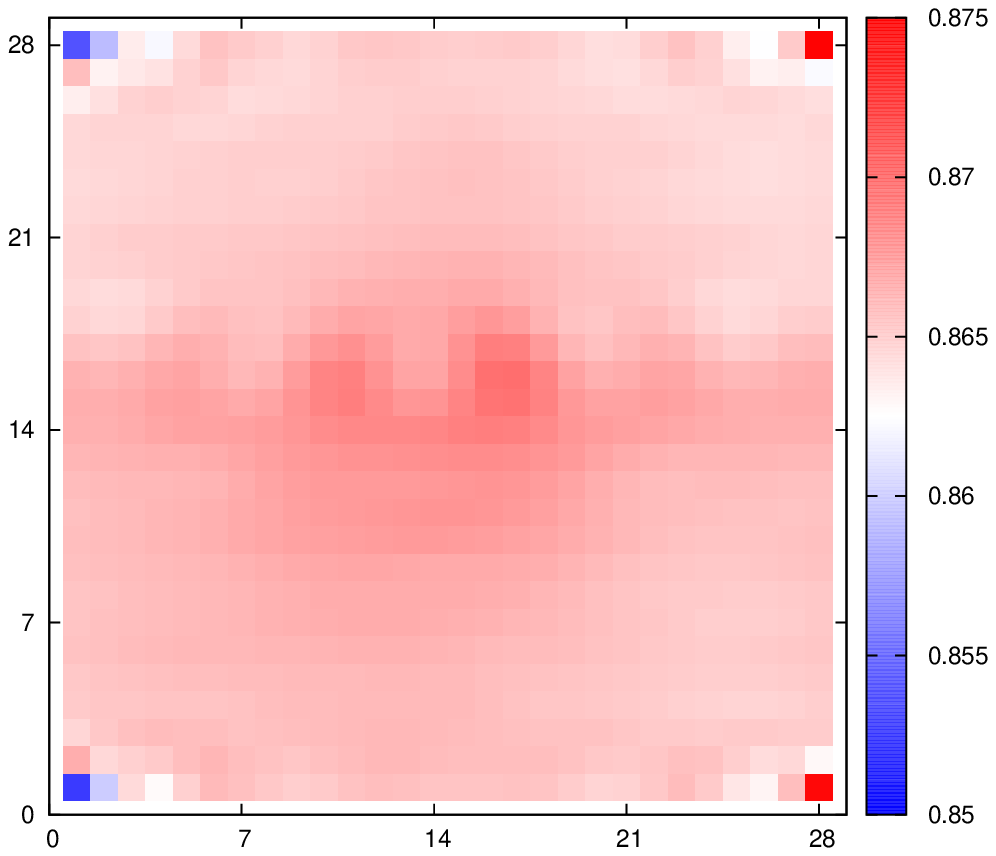}}
\subfigure[]{\includegraphics[width=120pt]{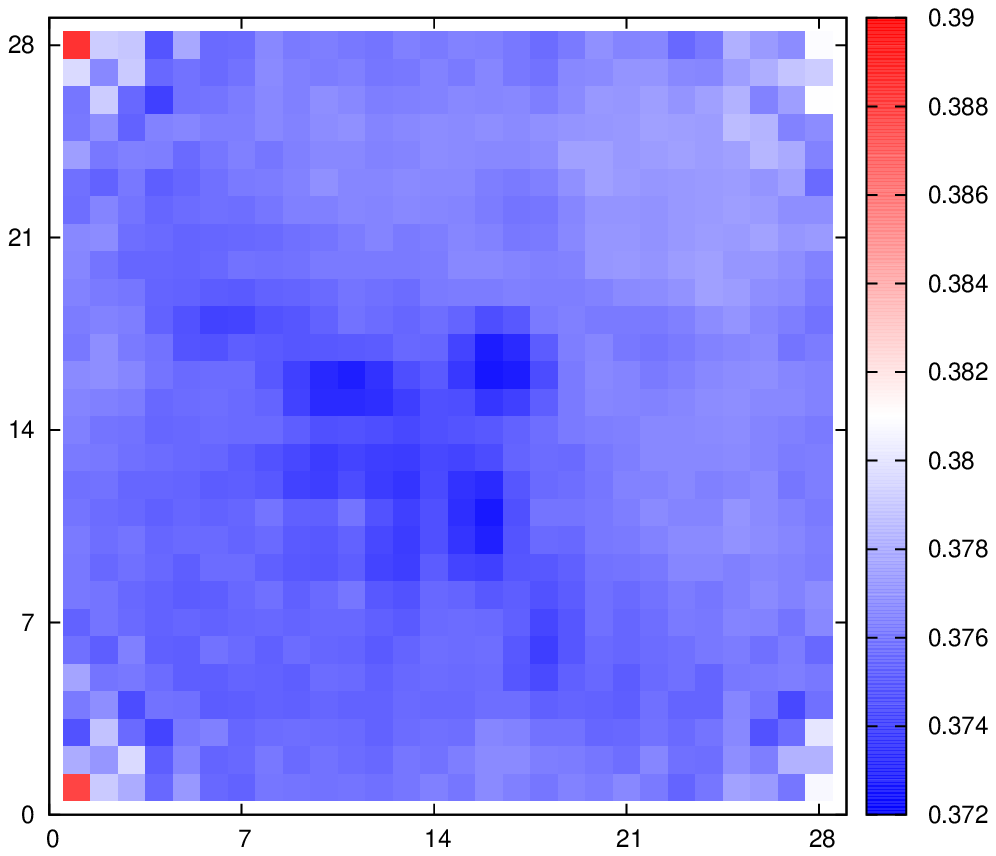}}
\subfigure[]{\includegraphics[width=120pt]{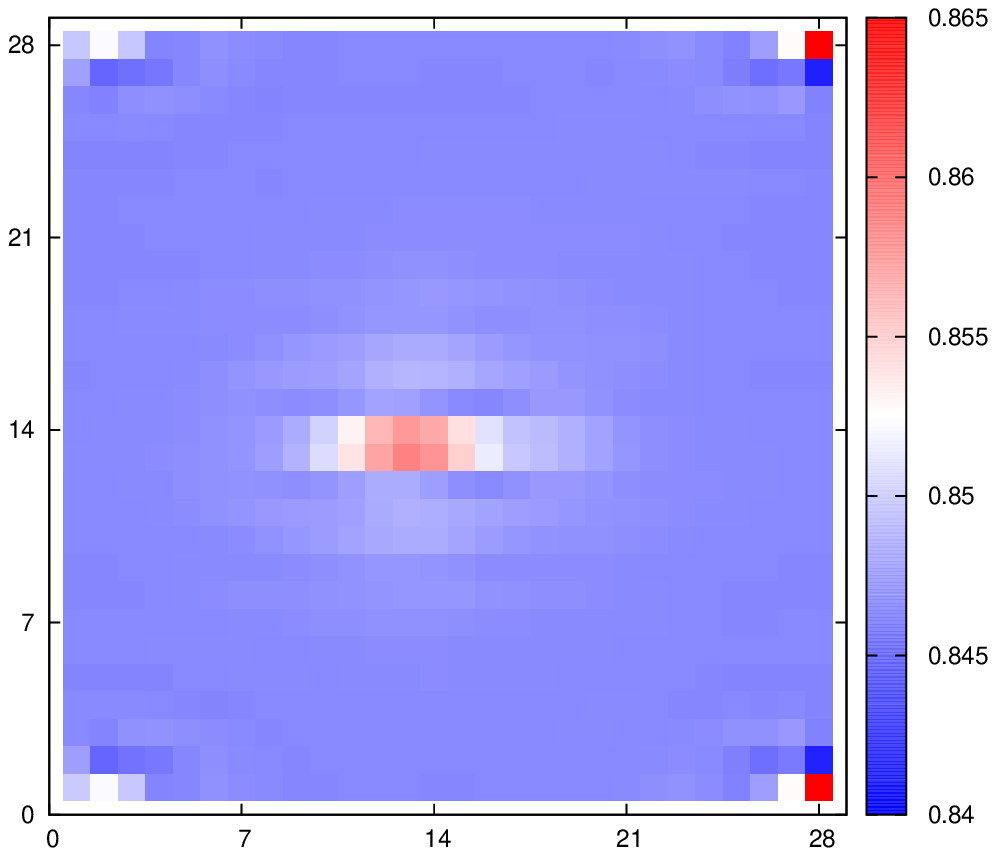}}
\subfigure[]{\includegraphics[width=120pt]{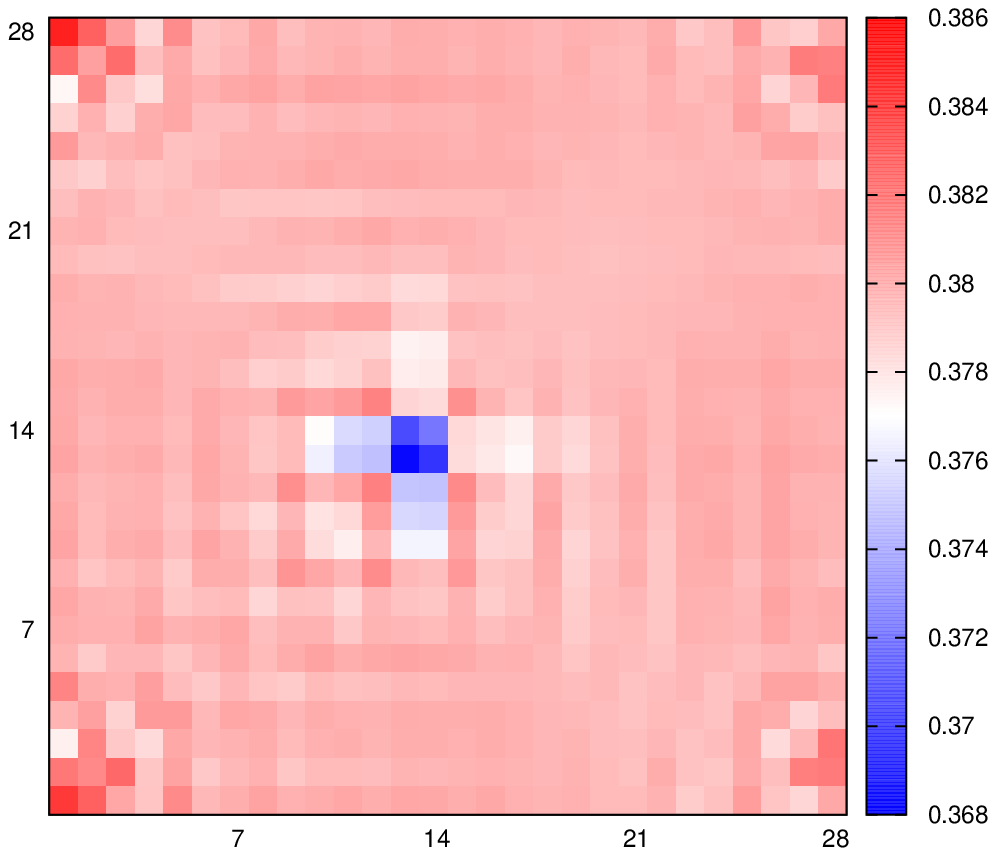}}
\subfigure[]{\includegraphics[width=120pt]{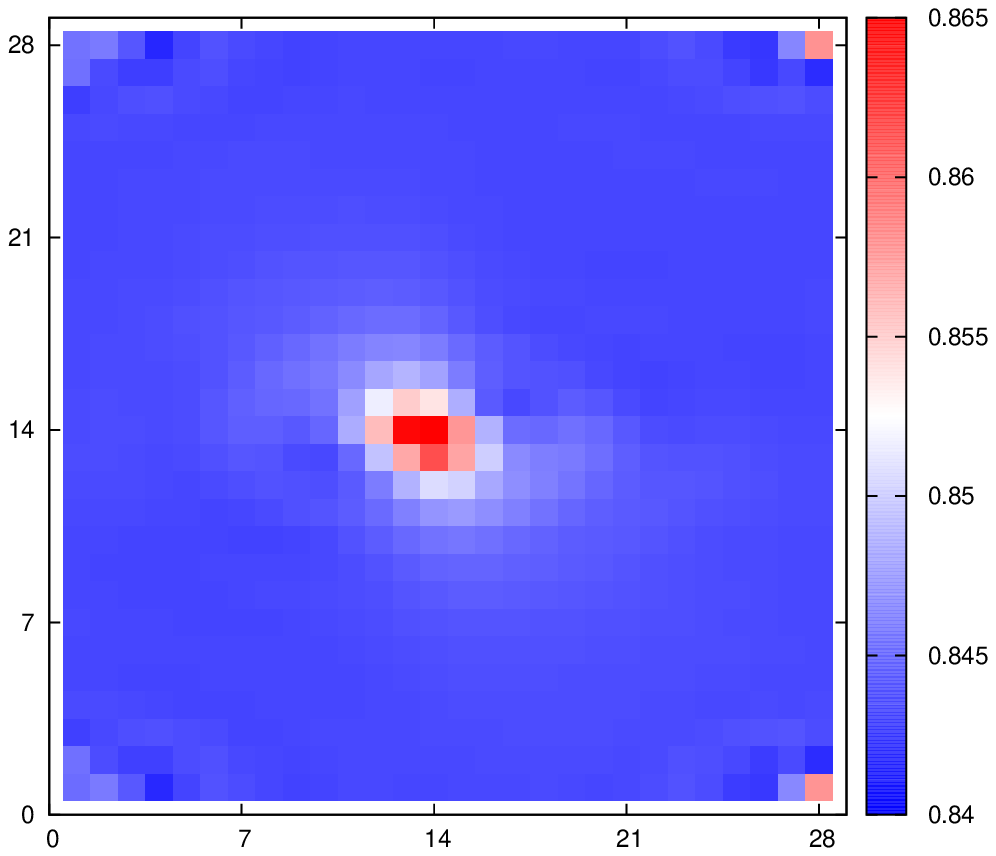}}
\subfigure[]{\includegraphics[width=120pt]{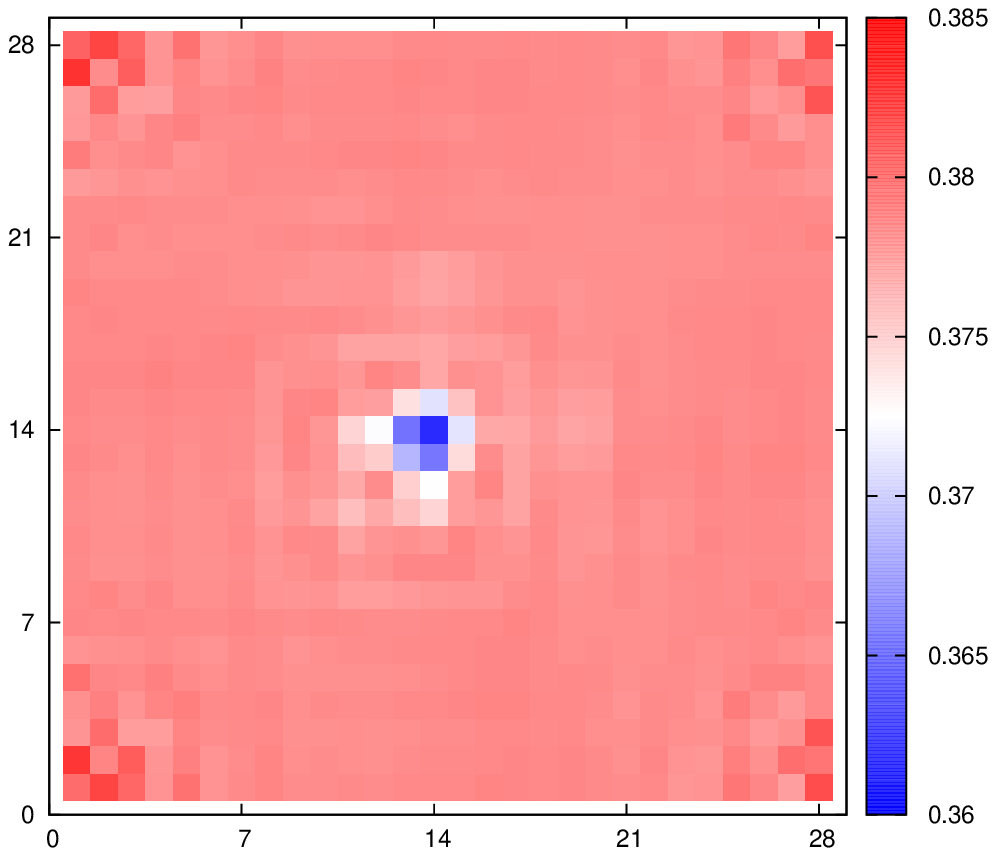}}
\caption{(color online) Orbital-resolved electron density for
$d_{xz}$ and $d_{xy}$ orbitals for $s$ wave (a) and (b) , anisotropic $s$
wave (c) and (d), and $d_{x^2-y^2}$ wave (e) and (f) vortices,
respectively. Electron density for $d_{yz}$ orbital in the cases of
different pairing symmetries are same as $d_{xz}$ orbital.}
\label{fig16}
\end{center}
\end{figure}

\begin{figure*}[]
\begin{center}
%\subfigure[]{\includegraphics[width=300pt]{fig17a.eps}}
\subfigure[]{\includegraphics[width=300pt]{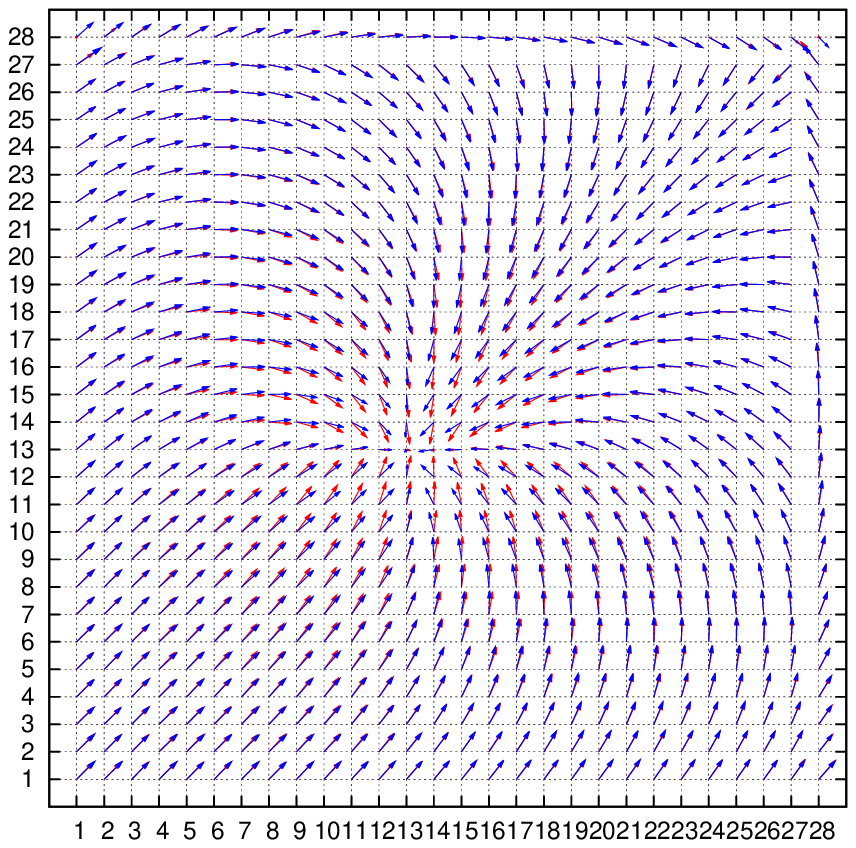}}
\subfigure[]{\includegraphics[width=300pt]{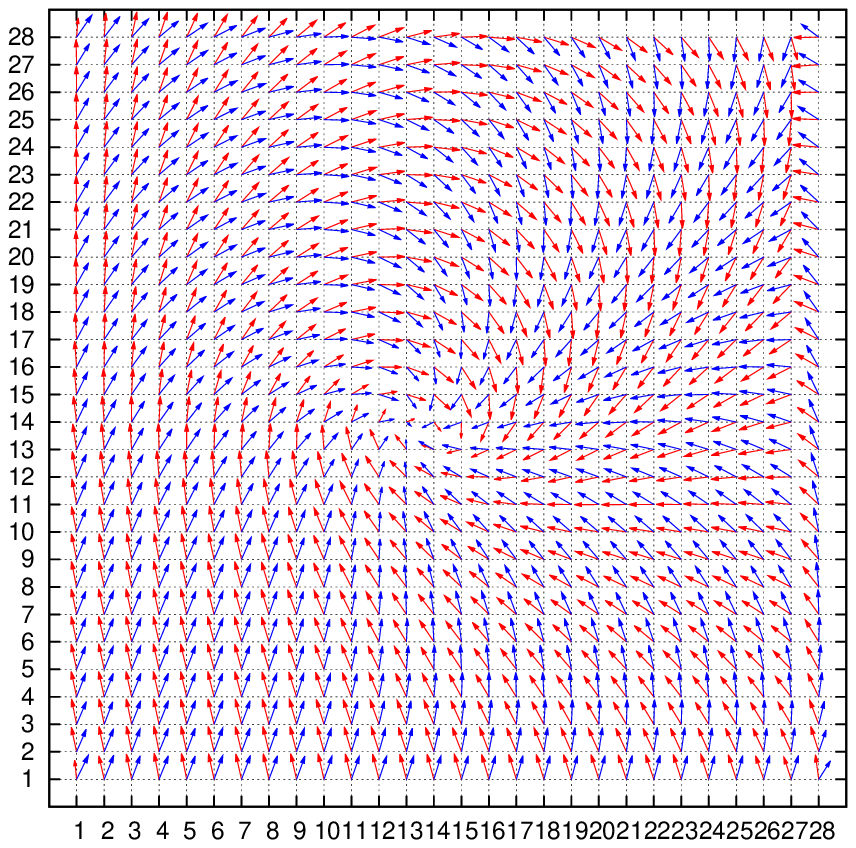}}
\caption{(color online) Phase difference between $d_{xz}$(red arrows) and $d_{xy}$(blue arrows) orbital-resolved vortices for anisotropic s wave (a) and $d_{x^2-y^2}$ wave (b) pairing states. We note that $d_{yz}$ orbital vortices show the same symmetry as $d_{xz}$ does and the results of which is omitted. This phase difference is trivial in (a) due to same $G_5$ symmetry and is smaller than $\pi/4$ from our numerical results due to spatial anisotropy of t$_{2g}$ orbitals in (b). The arrows are amplified from original data to obtained enough resolution.}
\label{fig17}
\end{center}
\end{figure*}

Results of $d_{x^2-y^2}$ wave vortices are different from $A_{1g}$ vortices discussed above in many aspects. The orbital anisotropy dominates the vortex structures. Fig. \ref{fig10} and \ref{fig11} show the amplitudes and phase distribution of $d_{x^2-y^2}$ wave pairing bonds for each orbitals. It has been pointed out in previous section that the symmetry of band structure gives constraints to symmetry of pairing states. A strong
hybridization of $d_{xz}$ and $d_{yz}$ orbitals, as shown in PDOS in Fig. \ref{fig1}, results in a re-defined $d_{x^2-y^2}$ wave pairing state, as shown in Fig. \ref{fig12}, since the wavefunctions of these two orbitals transform under action of generator $C_{4z}$ as
\begin{flalign}
&C_{4z}|d_{xz}\rangle = |d_{yz}\rangle \\ \nonumber
&C_{4z}|d_{yz}\rangle = -|d_{xz}\rangle
\end{flalign}
while $d_{xy}$ orbital does not mix with them under such a transformation. Here we give an example of numerical results of order parameters for each orbitals on site (3,3), as shown in Table \ref{table:op}. In zero-field case, phase difference of $e^{i\pi}$ is observed between $\pi_{x}$ and $\pi_{y}$, $\sigma_{x}$ and $\sigma_{y}$ bonds, which are defined on different orbitals, whereas in vortex states, such a phase will undergo a gauge modification which is induced by magnetic field. The winding structures shown in Fig. \ref{fig10} and \ref{fig11} for different orbitals share this common feature for all order parameters defined on entire magnetic unit cell. For $d_{xy}$ orbital, $G^{*}_{5}$ vortices which are defined on pairing bonds $\Delta_{xy}(\hat{x})$ and $\Delta_{xy}(\hat{y})$ are of sink- and source-type, respectively, because the order parameters change sign as they transform according to $B_{1g}$ irreducible unitary representation. In the presence of magnetic field, the sign change of $d_{x^2-y^2}$ wave pairing symmetry, along with the orbital-hybridized order parameters together give rise to a $G^{*}_{6}$ winding structure for $d_{xz/yz}$ orbitals, which seems like a solenoidal vector field. Such phase difference has been observed between $\Delta_{xz}(\sigma_{x})$ as shown in Fig. \ref{fig10} (b) and $\Delta_{yz}(\sigma_{y})$ as shown in Fig. \ref{fig10} (h), and also between $\Delta_{xz}(\pi_{y})$ as shown in Fig. \ref{fig10} (d) and $\Delta_{yz}(\pi_{x})$ as shown in Fig. \ref{fig10} (f). Among 17(positive) in-gap states associated with orbital-resolved $d_{x^2-y^2}$ wave vortices as shown in Fig. \ref{fig13}, the wavefunctions of eigenvalue $|\epsilon_{2353\uparrow}\rangle$ for $d_{xz}$ and $d_{yz}$ orbitals, and $|\epsilon_{2354\uparrow}\rangle$ for $d_{xy}$ orbital are shown in Fig. \ref{fig14}. The particle-hole asymmetry is evidently for $d_{xz}$ and $d_{yz}$ orbitals in that the bound states have three peaks for particle part and two peaks for hole part. The most localized vortex bound state has been observed for $d_{xy}$ orbital for particle part. The discontinuity of phase distribution on boundary of magnetic unit cell is also observed in $d_{x^2-y^2}$ wave vortices due to next nearest neighbor site pairing.

In order to have an understanding of distinction of vortex states between different pairing symmetries, we compare the orbital-resolved LDOS along off-diagonal line approaching vortex core and then away from it. Fig. \ref{fig15} (a) shows results for isotropic $s$ wave, where vortices of $d_{xz}$ and $d_{yz}$ orbitals pinning at site (15,15) are characterized by symmetrically located two peaks, while the vortex of the $d_{xy}$ orbital shows single peak. The two peaks start to shrink towards Fermi level from site (7,7) and then transit back to SC coherence peak at site (19,19), therefore the isotropic $s$ wave vortices have a relative large core region. Another characteristic of $s$ wave vortices is that the LDOS shows no Landau oscillation due to on-site pairing. However, since the wavefunctions of all the in-gap states for both positive and negative eigenstates are not localized, such vortex states may not be favored in FeSe superconductor. Additionally, the particle-hole symmetry protects electron density from accumulating or losing in the vortex core region as shown in Fig. \ref{fig16} (a) and (b).

For anisotropic $s$ wave vortices, an oscillation in LDOS for $d_{xy}$ orbital has been observed, as shown in Fig. \ref{fig15} (b). The LDOS at the Fermi level varies alternately from zero at site (3,3) to a finite value, and then oscillates until being stabilized at the core center. At site (13,13) and (14,14) the core states always manifest themselves as double peaks, which is different from the results of isotropic $s$ and $d_{x^2-y^2}$ wave vortices. Such an alternating appearance of bound states at Fermi level may come from the fact that for $d_{xz}$ and $d_{yz}$ orbitals, as shown in Fig. \ref{fig16} (c), there are charge density accumulations, while for $d_{xy}$ orbital electron density is suppressed inside the core region, as shown in Fig. \ref{fig16} (d).

Finally, Fig. \ref{fig15} (c) shows LDOS of $d_{x^2-y^2}$ wave vortices. It has been found that for $d_{xz/yz}$ orbitals, the vortex bound states are exactly localized at site (14,14), with stable SC coherence locating at around $\pm0.05$ eV, and for $d_{xy}$ orbital the core region includes site (13,13). Similarly to the cases of anisotropic $s$ wave vortices, charge accumulation on $d_{xz/yz}$ orbitals and loss on $d_{xy}$ orbital have been observed as shown in Fig. \ref{fig16} (e) and (f), which indicates signature of charged vortex core states. However, no particle density oscillation appears in LDOS spectrum. The superposition of in-gap bound states at site (14,14) in Fig. \ref{fig15} (c) contributed from different orbitals reproduces a peak at the center of a vortex, which resembles the results of STM observation as shown in Fig. \ref{fig15} (d) \cite{C. L. Song}. The fact that the oscillation of LDOS in the case of anisotropic $s$ wave vortices is not observed in STM measurement, and the bound sates of isotropic $s$ wave vortices are extended makes us conclude that the vortex structures observed by STM may be of $d_{x^2-y^2}$ wave feature.

We have noted that the self-consistent calculation gives different winding structures of vortex states with respect to different pairing symmetries. However, isotropic $s$ and anisotropic $s$ wave vortices share a common winding structure, which is characterized by a sink-type core state. But in the case of $d_{x^2-y^2}$ wave pairing, vortices contributed from $d_{xz/yz}$ orbitals show a phase distribution as a solenoidal vector field, whereas $d_{xy}$ orbital shows sink- and source-type winding structures. Topologically, all these vortices correspond homotopy group $\pi_{1}[U(1), x_{0}]=\mathbb{Z}$. As shown in Table \ref{table:wd}, the orbital-resolved $s$ and anisotropic $s$ wave vortices belong to same symmetry group $G_5$\cite{M.Ozaki1}, and $d_{x^2-y^2}$ wave pairing symmetry has $d_{xz/yz}$ orbital vortices belonging to $G^{*}_{6}$ group and $d_{xy}$ orbital vortices $G^{*}_{5}$. Such results reveal that the local surgery, i.e., the continuous transformation between element within same homotopic class, is actually carried out by a gauge transformation, or equivalently the co-representation transformation between $G^{*}_{5}$ and $G^{*}_{6}$\cite{M.Ozaki1}. The pairing bonds of orbital-resolved $d_{x^2-y^2}$ wave vortices defined on each orbitals have a phase difference which is smaller than $\pi$ in the vicinity of the vortex core. Far away from the vortex core, it approaches to $\pi$ as the usual $d_{x^2-y^2}$ wave pairing states in the case of zero magnetic field\cite{P.I.Sonininen}. We have noted that mathematically same reference point $i_{0}$ in real space can be mapped to difference reference points $x_{0}$ and $x_{1}$ in U(1) SC order parameter space which manifests themselves as different absolute phase values, while the homotopic classes generated by $x_{0}$ and $x_{1}$ correspond to same homotopy group $\pi_{1}[U(1), \forall x]=\mathbb{Z}$. This is why $G_{5}(G^{*}_{5})$ and $G_{6}(G^{*}_{6})$ subgroups have a local relative phase difference. In our numerical calculations, vortices in the case of anisotropic s wave pairing state, as shown in Fig. \ref{fig17} (a), show the same $G_5$ symmetry for different orbitals which results in a trivial phase difference. Minor phase differences appear in the center region of the magnetic unit cell due to the amplification of the lengths of the arrows when we plot the figure. From group theoretical derivation, there is a phase difference of $\pi/4$ between $G^{*}_{6}$ symmetry, defined on $d_{xz, yz}$ orbitals, and $G^{*}_{5}$ symmetry, defined on $d_{xy}$ orbital, respectively, in the case of $d_{x^2-y^2}$ wave pairing state. We have observed such a fixed phase difference from our numerical calculation as shown in Fig. \ref{fig17} (b). The observed phase difference is smaller than $\pi/4$ due to spatial anisotropy of $t_{2g}$ orbital wavefunctions. As a stable topological defect, one remarkable phenomenon is that the fixed relative phase difference is essentially a signature of all the order parameters defined on the entire magnetic unit cell which is in reality feature originated from topological property of U(1) gauge field. Physically, even though we have only included the intra-orbital pairings, the inter-orbital hoppings between $d_{xz/yz}$ and $d_{xy}$ orbitals are responsible for this phase lock-in phenomenon. From a viewpoint of quasiparticle interference, the orbital degree of freedom actually gives rise to an orbital-resolved interfered phase distribution. Without loss of generality we propose that such a phase difference between $d_{xz/yz}$ and $d_{xy}$ orbital vortices can in principle be observed experimentally which is independent upon specific gauge choice and consequently a physical manifestation of $d_{x^2-y^2}$ wave pairing states. We have confirmed that $d_{xz/yz}$ vortices always have $G^{*}_{6}$ symmetry even if we carry out an artificial gauge transformation where the relative phase of $d_{xz/yz}$ and $d_{xy}$ orbital hoppings in band structure are changed as
\begin{eqnarray}
&t_{\sigma\sigma}(i\alpha,j\beta)\rightarrow
t_{\sigma\sigma}(i\alpha,j\beta)e^{i\theta_{\alpha\beta}}
\end{eqnarray}
where $\theta_{\alpha\beta}$ is set to $\frac{\pi}{4}$ or $-\frac{3\pi}{4}$. which is consistent with the co-representation transformaton\cite{M.Ozaki1}. The resultant winding pattern of $d_{xz/yz}$ orbital vortices remain unchanged, while $d_{xy}$ orbital vortex changes obviously. It turns out that if we set an equal on-site atomic energy, such phase difference of $\frac{\pi}{4}$ disappears.

\section{Summary}\label{sec5}

In summary, using a three-orbital model, we present a comprehensive investigation of single vortex core states in FeSe superconductors by means of BdG theory. The numerical results have been classified by invariant subgroups of magnetic translation group. It turns out that isotropic $s$ and anisotropic $s$ wave pairing symmetries give rise to $G_{5}$ vortex states. $G^{*}_{6}$ vortex states are obtained for $d_{xz/yz}$ orbitals due to orbital hybridization, and $G^{*}_{5}$ vortex states for $d_{xy}$ orbital in the case of $d_{x^2-y^2}$ wave pairing. By analyzing behavior of orbital-resolved quasi-particle wavefunctions and LDOS, and by comparing the results with STM observation, we propose that $d_{x^2-y^2}$ wave vortices are most likely candidate. The phase difference of $\frac{\pi}{4}$ in terms of winding structures between hybridized $d_{xz/yz}$ orbitals and $d_{xy}$ orbital can also be testified experimentally as a signature of $d_{x^2-y^2}$ wave pairing symmetry in FeSe superconductors.

\section{Acknowledgement}
We thank Y. Chen, Z. J. Yao, H. L. Pang and Z. Z. Yu for inspiring discussions. We acknowledge financial support from The Research Grant Council, University Grant Committee, Hong Kong via GRF Grant No. 706809, National Basic Research Program of China, No. 2014CB921203, and National Science Foundation of China, Grant No. 11274269.

\end{document}